\documentclass[journal,10pt]{IEEEtran}

\usepackage{lipsum}
\usepackage{wrapfig}
\usepackage[usenames, dvipsnames]{color}
\usepackage{graphicx}
\usepackage{pdfpages}
\usepackage{epstopdf}	
\usepackage{amsmath}
\usepackage{mathrsfs}
\usepackage{longtable}
\usepackage{enumerate}
\usepackage{algorithmic}
\usepackage{multirow}
\usepackage{cite}
\newcommand{\ignore}[1]{}
\usepackage{caption}
\usepackage{fix-cm}    
\usepackage{subfig}
\makeatletter
\newcommand\HUGE{\@setfontsize\Huge{50}{60}}
\makeatother  
\ignore{
\makeatletter
\def\ps@headings{%
\def\@oddhead{\mbox{}\scriptsize\rightmark \hfil \thepage}%
\def\@evenhead{\scriptsize\thepage \hfil \leftmark\mbox{}}%
\def\@oddfoot{}%
\def\@evenfoot{}}
\makeatother
\pagestyle{headings}
}

\renewcommand{\baselinestretch}{1}
\newtheorem{theorem}{Theorem}
\newtheorem{property}{Property}

\newtheorem{condition}{Condition}

\newtheorem{lemma}{Lemma}

\newtheorem{problem}{Problem}
\newcommand{\argmax}{\operatornamewithlimits{argmax}}
\begin{document}

\setlength{\belowdisplayskip}{3pt}

\title{\LARGE \bf Achieving Arbitrary
Throughput-Fairness Trade-offs in the
Inter Cell Interference Coordination
with Fixed Transmit Power Problem}

\author{Vaibhav Kumar Gupta, and Gaurav S. Kasbekar}

\IEEEoverridecommandlockouts
%\IEEEpubid{\makebox[\columnwidth]{978-1-4244-8953-4/11/\$26.00~\copyright~2015 IEEE \hfill} \hspace{\columnsep}\makebox[\columnwidth]{ }} 

\maketitle
{\renewcommand{\thefootnote}{} 
\footnotetext{V. K. Gupta and G. S. Kasbekar are with Department of Electrical
Engineering, Indian Institute of Technology (IIT) Bombay. Their email
addresses are \{vaibhavgupta, gskasbekar\}@ee.iitb.ac.in.

A preliminary version of this paper appeared in the proceedings of the NETGCOOP 2018 conference~\cite{netgcoop}.
}
\vspace{-1em}
\begin{abstract}
We study the problem of  inter cell interference coordination (ICIC) with fixed transmit power in OFDMA-based cellular networks, in which each base station (BS) needs to decide as to which subchannel, if any, to allocate to each of its associated mobile stations (MS) for data transmission. In general, there exists a trade-off between the total throughput (sum of throughputs of all the MSs) and fairness under the allocations found by resource allocation schemes. We introduce the concept of $\tau-\alpha-$fairness by modifying the  concept of $\alpha-$fairness, which was earlier proposed in the context of designing fair end-to-end window-based congestion control protocols for packet-switched networks. The concept of $\tau-\alpha-$fairness allows us to achieve arbitrary trade-offs between the total throughput and degree of fairness by selecting an appropriate value of $\alpha$ in $[0,\infty)$. We show that for every $\alpha \in [0,\infty)$ and every $\tau > 0$, the problem of finding a $\tau-\alpha-$fair allocation  is NP-Complete. Further, we show that for every $\alpha \in [0, \infty)$, there exist thresholds such that if the potential interference levels experienced by each MS on every subchannel are above the threshold values, then the problem can be optimally solved in polynomial time by reducing it to the bipartite graph matching problem. Also, we propose a simple, distributed subchannel allocation algorithm for the ICIC problem, which is flexible, requires a small amount of time to operate, and requires information exchange among only neighboring BSs. We investigate via simulations as to how the algorithm parameters should be selected so as to achieve any desired trade-off between the total throughput and fairness. 
\end{abstract}
\vspace{-1em}
\section{Introduction}
\label{introduction}
%\vspace{-.5em}
The Long Term Evolution (LTE) - Advanced cellular system, which is a 4G technology that is being extensively deployed throughout the world, relies on Orthogonal Frequency Division Multiple Access (OFDMA) technology~\cite{RF:ghosh:fundamentals:of:lte}. Often, an OFDMA-based cellular network is deployed with frequency reuse factor one, \emph{i.e.,} the entire available frequency band can be potentially used in all the cells. Also, the dense deployment of small sized cells in 4G systems to increase the system capacity results in non-negligible inter cell interference~\cite{RF:Kosta},~\cite{RF:eICIC:lopezperez}. 4G can also support a large number of mobile devices simultaneously, which generate high data traffic in each cell, and this results in heavy inter cell interference~\cite{RF:Kosta},~\cite{RF:eICIC:lopezperez}. Therefore, how to combat inter cell interference in these systems is an important question.  Moreover, although it is expected that in 5G cellular networks, \emph{mmWave} spectrum will be used, on which communication will take place using highly directional antennas, which reduces the amount of inter cell interference, it is likely that lower-frequency bands will continue to be used in the future (\emph{e.g.,} to achieve wide coverage, support high mobility users etc), on which a large amount of inter cell interference can potentially take place~\cite{Andrews20141065}.

Static and dynamic schemes are the two broad categories of interference avoidance techniques. \emph{Inter cell interference coordination (ICIC)} is a prime class of dynamic interference avoidance schemes, which can be further categorized into the schemes using \emph{variable transmit power} and \emph{fixed transmit power} allocations on subchannels~\cite{RF:Kosta}. The ICIC problem with fixed transmit power allocation on subchannels is the focus of this work. In this problem, each base station (BS), if a given subchannel is assigned to a mobile station (MS) within its cell, transmits with fixed power on the assigned subchannel and does not transmit on subchannels that are not assigned to any MS in its cell. Therefore, the problem translates into a problem of deciding as to which MS, if any,  to allocate each available subchannel to in each cell. Note that typically in each cell, some of the subchannels are not assigned to any MS in order to limit the inter cell interference.  %In our prior work~\cite{Report2017}, we have investigated the same problem with the goal of maximizing the total throughput, \emph{i.e.,} the sum of throughputs of all the MSs in the system. 

Most of the proposed resource allocation schemes to address the ICIC problem consider maximizing the total throughput, \emph{i.e.,} the sum of throughputs of all the MSs in the system, while completely neglecting the aspect of fairness~\cite{RF:eICIC:lopezperez},~\cite{BinSediq2},~\cite{Report2017},~\cite{Kosta2012}, \cite{RF:Rahman}, \cite{RF:wei:Yassin},~\cite{mert}. In the context of cellular systems, fairness means that each MS, irrespective of its channel gain (which is a measure of the quality of the channel from the BS to the MS), has an equal chance of being allocated each of the available subchannels,  \emph{i.e.,} no MS is preferred over the other MSs while allocating a subchannel in the system. Maximization of the total throughput results in high throughput of the MSs with good channel gain values; however, this is at the cost of low throughput of the MSs with poor channel gain values such as MSs at the cell boundaries~\cite{HTch}. However, one of the objectives of 4G systems is to offer good data rates to the MSs at the cell boundaries~\cite{RF:Rahman}.
% Therefore, the objective of maximizing the sum of throughputs favours the user with good channel condition, but would result in disfavour of cell edge users~\cite{HTch},~\cite{Jmo}. 
On the other hand, if lower (respectively, higher) throughputs were assigned to MSs with good (respectively, poor) channel gains, then it would lead to better fairness, but at the expense of a decrease in the total throughput. So, there exists a trade-off between the total throughput and fairness of resource allocation schemes~\cite{BinSediq}.
Motivated by this fact, our objective in this paper is to formulate the problem of achieving different trade-offs between the total throughput and fairness, study its complexity and design a distributed resource allocation algorithm to solve it.

We use \emph{Jain's fairness index}, which was proposed in~\cite{Jain} and has been extensively used in the networking literature, \emph{e.g.,} in~\cite{HTch},~\cite{BinSediq},~\cite{Sheikh}, as a fairness metric. One way to optimize the total throughput-fairness trade-off in cellular systems is to allocate resources such that the total throughput is maximized subject to the constraint that the throughput of each MS must exceed some predefined lower bound~\cite{HTch}. The trade-off can be optimized by varying the values of these lower bounds over the set of achievable rates. Another way is to use the $\alpha-$fair scheme, which was originally proposed in the context of designing fair end-to-end window-based congestion control protocols for packet-switched networks~\cite{Jmo}.
 In the $\alpha-$fair scheme, a parametric objective function, which is a function of the throughputs of the users and a parameter $\alpha$, is maximized. The parameter $\alpha$ is varied to achieve the required trade-off between the total throughput and fairness. For instance, the maximum total throughput (and minimum degree of fairness) is obtained when $\alpha=0$. Similarly, proportional fairness~\cite{Kelly} and max-min fairness~\cite{Gall} correspond to $\alpha=1$ and $\alpha=\infty$ respectively. In general, the degree of fairness (respectively, total throughput) increases (respectively, decreases) as $\alpha$ increases~\cite{Tlan}. 

In this paper, we adapt the concept of $\alpha-$fairness to the problem of ICIC with fixed transmit power, and show via simulations in Section~\ref{SC:simulations} that when the adapted $\alpha$-fair scheme is used to find a subchannel allocation, the Jain's fairness index increases and the total throughput decreases with $\alpha$. Thus, the adapted $\alpha-$fair scheme provides any degree of fairness by choosing an appropriate value of $\alpha$ in $[0, \infty)$, which is not possible in the scheme using predefined lower bounds~\cite{HTch} (see the previous paragraph). In addition, there is no clear procedure to select the lower bounds on the throughput of each MS in the latter scheme. In contrast, no lower bounds on the throughputs of MSs need to be selected in the adapted $\alpha-$fair scheme, which makes its implementation simpler than that of the scheme that uses predefined lower bounds. However, the concept of $\alpha-$fairness in~\cite{Jmo} has to be modified by introducing a new parameter $\tau > 0$ in the original parametric objective function. If the original parametric objective function were directly used in our context without change, the following problem would arise. If no subchannel is allocated to a MS (\emph{e.g.}, when the number of available subchannels is small relative to the number of MSs), its throughput is 0; this makes the value of the originally defined parametric objective function of the system $-\infty$ for $\alpha > 1$. Therefore, we introduce the concept of \emph{$\tau-\alpha-$fairness}, which is a modification of the aforementioned $\alpha-$fairness, and we define a new parametric objective function in Section~\ref{SC:model:objective}, which is a function of both $\alpha$ and $\tau$.  
We prove that the problem of finding a $\tau-\alpha-$fair allocation in the ICIC with fixed transmit power problem is NP-Complete for all values of  $\alpha$ in the range $[0,\infty)$ and for all $\tau > 0$ (see Section~\ref{SSC:np:completeness}). 

Next, we address the question of finding conditions under which the problem of finding a $\tau-\alpha-$fair allocation in the ICIC with fixed transmit power problem is solvable in polynomial time. 
Interestingly, it turns out that for every $\alpha \in [0, \infty)$, there exist thresholds such that if the potential interference levels experienced by each MS on every subchannel are above the threshold values, then the problem can be \emph{optimally solved in polynomial time by reducing it to the bipartite graph matching problem}~\cite{RF:kleinberg:algorithm} (see Section~\ref{polynomial time solution}). Also, the above threshold values are decreasing functions of the transmit power level of each BS. The above result implies that for a scenario in which BSs are densely deployed in an area and transmit with high power, the problem of finding a $\tau-\alpha-$fair allocation in the ICIC with fixed transmit power problem can be optimally solved in polynomial time. This is a surprising result since the above problem 
is NP-Complete in general (see the previous paragraph).

Next, we propose a simple distributed subchannel allocation
algorithm for the ICIC with fixed transmit power problem (see Section~\ref{SC:algorithms}) and investigate as to how the algorithm parameters should be selected so as to achieve a desired trade-off between the total throughput and fairness, via simulations. The proposed  algorithm is flexible, requires a small amount of time to operate, and requires information exchange among only neighboring BSs. 
%our simulation results show that the performance of the algorithm improves as the numbers of directly connected BSs of different BSs in the network increase (see Section~\ref{performance of distributed}). The simulated annealing (SA) based algorithm is centralized and can be implemented if a central entity (\emph{e.g.}, radio network controller~\cite{RF:Li:Downlink}) to which all the BSs in the network are directly connected is available. This algorithm allows a trade-off between quality of the obtained solution and execution time  by means of an appropriate choice of parameters. Also, we compare the performance of the BR algorithm, greedy algorithm and SA algorithm via simulations in Section~\ref{performance comparison}. Our simulation results show that the total throughput obtained by the BR algorithm is very small compared to those obtained using the SA and greedy algorithms; however, the execution time of the BR algorithm is much smaller than those of the latter two algorithms. Finally, the greedy algorithm outperforms the SA algorithm and uses only a small fraction of the number of computations in dense cellular networks (see Section~\ref{performance comparison}).

The rest of this paper is organized as follows. In Section~\ref{lit}, we review related research literature. We describe the system model and problem formulation in Section~\ref{SC:model:objective}. The complexity of the problem is analyzed and conditions under which it is polynomial time solvable are derived in Section~\ref{SC:complexity}. A distributed algorithm to solve the problem is presented in Section~\ref{SC:algorithms}. We present simulation results in Section~\ref{SC:simulations}, and provide conclusions and directions for future research in Section~\ref{conc}.

\section{Related Work}
\label{lit}
Resource allocation algorithms for the ICIC problem were proposed in~\cite{RF:eICIC:lopezperez},~\cite{BinSediq2},~\cite{Report2017},~\cite{Kosta2012}, \cite{RF:Rahman}, \cite{RF:wei:Yassin},~\cite{mert},~\cite{yoon}, but the aspect of fairness was not considered. The ICIC problem with the objective of maximizing the total throughput of a multi-cell system with multiple subchannels was investigated in our prior work~\cite{Report2017}, the problem was proved to be NP-Complete, and a set of conditions under which the problem can be solved in polynomial time were derived. However, in contrast to this paper, the aspect of fairness was not studied in~\cite{Report2017}.

We now review the existing literature on resource allocation that considers the fairness aspect in cellular systems. The authors of~\cite{BinSediq} proposed two multi-user resource allocation schemes to achieve an optimal system efficiency-fairness trade-off. For these schemes to apply, the user's benefit set must satisfy the monotonic trade-off property in which the Jain's fairness index decreases with the increase in the system efficiency beyond a threshold value. In contrast, our proposed scheme does not require such a monotonic trade-off condition to be satisfied. A two-stage resource allocation algorithm for achieving fair cell-edge performance was proposed in~\cite{Kim}. However, in~\cite{Kim},  only the cell-edge MSs and the interference caused only by the dominant BS were considered. In contrast, in this paper, we consider all the MSs, and the interference to an MS caused by all the  BSs, except the one serving the MS, which transmit over the subchannel used by the MS. A resource allocation algorithm for an OFDMA based single cell multicast system with proportional fairness was proposed in~\cite{jiang}. A waterfilling cumulative distribution function based scheduling  scheme for uplink transmissions, which provides fair resource sharing, in a single cell cellular network was proposed in~\cite{Xin}. The authors of~\cite{Shen} formulated the fair resource allocation problem in a single cell system as a mixed integer problem and proposed two suboptimal algorithms, for chunk allocation and power allocation, respectively. However,~\cite{jiang},~\cite{Xin} and~\cite{Shen} consider a single cell system; in contrast, we consider a multi-cell system in this paper.

A joint user association and ICIC problem was formulated as a utility maximization problem and an iterative algorithm was proposed to solve it in~\cite{Miki}. A logarithmic utility function was used to obtain a proportional fair solution, which is similar to the case $\alpha=1$ in our work. However, no schemes were provided to achieve different trade-offs between the total throughput and level of fairness. In contrast, in this paper, we provide a resource allocation scheme that can be used to achieve arbitrary trade-offs between the total throughput and level of fairness. In~\cite{pastore}, an analytical framework was proposed to investigate fairness-throughput trade-offs in the context of non-orthogonal multiple access (NOMA) downlink broadcasting in cellular networks. The ratio of weak user to strong user throughput is used as a fairness metric in~\cite{pastore}. The authors of~\cite{mik} proposed a distributed optimization scheme for joint user association and ICIC  with the proportional fairness criterion in small cell deployments. A fair distributed resource allocation algorithm to achieve a high total throughput in heterogeneous networks was proposed in~\cite{xhuang}. To ensure fairness, users in BSs that have low satisfaction degrees (ratio of number of channels currently allocated and number of required channels) and high traffic requirement levels are preferentially allocated channels. Also, in~\cite{jiang} (respectively,~\cite{Shen}), an allocation algorithm was proposed to provide proportional fairness (respectively, max-min fairness), which corresponds to the case $\alpha=1$ (respectively, $\alpha=\infty$), of the scheme proposed in our work. In contrast to~\cite{jiang},~\cite{Shen},~\cite{Miki},~\cite{pastore},~\cite{mik} and~\cite{xhuang}, a modification of the $\alpha-$fairness criterion is considered in this paper. Note that we consider all the values of $\alpha$ in $[0, \infty)$, which correspond to different trade-offs between the total throughput and fairness.

A semi-centralized joint cell muting and user scheduling scheme for interference coordination with temporal fairness in multi-cell networks was proposed in~\cite{shahsavari}. Downlink transmission over a single subchannel was considered. In contrast, multiple subchannels are considered in this paper. The authors of~\cite{jin} studied a problem similar to that in this paper, but in the context of a system consisting of resources (CPUs) and users instead of OFDMA based cellular networks. The dominant $\alpha-$fairness concept was proposed and the trade-off between fairness and efficiency was studied. 

 To the best of our knowledge, \emph{our work is the first to formulate the ICIC with fixed transmit power problem with the goal of  achieving arbitrary trade-offs between the total throughput and fairness};  in addition, we characterize the complexity of this problem, derive conditions under which the problem is polynomial time solvable, propose a distributed algorithm to solve it and evaluate its performance via simulations.
%The rest of the paper is organized as follows. We describe our system model and problem formulation in Section~\ref{SC:model:objective}. The proof of the NP-Completeness of the defined problem is provided in Section~\ref{SC:complexity}. A distributed $\tau-\alpha-$fair subchannel allocation algorithm is proposed in Section~\ref{SC:algorithms}. Simulation results are provided in Section~\ref{SC:simulations}. Finally, concluding remarks are provided in Section~\ref{conc}.
\vspace{-1em}
\section{System Model, Problem Definition and Background}
\label{SC:model:objective}
%\vspace{-1em}
%\vspace{-.5em}
We  consider an OFDMA based cellular system in which there are multiple cells; in each cell, a base station (BS) serves the mobile stations (MS) in the cell. The available frequency band (channel) is divided into multiple \emph{subchannels}; each subchannel has equal bandwidth.  Let the set of all BSs and the set of all available subchannels be denoted by $\mathcal{B}= \{1, \ldots, K\}$ and $\mathcal{N} = \{1, \ldots, N\}$ respectively. The cardinality of a set $A$ is denoted by $|A|$. Suppose frequency reuse factor one is used, which implies that any subset of the BSs in $\mathcal{B}$ may use the same subchannel in $\mathcal{N}$ simultaneously. Let $\mathcal{M}_a$ represent the set of all the MSs associated with BS $a \in \mathcal{B}$ and let $|\mathcal{M}_a|= M_a$. Similarly,  the set of all MSs in the system is represented by $\mathcal{M} = \cup_{a \in \mathcal{B}} \mathcal{M}_a$. Therefore, the total number of MSs in the system is given by $M = \sum_{a \in \mathcal{B}} M_a$. Whenever two or more BSs simultaneously allocate a given subchannel to one of their associated MSs, it results in inter cell interference. Note that typically in each cell, some of the subchannels are not assigned to any MS in order to limit the inter cell interference. The example in Fig.~\ref{mod} illustrates the model.  
%\begin{figure}[!hbt]
%\centering
%\resizebox{0.8\columnwidth}{!}{\includegraphics{model.pdf}}
%%\includegraphics[width=10cm, height=4cm]{qunzgain1.eps}
%\vspace{-1em}
%\caption{{Example illustrates association of base stations and mobile stations in a cellular system and inter cell interference phenomenon.}}
%\vspace{-2em}
%\label{mod}
%\end{figure}
%\vspace{-1cm}
%\begin{figure}
%%\label{mod}
%\centering
%\begin{minipage}{0.47\textwidth}
%  \centering
%  \includegraphics[width=0.9\linewidth]{model.pdf}
%  \captionof{figure}{In the example in the figure, there are two subchannels; let $\{1,2\}$ be the two subchannels in $\mathcal{N}$.  Subchannel $1$ is allocated to the $1^{st}$, $4^{th}$ and $8^{th}$ MSs, and subchannel $2$ is allocated to the $2^{nd}$, $6^{th}$ and $9^{th}$ MSs as shown by the different arrows in the figure. }
%  \label{mod}
%\end{minipage}%
%\hspace{1em}
%\begin{minipage}{0.47\textwidth}
%  \centering
%  \includegraphics[width=0.9\linewidth]{flow.pdf}
%  %\vspace{-1em}
%  \captionof{figure}{The figure illustrates a network with multiple flows over it. There are four nodes, $S_0, \ldots, S_3$, five links, $L_1, \ldots, L_5$, and three flows, $f_0, f_1, f_2$, in the network.}
%  \label{flow}
%\end{minipage}
%\end{figure}
 \begin{figure}
\centering
\includegraphics[width=0.8\linewidth]{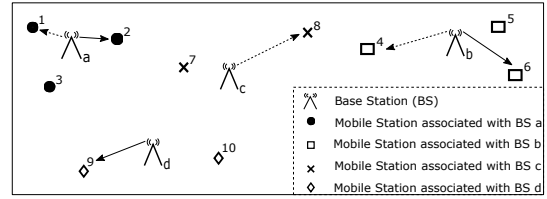}
  \captionof{figure}{In the example in the figure, there are two subchannels; let $\{1,2\}$ be the two subchannels in $\mathcal{N}$.  Subchannel $1$ is allocated to the $1^{st}$, $4^{th}$ and $8^{th}$ MSs, and subchannel $2$ is allocated to the $2^{nd}$, $6^{th}$ and $9^{th}$ MSs as shown by the different arrows in the figure. }
  \label{mod}
  \vspace{-1em}
\end{figure}

We consider the problem of subchannel allocation to MSs for downlink transmissions (\emph{i.e.}, transmissions from BSs to MSs) in a given time slot.
% The traffic demands of various MSs in the system in the considered time slot can be modelled as follows. Let the \emph{weight} of MS $j$ within the cell of BS $a$ is a positive real number and is denoted by $w_{a,j}$, $a \in \mathcal{B}, j \in \mathcal{M}_a$. The urgency of the downlink traffic meant for MS $j$ is modelled by the weight. Different kinds of traffic requirements can be modelled by allocating different weights to the MSs, \emph{e.g.}, elastic traffic and delay-sensitive traffic can be modelled by assigning relatively low and high weights to the intended MSs respectively. 
Let
%\vspace{-1em}
\begin{equation}
%\label{EQ:TH:PB:ICIC:fixed:powers:np:complete:eq1}
z_{a,j}^n = \left\{ 
\begin{array}{ll}
1, & \mbox{if MS}\; j \in \mathcal{M}_a \; \mbox{is assigned subchannel}\; n, \\
0, & \mbox{otherwise}. \\
\end{array}
\right. 
\end{equation}
%\vspace{-1em}
The complete allocation is denoted by $\mathbf{Z} = \{z_{a,j}^n: a \in \mathcal{B}, j \in \mathcal{M}_a, n \in \mathcal{N} \}$. Let 
\vspace{-1em}
\begin{equation}
\label{EQ:yin:definition}
y_a^n = \sum_{j \in \mathcal{M}_a} z_{a,j}^n. 
\end{equation}
Intra-cell interference can be avoided by introducing the constraint that any subchannel $n$ cannot be allocated to more than one MS within a cell; thus, we obtain the constraint:
%\vspace{-1em} 
\begin{equation}
\label{EQ:yin:binary}
y_a^n \in \{0,1\}, \ \forall a \in \mathcal{B}, n \in \mathcal{N}. 
\end{equation}
Also, $y_a^n$ equals $1$ if subchannel $n$ is assigned to one of the MSs in $\mathcal{M}_a$, else $0$.
%\vspace{-1em}
%\vspace{-1em}
Any given BS $a \in \mathcal{B}$ transmits on a subchannel $n \in \mathcal{N}$ with \emph{fixed} power $P$ if $z_{a,j}^n = 1$ for some $j \in \mathcal{M}_a$; else transmits with power $0$. Assume that the noise power spectral density is $N_0$. Let each subchannel $n \in \mathcal{N}$ be an approximately flat fading channel; that is, the coherence bandwidth is larger than the subchannel bandwidth~\cite{Rapp}. Let $H_{a,j}^n$ denote the channel gain (which is a measure of the channel quality) from BS $a$ to MS $j$ on subchannel $n$; we assume that the channel gain values $\{H_{a,j}^n: a \in \mathcal{B}, j \in \mathcal{M}, n \in \mathcal{N}\}$ remain unchanged during the considered time slot. Orthogonal cell-specific reference signals can be used to estimate the channel gain values $\{H_{a,j}^n: a \in \mathcal{B}, j \in \mathcal{M}, n \in \mathcal{N}\}$~\cite{RF:Kosta}. Hence, we assume that the channel gain values $\{H_{a,j}^n: j \in \mathcal{M}, n \in \mathcal{N}\}$ are known to BS $a$.     

Consider an allocation $\mathbf{Z} = \{z_{a,j}^n: a \in \mathcal{B}, j \in \mathcal{M}_a, n \in \mathcal{N} \}$. If $\mathbf{Z}$ satisfies \eqref{EQ:yin:definition} and \eqref{EQ:yin:binary}, it is called a \emph{feasible} allocation. Given a \emph{feasible} allocation $\mathbf{Z}$, the total throughput of all the MSs in the network is given by:
\begin{equation}
U(\mathbf{Z}) = \sum_{a \in \mathcal{B}} \sum_{j \in \mathcal{M}_a} \sum_{n \in \mathcal{N}} z_{a,j}^n \log \left( 1 + \frac{P H_{a,j}^n}{P \displaystyle \sum_{i \in \mathcal{B}, i \neq a} H_{i,j}^n y_i^n + N_0} \right).  
\label{EQ:objective}
\end{equation}
In \eqref{EQ:objective}, the throughput of the channel from BS $a$ to MS $j$ is calculated using the Shannon capacity formula for each $a \in \mathcal{B}$ and $j \in \mathcal{M}_a$~\cite{Tse}; in particular, the second term inside the $\log(\cdot)$ is the Signal to Interference and Noise ratio on subchannel $n$ from BS $a$ to MS $j$. As a normalization, we assume that each subchannel has unit bandwidth.      

For future use, suppose the throughput of MS $j \in \mathcal{M}_a$ is denoted as follows: 
%\vspace{-1em}
\begin{equation}
\label{EQ:UnZ}
U_j(\mathbf{Z}) =  \sum_{n \in \mathcal{N}} z_{a,j}^n \log \left( 1 + \frac{P H_{a,j}^n}{P \displaystyle \sum_{i \in \mathcal{B}, i \neq a} H_{i,j}^n y_i^n + N_0} \right).  
\end{equation}
Note that:
%\vspace{-1em}
\begin{equation}
\label{EQ:UZ:eq:sum:UnZ}
U(\mathbf{Z}) = \sum_{a \in \mathcal{B}}\sum_{j \in \mathcal{M}_a} U_j(\mathbf{Z})=\sum_{j \in \mathcal{M}} U_j(\mathbf{Z}) .
\end{equation}

The notion of $\alpha-$fair allocation was introduced in the context of multiple flows over a network having multiple nodes and links~\cite{Jmo} as illustrated by the example in Fig.~\ref{flow}. The capacity of each link is finite and fixed. Each flow traverses a path that consists of multiple links and transmits at some flow rate. The concept of $\alpha-$fair allocation was introduced to address the problem of how the bandwidths of the links in the network can be shared in a fair manner among the different flows~\cite{Jmo}. 
%\vspace{-2em}
\begin{figure}
  \centering
  \includegraphics[width=0.8\linewidth]{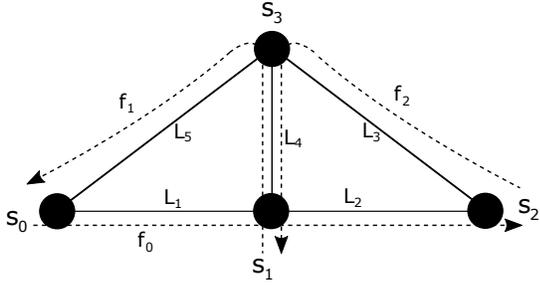}
  %\vspace{-1em}
  \captionof{figure}{The figure illustrates a network with multiple flows over it. There are four nodes, $S_0, \ldots, S_3$, five links, $L_1, \ldots, L_5$, and three flows, $f_0, f_1, f_2$, in the network.}
  \vspace{-1.5em}
  \label{flow}
\end{figure}
%\vspace{-1em}
Suppose $\mathcal{S}$,  $\mathcal{L}$ and $\mathcal{F}$ are the sets of all the nodes, links and flows respectively, in a network (see Fig.~\ref{flow}). Let $x_r \geq 0$ be the flow rate of flow $r \in \mathcal{F}$ and let $\mathbf{X}=\{x_r : r \in \mathcal{F}\}$ represent the flow rate vector. For $\alpha > 0$, the utility of a flow $r$ is defined as $U_r^{\alpha}(x_r) = 
\log(x_r)$  if  $\alpha = 1$ and $U_r^{\alpha}(x_r) = \frac{x_r^{1-\alpha}}{1-\alpha}$ if $\alpha \neq 1$. The flow rate vector $\mathbf{X}$ which maximizes $\sum_{r \in \mathcal{F}} U_r^{\alpha}(x_r)$, \emph{i.e.,} the total utility of all the flows, such that the sum of the flow rates through any link does not exceed its capacity, is known as the $\alpha-$fair allocation. In general, the degree of fairness  (respectively, total throughput) under the $\alpha$-fair allocation increases (respectively, decreases) as $\alpha$ increases~\cite{Tlan}. 

In the model in this paper, the users of the network are MSs, in contrast to the above model where the users are the various flows. If the above definition of $\alpha-$fairness were directly used in our context without change, \emph{i.e.}, if we defined the $\alpha$-fair allocation to be the feasible allocation $\mathbf{Z}$ that maximizes $\sum_{j \in \mathcal{M}}  \log(U_{j}(\mathbf{Z}))$ if $\alpha = 1$ and $\sum_{j \in \mathcal{M}}  \frac{U_{j}(\mathbf{Z})^{1-\alpha}}{1-\alpha}$ if $\alpha \neq 1$, the following problem would arise. If no subchannel is allocated to a MS $j$, its throughput, $U_{j}(\mathbf{Z})$, is 0 (see~\eqref{EQ:UnZ}); this makes $\sum_{l \in \mathcal{M}}  \frac{U_{l}(\mathbf{Z})^{1-\alpha}}{1-\alpha} = -\infty$ for $\alpha > 1$ since $\frac{U_{j}(\mathbf{Z})^{1-\alpha}}{1-\alpha} = -\infty$ for $U_{j}(\mathbf{Z}) = 0$ and $\alpha > 1$. Note that this is a potentially commonly arising situation in practice, \emph{e.g.,} some of the MSs would not be assigned any subchannels when the number of subchannels is small relative to the number of MSs in the network.  To avoid this situation, we define a modified $\alpha-$fair utility function by incorporating a positive number $\tau$. Specifically, for a given $\alpha \in [0,\infty ), \tau > 0$ and a feasible allocation $\mathbf{Z}$, we define the $\tau-\alpha-$fair utility function of the system as follows:
%\vspace{-1em}
\begin{equation}
\label{EQ:objective1}
U_{\alpha,\tau}(\mathbf{Z})=\left\{ 
\begin{array}{ll}
\sum_{j \in \mathcal{M}} \log (\tau+U_j(\mathbf{Z})), & \mbox{if} \; \;\alpha=1, \\
\sum_{j \in \mathcal{M}}  \frac{(\tau+U_{j}(\mathbf{Z}))^{1-\alpha}}{1-\alpha}, & \mbox{if } \alpha \neq 1. \\
\end{array}
\right.
\end{equation}
Suppose the set of all possible feasible allocations is denoted by $\mathcal{Z}$. We define a $\tau-\alpha-$fair allocation to be a feasible allocation $\mathbf{Z} \in \mathcal{Z}$ that maximizes the function $U_{\alpha,\tau}(\mathbf{Z})$ in~\eqref{EQ:objective1}. Our goal is to find a $\tau-\alpha-$fair allocation:
\begin{problem}
\label{PB:ICIC:fixed:powers}
Find a $\tau-\alpha-$fair allocation.   
\end{problem}

Our simulations (see Section~\ref{SC:simulations}) show that by solving Problem~\ref{PB:ICIC:fixed:powers} with a fixed value $\tau > 0$ and different values of $\alpha \in [0, \infty)$, allocations that achieves various trade-offs between the total throughput and degree of fairness can be obtained. Also, the question of how the value of $\tau$ in Problem~\ref{PB:ICIC:fixed:powers} should be selected is addressed in Section~\ref{SC:simulations}.
%\vspace{-1em}

\section{Complexity Analysis}
\label{SC:complexity}
%\vspace{-.5em}
\subsection{NP-Completeness of Problem~\ref{PB:ICIC:fixed:powers}}
\label{SSC:np:completeness}
In this section, we show that for each $\alpha \in [0, \infty)$ and $\tau > 0,$ Problem~\ref{PB:ICIC:fixed:powers} is NP-Complete. The decision version associated with Problem~\ref{PB:ICIC:fixed:powers} is: for a given number $T$, can we find a feasible allocation $\mathbf{Z}$ which satisfies the condition $U_{\alpha,\tau}(\mathbf{Z}) \geq T$? The following result shows that (the decision version of) Problem~\ref{PB:ICIC:fixed:powers} is NP-Complete. 

\begin{theorem}
\label{TH:PB:ICIC:fixed:powers:np:complete}
 For each $\alpha \in [0,\infty)$ and $\tau > 0$, Problem~\ref{PB:ICIC:fixed:powers} is NP-Complete.  
\end{theorem}
\begin{IEEEproof}
For any allocation $\mathbf{Z}$, it is possible to verify  in polynomial time whether $\mathbf{Z}$ is feasible or not using \eqref{EQ:yin:definition} and \eqref{EQ:yin:binary}. Also, we can calculate $U_{\alpha,\tau}(\mathbf{Z})$ using \eqref{EQ:objective1} and verify whether $U_{\alpha,\tau}(\mathbf{Z}) \geq T$ in polynomial time. Hence, Problem~\ref{PB:ICIC:fixed:powers} lies in class NP~\cite{RF:kleinberg:algorithm}. 

Next, we show the NP-Completeness of Problem~\ref{PB:ICIC:fixed:powers} by reducing the Maximum Independent Set (MIS) problem, which is known to be NP-Complete~\cite{RF:kleinberg:algorithm}, to Problem~\ref{PB:ICIC:fixed:powers} in polynomial time, \emph{i.e.}, we show that MIS $<_p$ Problem~\ref{PB:ICIC:fixed:powers}. Consider the following instance of the MIS problem: we are given an undirected graph $G = (V, E)$, in which $V$ and $E$ are vertex set and edge set respectively, and a positive integer $k$. Does there exist an independent set of size at least $k$ in $G$?

From the above instance, a particular instance of Problem~\ref{PB:ICIC:fixed:powers} is generated as follows: suppose that only one subchannel$^1$~\footnote{$^1$For simplicity, we discard the superscript $n$ (subchannel number) in the remaining proof.}  is available (\emph{i.e.}, $N = 1$). Let $ \mathcal{B}= V$, \emph{i.e.,} corresponding to each node $a \in V$, there is a BS $a \in \mathcal{B} $. Also, there is $1$ MS associated with each BS (\emph{i.e.}, $M_a = 1$ for all $a \in \mathcal{B}$).  Let $j_a$ denote the MS associated with BS $a$ and the edge connecting the two distinct nodes $u$ and $v$ is denoted by $(u,v)$. 

Consider the above generated instance of Problem~\ref{PB:ICIC:fixed:powers} and let $N_o=P$. Suppose the channel gains are modelled as follows:
\begin{equation}
\label{gain1}
H_{u,j_u}= 2, \; \forall u \in V 
\end{equation} 
%\vspace{-2em}
\begin{equation}
\label{gain2}
H_{u,j_v} = \left\{ 
\begin{array}{ll}
\infty, \mbox{if} \; (u,v)\in E, u \neq v, \\
0, & \hspace*{-3cm}\mbox{else}. \\
\end{array}
\right. 
\end{equation}
%The transmission data rate of BS a is given by:
%\begin{equation}
%\label{EQ:UnZ}
%R_a = z_{a,j_a} \log \left( 1 + \frac{P H_{a,j_a}^n}{P \displaystyle \sum_{i \in \mathcal{B}, i \neq a} H_{i,j_a}^n y_i^n + N_0} \right).  
%\end{equation}
Consider an allocation $\mathbf{Z} = \{ z_{u,j_u} \in \{0,1\}: u \in V\}$ in the generated instance of Problem~\ref{PB:ICIC:fixed:powers}. Since $M_u = 1$ for all $u \in V$, it implies that $y_u \in \{0,1\}$ for all $u \in V$; so constraints \eqref{EQ:yin:definition} and \eqref{EQ:yin:binary} are satisfied. Thus, every allocation $\mathbf{Z} = \{ z_{u,j_u} \in \{0,1\}: u \in V\}$ is feasible in the generated instance.

Now, we divide the proof into three cases depending on the value of $\alpha$.\\
\textbf{\emph{Case (i): $\alpha<1$:}}

The utility of the system under $\tau-\alpha-$fair allocation is calculated by \eqref{EQ:objective1} for $\alpha<1$. In the generated instance of Problem~\ref{PB:ICIC:fixed:powers}, we want to verify whether there exists a (feasible) allocation $\mathbf{Z}$ which satisfies $U_{\alpha,\tau}(\mathbf{Z}) \geq \frac{k}{(1-\alpha)}\left( \tau+\log\left(3 \right)\right)^{1-\alpha}+\frac{|V|-k}{1-\alpha}\tau^{1-\alpha}$? Our claim is that the answer is yes if and only if an independent set of size at least $k$ exists in $G$. To show sufficiency, suppose an independent set, $I$, of size $k^{\prime} \geq k$ exists in $G$. Then by \eqref{EQ:objective1}, \eqref{gain1} and \eqref{gain2}, the following allocation:
\vspace{-.5em}
\begin{equation}
\label{eq15}
z_{u,j_u} = \left\{ 
\begin{array}{ll}
1, & \mbox{if } u \in I, \\
0, & \mbox{else}, \\
\end{array}
\right. 
\end{equation}
has utility $\frac{k^{\prime}}{(1-\alpha)}\left( \tau+\log\left(3 \right)\right)^{1-\alpha}+\frac{|V|-k^{\prime}}{1-\alpha}\tau^{1-\alpha} \geq \frac{k}{(1-\alpha)}\left( \tau+\log\left(3 \right)\right)^{1-\alpha}+\frac{|V|-k}{1-\alpha}\tau^{1-\alpha}$ since $k^{\prime} \geq k$, which shows sufficiency. To show necessity, suppose that an allocation $\mathbf{Z}   = \{ z_{u,j_u} \in \{0,1\}: u \in V\}$ exists such that:
\vspace{-.5em}
\begin{equation}
\label{eq11}
U_{\alpha,\tau}(\mathbf{Z}) \geq \frac{k}{(1-\alpha)}\left( \tau+\log\left(3 \right)\right)^{1-\alpha}+\frac{|V|-k}{1-\alpha}\tau^{1-\alpha}
\end{equation}  
and let $I = \{u \in V: z_{u,j_u} = 1\}$. If two nodes $u, v \in I$ are connected by an edge, then by \eqref{EQ:objective1}, \eqref{gain1} and \eqref{gain2}, it follows that $\frac{(\tau+U_{u}(\mathbf{Z}))^{1-\alpha}}{1-\alpha} = \frac{(\tau+U_{v}(\mathbf{Z}))^{1-\alpha}}{1-\alpha}=  \frac{\tau^{1-\alpha}}{(1-\alpha)}$, which are the same as when both $u$ and $v$ are not allocated a subchannel. By this fact and by \eqref{EQ:objective1}, it follows that: 
\vspace{-.5em}
\begin{equation}
\label{eq12}
U_{\alpha,\tau}(\mathbf{Z}^{\prime}) = U_{\alpha,\tau}(\mathbf{Z}),
\end{equation}
\vspace{-1em}
where allocation $\mathbf{Z}^{\prime}$ is given as follows:
\begin{equation}
z_{u,j_u}^{\prime} = \left\{ 
\begin{array}{ll}
1, & \mbox{if } u \in I^{\prime}, \\
0, & \mbox{else}, \\
\end{array}
\right. 
\end{equation}
and $I^{\prime}$ is the independent set derived from $I$ by excluding all node pairs having an edge between them. Let $|I^{\prime}| = k^{\prime}$. Then: 
\vspace{-.5em}
\begin{equation}
\label{eq14}
U_{\alpha,\tau}(\mathbf{Z}^{\prime})  = \frac{k^{\prime}}{(1-\alpha)}\left( \tau+\log\left(3 \right)\right)^{1-\alpha}+\frac{|V|-k^{\prime}}{1-\alpha}\tau^{1-\alpha}
\end{equation}
by \eqref{EQ:objective1}, \eqref{gain1} and \eqref{gain2}. By \eqref{eq11}, \eqref{eq12} and \eqref{eq14}, we get:
\begin{eqnarray}
 &&\hspace{-.2cm}\frac{k^{\prime}}{(1-\alpha)}\left( \tau+\log\left(3 \right)\right)^{1-\alpha}+\frac{|V|-k^{\prime}}{1-\alpha}\tau^{1-\alpha}  \nonumber\\
&\hspace{.5cm}& \geq\frac{k}{(1-\alpha)}\left( \tau+\log\left(3 \right)\right)^{1-\alpha}+\frac{|V|-k}{1-\alpha}\tau^{1-\alpha}
\end{eqnarray}
%& \frac{k^{\prime}}{(1-\alpha)}\left( \tau+\log\left(3 \right)\right)^{1-\alpha}+\frac{|V|-k^{\prime}}{1-\alpha}\tau^{1-\alpha}& \\
%& \geq \frac{k}{(1-\alpha)}\left( \tau+\log\left(3 \right)\right)^{1-\alpha}+\frac{|V|-k}{1-\alpha}\tau^{1-\alpha} &
%\end{eqnarray*}   
%\normalsize
\begin{equation}
\implies (k^{\prime}-k) \left[\left( \tau+\log\left(3 \right)\right)^{1-\alpha}-\tau^{1-\alpha}\right]\geq 0 \hspace{2em}( \mbox{as} \;\; \alpha <1)
\end{equation}
So $k^{\prime} \geq k$. Hence, necessity holds as an independent set of size at least $k$ exists in $G$. \\
%The proof for this case follows on the same lines as for $\alpha=0$ in our previous work~\cite{Report2017}, because for the considered instance of Problem~\ref{PB:ICIC:fixed:powers}, if there exists an allocation $\mathbf{Z}$ for which the condition $U(\mathbf{Z}) \geq k \log \left( 1 + \frac{P}{N_0} \right)$ holds, then the allocation $\mathbf{Z}$ also satisfies the condition $U_{\alpha,\tau}(\mathbf{Z}) \geq k \left( \tau+\log \left( 1 + \frac{P}{N_0} \right)\right)^{(1-\alpha)}$.  
% $\alpha-$fair utility function $U_{\alpha,\tau}(\mathbf{Z})$ in \eqref{EQ:objective1} is a monotonic increasing function of the utility function considered in previous work~\cite{Report2017}.
\textbf{\emph{Case (ii): $\alpha=1$:}}
 
The utility of the system under an allocation $\mathbf{Z}$ is calculated by \eqref{EQ:objective1} for $\alpha=1$.
%\begin{equation}
%\label{EQ:TH:PB:ICIC:fixed:powers:np:complete:eq3}
%U(\mathbf{Z})=\sum_{u\in B} \log \left( \tau+R_u \right)
%\end{equation} 
In the generated instance of Problem~\ref{PB:ICIC:fixed:powers}, we want to verify whether there exists a (feasible) allocation $\mathbf{Z}$ which satisfies $U_{\alpha,\tau}(\mathbf{Z}) \geq k \log \left( \tau+\log(3) \right) + (|V|-k)\log(\tau)$? Our claim is that the answer is yes if and only if an independent set of size at least $k$ exists in $G$. To show sufficiency, suppose an independent set, $I$, of size $k^{\prime} \geq k$ exists in $G$. Then by \eqref{EQ:objective1}, \eqref{gain1} and \eqref{gain2}, the following allocation:
\vspace{-.5em}
\begin{equation}
%\label{EQ:TH:PB:ICIC:fixed:powers:np:complete:eq1}
z_{u,j_u} = \left\{ 
\begin{array}{ll}
1, & \mbox{if } u \in I, \\
0, & \mbox{else}, \\
\end{array}
\right. 
\end{equation}
has utility $k^{\prime} \log \left( \tau+\log(3) \right) + (|V|-k^{\prime})\log(\tau) \geq k \log \left( \tau+\log(3) \right) + (|V|-k)\log(\tau)$ since $k^{\prime} \geq k$, which shows sufficiency. To show necessity, suppose that an allocation $\mathbf{Z}   = \{ z_{u,j_u} \in \{0,1\}: u \in V\}$ exists such that:
\vspace{-.5em}
\begin{equation}
\label{eq5}
U_{\alpha,\tau}(\mathbf{Z}) \geq k \log \left(\tau+ \log(3) \right) + (|V|-k)\log(\tau), 
\end{equation}  
and let $I = \{u \in V: z_{u,j_u} = 1\}$. If two nodes $u, v \in I$ are connected by an edge, then by \eqref{EQ:objective1}, \eqref{gain1} and \eqref{gain2} it follows that $ \log(\tau+U_v(\mathbf{Z}))=\log(\tau+U_u(\mathbf{Z}))= \log(\tau)$, which are the same as when both $u$ and $v$ are not allocated a subchannel. By this fact and \eqref{EQ:objective1}, it follows that: 
\vspace{-.5em}
\begin{equation}
\label{eq6}
U_{\alpha,\tau}(\mathbf{Z}^{\prime}) = U_{\alpha,\tau}(\mathbf{Z}),
\end{equation}
\vspace{-.5em}
where allocation $\mathbf{Z}^{\prime}$ is given as follows:
\begin{equation}
z_{u,j_u}^{\prime} = \left\{ 
\begin{array}{ll}
1, & \mbox{if } u \in I^{\prime}, \\
0, & \mbox{else}, \\
\end{array}
\right. 
\end{equation}
and $I^{\prime}$ is the independent set derived from $I$ by excluding all node pairs having an edge between them. Let $|I^{\prime}| = k^{\prime}$. Then: 
\begin{equation}
\label{eq8}
U_{\alpha,\tau}(\mathbf{Z}^{\prime})  = k^{\prime} \log \left(\tau+ \log(3) \right) + (|V|-k^{\prime})\log(\tau)
\end{equation}
by \eqref{EQ:objective1}, \eqref{gain1} and \eqref{gain2}. By \eqref{eq5}, \eqref{eq6} and \eqref{eq8} we get:
\begin{eqnarray}
&& \hspace{-0.2cm} k^{\prime} \log \left(\tau+ \log(3) \right) + (|V|-k^{\prime})\log(\tau) \nonumber\\
&& \geq k \log \left(\tau+ \log(3) \right) + (|V|-k)\log(\tau)
\end{eqnarray}

%& k^{\prime} \log \left(\tau+ \log(3) \right) + (|V|-k^{\prime})\log(\tau) &\\
%&\geq k \log \left(\tau+ \log(3) \right) + (|V|-k)\log(\tau)&
%\end{eqnarray*}
\begin{equation}
\implies (k^{\prime}-k) \log\left(1+ \frac{\log(3)}{\tau} \right)\geq 0
\end{equation}
%\end{eqnarray*}
So $k^{\prime} \geq k$. Hence, necessity holds as an independent set of size at least $k$ exists in $G$.\\
\textbf{\emph{Case (iii): $\alpha>1$:}}

%The proof for this case is exactly on the same lines as in \emph{Case A}.
%Consider the generated instance of the Problem~\ref{PB:ICIC:fixed:powers}. Let the channel gains are modelled as follows:
%\begin{equation}
%\label{eq14}
%H_{u,j_v} = \left\{ 
%\begin{array}{ll}
%\infty, & \mbox{if }(u,v) \in E, v \neq u \\\
%0, & \mbox{else}, \\
%\end{array}
%\right. \ 
%\end{equation}
%
%Consider an allocation $\mathbf{Z} = \{ z_{u,j_u} \in \{0,1\}: u \in V\}$ in the generated instance of Problem~\ref{PB:ICIC:fixed:powers}. Since, $M_u = 1$ for all $u \in V$, it implies that $y_u \in \{0,1\}$ for all $u \in V$. Thus, every allocation $\mathbf{Z} = \{ z_{u,j_u} \in \{0,1\}: u \in V\}$ is feasible in the generated instance. 
The utility of the system under an allocation $\mathbf{Z}$ is calculated by \eqref{EQ:objective1} for $\alpha>1$.
In the generated instance of Problem~\ref{PB:ICIC:fixed:powers}, we want to verify whether there exists a (feasible) allocation $\mathbf{Z}$ which satisfies $U_{\alpha,\tau}(\mathbf{Z}) \geq \frac{k}{(1-\alpha)}\left( \tau+\log\left(3 \right)\right)^{1-\alpha}+\frac{|V|-k}{1-\alpha}\tau^{1-\alpha}$? Our claim is that the answer is yes if and only if an independent set of size at least $k$ exists in $G$. To show sufficiency, suppose an independent set, $I$, of size $k^{\prime} \geq k$ exists in $G$. Then by \eqref{EQ:objective1}, \eqref{gain1} and \eqref{gain2}, the following allocation:
\begin{equation}
\label{eq15_1}
z_{u,j_u} = \left\{ 
\begin{array}{ll}
1, & \mbox{if } u \in I, \\
0, & \mbox{else}, \\
\end{array}
\right. 
\end{equation}
has utility $\frac{k^{\prime}}{(1-\alpha)}\left( \tau+\log\left(3 \right)\right)^{1-\alpha}+\frac{|V|-k^{\prime}}{1-\alpha}\tau^{1-\alpha} \geq \frac{k}{(1-\alpha)}\left( \tau+\log\left(3 \right)\right)^{1-\alpha}+\frac{|V|-k}{1-\alpha}\tau^{1-\alpha}$ since $k^{\prime} \geq k$, which shows sufficiency. To show necessity, suppose that an allocation $\mathbf{Z}   = \{ z_{u,j_u} \in \{0,1\}: u \in V\}$ exists such that:
\begin{equation}
\label{eq16}
U_{\alpha,\tau}(\mathbf{Z}) \geq \frac{k}{(1-\alpha)}\left( \tau+\log\left(3 \right)\right)^{1-\alpha}+\frac{|V|-k}{1-\alpha}\tau^{1-\alpha}
\end{equation}  
and let $I = \{u \in V: z_{u,j_u} = 1\}$. If two nodes $u, v \in I$ are connected by an edge, then by \eqref{EQ:objective1}, \eqref{gain1} and \eqref{gain2}, it follows that $\frac{(\tau+U_{u}(\mathbf{Z}))^{1-\alpha}}{1-\alpha} = \frac{(\tau+U_{v}(\mathbf{Z}))^{1-\alpha}}{1-\alpha}=  \frac{\tau^{1-\alpha}}{(1-\alpha)}$, which are the same as when both $u$ and $v$ are not allocated a subchannel. By this fact and by \eqref{EQ:objective1}, it follows that: 
\begin{equation}
\label{eq17}
U_{\alpha,\tau}(\mathbf{Z}^{\prime}) = U_{\alpha,\tau}(\mathbf{Z}),
\end{equation}
where allocation $\mathbf{Z}^{\prime}$ is given as follows:
\begin{equation}
z_{u,j_u}^{\prime} = \left\{ 
\begin{array}{ll}
1, & \mbox{if } u \in I^{\prime}, \\
0, & \mbox{else}, \\
\end{array}
\right. 
\end{equation}
and $I^{\prime}$ is the independent set derived from $I$ by excluding all node pairs having an edge between them. Let $|I^{\prime}| = k^{\prime}$. Then: 
\begin{equation}
\label{eq19}
U_{\alpha,\tau}(\mathbf{Z}^{\prime})  = \frac{k^{\prime}}{(1-\alpha)}\left( \tau+\log\left(3 \right)\right)^{1-\alpha}+\frac{|V|-k^{\prime}}{1-\alpha}\tau^{1-\alpha}
\end{equation}
by \eqref{EQ:objective1}, \eqref{gain1} and \eqref{gain2}. By \eqref{eq16}, \eqref{eq17} and \eqref{eq19}, we get:
%\scriptsize
\begin{eqnarray}
&& \hspace{-0.2cm}\frac{k^{\prime}}{(1-\alpha)}\left( \tau+\log\left(3 \right)\right)^{1-\alpha}+\frac{|V|-k^{\prime}}{1-\alpha}\tau^{1-\alpha} \nonumber\\
&& \geq \frac{k}{(1-\alpha)}\left( \tau+\log\left(3 \right)\right)^{1-\alpha}+\frac{|V|-k}{1-\alpha}\tau^{1-\alpha}
\end{eqnarray}
%& \frac{k^{\prime}}{(1-\alpha)}\left( \tau+\log\left(3 \right)\right)^{1-\alpha}+\frac{|V|-k^{\prime}}{1-\alpha}\tau^{1-\alpha}& \\
%& \geq \frac{k}{(1-\alpha)}\left( \tau+\log\left(3 \right)\right)^{1-\alpha}+\frac{|V|-k}{1-\alpha}\tau^{1-\alpha} &
%\end{eqnarray*}   
%\normalsize

\begin{equation}
\implies (k^{\prime}-k) \left[\left( \tau+\log\left(3 \right)\right)^{1-\alpha}-\tau^{1-\alpha}\right]\leq 0 \hspace{2em}( \mbox{as} \;\; \alpha >1)
\end{equation}
So $k^{\prime} \geq k$. Hence, necessity holds as an independent set of size at least $k$ exists in $G$. The result follows.
\end{IEEEproof}
\vspace{-1em}

\subsection{Conditions For Polynomial Time Solvability of Problem~\ref{PB:ICIC:fixed:powers}} \label{polynomial time solution}
Throughout this subsection, assume that $\tau > 0$ and $H_{i,j}^n > 0$ for all $i \in \mathcal{B}, j \in \mathcal{M}, n \in \mathcal{N}$. However, note that the latter is a mild assumption since the channel gains $H_{i,j}^n$ are allowed to be arbitrarily small.

For BS $a \in \mathcal{B}$, MS $j \in \mathcal{M}_a$ and subchannel $n \in \mathcal{N}$, let:
\begin{equation}
\label{eta}
\eta(a,j,n) = \frac{P H_{a,j}^n}{N_0}, 
\end{equation}
and
\begin{equation}
\label{EQ:beta}
\beta(a,j,n) = \frac{\min_{b \in \mathcal{B} \backslash \{ a \}} H_{b,j}^n}{H_{a,j}^n}. 
\end{equation}
Consider the following conditions:
\begin{condition}
\label{cond1}
\begin{equation}
\tau < \alpha \;\;\mbox{and,} \nonumber
\end{equation}

\begin{equation}
\beta(a,j,n) \geq  \max\left(\frac{\alpha-1}{\tau-\frac{\left(\tau+\log \left(1 + \eta(a,j,n) \right)\right)^{1-\alpha}}{\tau^{-\alpha}}},\frac{\alpha}{\tau}+\frac{1}{\eta(a,j,n)}\right),
\end{equation}
for all $a \in \mathcal{B}$, $j \in \mathcal{M}_a, n \in \mathcal{N}$. 
%for all $a \in \mathcal{B}$, $j \in \mathcal{M}_a, n \in \mathcal{N}$ and $\alpha > 0$. 
\end{condition}
\begin{condition}
\label{cond2}
\begin{equation}
\frac{\alpha(2^{(\alpha-1)}-1)}{\tau(\alpha-1)}>1,\; \log(1+\eta(a,j,n)) \leq \tau \; \mbox{and} \;\; \beta(a,j,n) \geq \nonumber
\end{equation}
\begin{equation}
 \max\left(\frac{\alpha-1}{\tau-\frac{\left(\tau+\log \left(1 + \eta(a,j,n) \right)\right)^{1-\alpha}}{\tau^{-\alpha}}},\frac{\alpha(2^{(\alpha-1)}-1)}{\tau(\alpha-1)}+\frac{1}{\eta(a,j,n)} \right), 
\end{equation}
for all $a \in \mathcal{B}$, $j \in \mathcal{M}_a, n \in \mathcal{N}$. 
\end{condition}
\begin{condition}
\label{cond3}
\begin{equation}
 \tau < 1 \; \; \mbox{and},\nonumber
\end{equation}
\begin{equation}
\beta(a,j,n) \geq \max\left(\frac{1}{\eta(a,j,n)}+ \frac{1}{\tau}, \frac{1}{\tau \log(1+ \frac{\log(1+\eta(a,j,n))}{\tau})}\right),
\end{equation}
for all $a \in \mathcal{B}$, $j \in \mathcal{M}_a, n \in \mathcal{N}$. 
\end{condition}
\begin{theorem}
For $\alpha \in (0,2)\backslash \{1\}$ (respectively, $\alpha \geq 2$, $\alpha =1$), an optimal solution to Problem~\ref{PB:ICIC:fixed:powers} can be found in $\mathcal{O}((M+N)^3)$ time using the algorithm in Fig.~\ref{algo}, if Condition~\ref{cond1} (respectively, Condition~\ref{cond2}, Condition~\ref{cond3}) is satisfied.
\label{thm_cond}
\end{theorem}

Now, we explain Theorem~\ref{thm_cond} and Conditions~\ref{cond1},~\ref{cond2} and~\ref{cond3}. For a given MS $j \in \mathcal{M}_a$ and subchannel $n$, ``crosstalk coefficients'' are the channel gains $H^{n}_{b,j}, \; \forall b \in \mathcal{B} \backslash \{a\}$. For a given MS $j \in \mathcal{M}_a$ and subchannel $n$, note that $\beta(a,j,n)$ is the ratio of the least crosstalk coefficient to the value of the channel gain, $H^{n}_{a,j}$,  from the BS, $a$, serving $j$ to $j$. Therefore, $\beta(a,j,n)$ is the minimum value of the potential~\textsuperscript{2}\footnote{$^2$We say ``potential" interference because an MS $j \in \mathcal{M}_a$ experiences interference only when subchannel $n$ is assigned to it and to an MS of BS $b \neq a$.} interference to MS $j$ on subchannel $n$ relative to the signal strength from BS $a$. Conditions~\ref{cond1},~\ref{cond2} and~\ref{cond3} hold when $\beta(a,j,n)$ is greater than a threshold value for all $a,\;j$ and $n$.  Hence, Theorem~\ref{thm_cond} says that \emph{Problem~\ref{PB:ICIC:fixed:powers} is polynomial time solvable if the potential interference levels are sufficiently high for all BSs, MSs and subchannels.} Moreover, the threshold interference levels in Conditions~\ref{cond1},~\ref{cond2} and~\ref{cond3} vary inversely with $\eta(a,j,n)$, and hence, from~\eqref{eta}, with the transmit power $(P)$ of each of the BSs. Therefore, Conditions~\ref{cond1},~\ref{cond2} and~\ref{cond3} become more relaxed as the transmit power $P$ increases.

In a practical scenario where BSs are densely deployed in an area (resulting in high crosstalk coefficients) and transmit with high power, Conditions~\ref{cond1},~\ref{cond2} and~\ref{cond3} would be satisfied in several of the time slots. In such a scenario, the algorithm in Fig.~\ref{algo} can be used to find an optimal solution to Problem~\ref{PB:ICIC:fixed:powers} in polynomial time. 

Next, a natural question arises as to whether, for a given value of $\alpha$,  there exist values of $\tau,\; \eta$ and $\beta$~\textsuperscript{3}~\footnote{$^3$ For simplicity, we have replaced $\eta(a,j,n)$ and $\beta(a,j,n)$ by $\eta$ and $\beta$ respectively.} that satisfy the applicable condition out of Conditions~\ref{cond1},~\ref{cond2} and~\ref{cond3}.
The answer is yes: Table~\ref{t1} illustrates, for different values of $\alpha$, some example values of $\tau,\; \eta$ and $\beta$ for which the applicable condition out of Conditions~\ref{cond1},~\ref{cond2} and~\ref{cond3} is satisfied and hence Problem~\ref{PB:ICIC:fixed:powers} can be solved in polynomial time by Theorem~\ref{thm_cond}. The values in Table~\ref{t1} have been obtained via numerical computations using the Matlab software.  
\begin{table}[h]
\renewcommand{\arraystretch}{1.2}
\centering
\caption{Example Parameter Values}
%\large
\begin{tabular}{|c|c|c|c|}
\hline 
$\alpha$ & $\eta$ & $\tau$ & $\beta$\\%[0.75ex]
\hline 
 1&$[10^2, 10^3]$ & $[0.99,1)$ & $\geq1.021$ \\%[0.75ex]
\hline
$[0.01, 2)\backslash \{1\}$ &  $[10^2, 10^3]$ & \begin{tabular}{@{}c@{}}$\alpha-k,$ \\ $\forall k \in [10^{-4}, 10^{-3}]$\end{tabular}  & $\geq 1.13$\\%[0.75ex]
\hline
$[2.7, 2.9]$  & $[30, 33]$ & \begin{tabular}{@{}c@{}}$\log(1+\eta)+k,$ \\  $\forall k \in [10^{-2}, 10^{-1}]$\end{tabular} & $\geq 1.25$\\%[0.75ex]
 \hline
 $[3.5, 3.7]$ & $(500, 600)$ & \begin{tabular}{@{}c@{}}$\log(1+\eta)+k,$\\ $\forall k \in [10^{-3}, 10^{-2}]$\end{tabular} & $\geq 1.22$\\%[0.75ex]
\hline
 $[4.4, 4.6]$ & $[790, 900]$ &  \begin{tabular}{@{}c@{}}$\log(1+\eta)+k,$\\ $\forall k \in [5, 5.5]$\end{tabular}& $\geq 1.22$\\
%[0.75ex]
 \hline
 $[5.3, 5.5]$ & $[810, 900]$ &  \begin{tabular}{@{}c@{}}$\log(1+\eta)+k,$\\ $\forall k \in [15, 16]$\end{tabular}& $\geq 1.22$\\
%[0.75ex]
 \hline
\end{tabular}
\label{t1}
\end{table}    

We provide the proof of Theorem~\ref{thm_cond} in the rest of this subsection. We divide the proof into two cases depending on the values of the crosstalk coefficients. Specifically, in Section~\ref{infinit}, we assume an idealized situation wherein all the crosstalk coefficients are $\infty$. Note that the crosstalk coefficients are not $\infty$ in practice. However, we assume them to be $\infty$ in Section~\ref{polynomial time solution}1 for ease of understanding, \emph{i.e.,} we assume that the  following condition holds:
\begin{condition}
Assume that the channel gains $H_{b,j}^n = \infty$ and $H_{a,j}^n$ are finite for all $a \in \mathcal{B}, j \in \mathcal{M}_a, n \in \mathcal{N}, b \neq a.$
\label{cond_inf}
\end{condition}
Subsequently, in Section~\ref{finite}, we consider the realistic case in which all the crosstalk coefficients are finite. We prove that  if the potential interference levels are  higher than the thresholds given in Conditions~\ref{cond1},~\ref{cond2} and~\ref{cond3}, then the optimal solution to Problem~\ref{PB:ICIC:fixed:powers} can be found in polynomial time using the algorithm in Fig.~\ref{algo}. Observe that the inequalities involving $\beta(a,j,n)$ stated  in Conditions~\ref{cond1},~\ref{cond2} and~\ref{cond3} are satisfied whenever Condition~\ref{cond_inf} holds, but Condition~\ref{cond_inf} is a much more relaxed condition than Conditions~\ref{cond1},~\ref{cond2} and~\ref{cond3}.

We now introduce some terminology and notations. Consider a weighted undirected graph $G =(\mathcal{V},\mathcal{E})$, where $\mathcal{V}$ (respectively, $\mathcal{E}$) is the set of nodes (respectively, set of edges), and each edge has a weight, which is a real number. Let $(i,j)$ denote the edge between  nodes $i,j \in \mathcal{V}$, where $i \neq j.$ A graph $G =(\mathcal{V},\mathcal{E})$ is \emph{bipartite} if $\mathcal{V} = \mathcal{V}_1 \cup \mathcal{V}_2$ such that $\mathcal{V}_1 \cap \mathcal{V}_2 = \emptyset$ and each edge in $\mathcal{E}$ is between a node in $\mathcal{V}_1$ and a node in $\mathcal{V}_2$~\cite{RF:kleinberg:algorithm}. A subset $\mathcal{E}_m \subseteq \mathcal{E}$ is known as a \emph{matching} if no two edges in $\mathcal{E}_m$ have a node in common~\cite{RF:kleinberg:algorithm}. The sum of the weights of the edges in a given matching $\mathcal{E}_m$ is known as the weight of the matching $\mathcal{E}_m$ and is denoted by $W(\mathcal{E}_m)$. The 
problem of obtaining a matching with the maximum weight in a bipartite graph is known as the \emph{bipartite matching problem}~\cite{RF:kleinberg:algorithm}.

\subsubsection{Infinite Crosstalk Coefficients}
\label{infinit}
\begin{theorem}
\label{TH:infinite:crosstalk:coefficients}
If Condition~\ref{cond_inf} is satisfied, then the optimal solution to Problem~\ref{PB:ICIC:fixed:powers} can be found in $\mathcal{O}((M+N)^3)$ time using the algorithm in Fig.~\ref{algo}.
\end{theorem}
\begin{IEEEproof}
Consider Problem~\ref{PB:ICIC:fixed:powers} with Condition~\ref{cond_inf}. We  show the equivalence of Problem~\ref{PB:ICIC:fixed:powers} with the bipartite matching problem. Let $\mathcal{V}_1 = \mathcal{M}$ (the set of
MSs) and  $\mathcal{V}_2=\mathcal{N}$ (the set of subchannels) in a bipartite graph. Consider an MS $j \in \mathcal{M}_a \subseteq \mathcal{M}$ and a subchannel $n \in \mathcal{N}$. Then the weight of the edge $(j,n)$ is defined as:
\begin{equation}
\label{wt}
W(j,n)=\left\{ 
\begin{array}{ll}
 \log \left(\tau+\log \left( 1 + \frac{P H_{a,j}^n}{ N_0} \right)\right), & \mbox{if} \; \;\alpha=1, \\
  \frac{\left(\tau+\log \left( 1 + \frac{P H_{a,j}^n}{ N_0} \right) \right)^{1-\alpha}}{1-\alpha}, & \mbox{if } \alpha \neq 1. \\
\end{array}
\right.
\end{equation}
Suppose $\mathcal{Z}^1 \subseteq \mathcal{Z}$ is the set of all feasible allocations in which no subchannel $n \in \mathcal{N}$ is assigned to two or more MSs, \emph{i.e.,} $\sum_{i \in \mathcal{B}} y_i^n \leq 1, \ \forall n \in \mathcal{N}$. Let the allocation $\mathbf{Z}(\mathcal{E}_m)$ corresponding to a matching, $\mathcal{E}_m$, in the above bipartite graph be as follows:
\begin{equation}
z_{a,j}^n = \left\{ 
\begin{array}{ll}
1, & \mbox{if } (j,n) \in \mathcal{E}_m, \\
0, & \mbox{else.}
\end{array}
\right. 
\end{equation}
Note that $\mathbf{Z}(\mathcal{E}_m) \in \mathcal{Z}^1$ since $\mathcal{E}_m$ is a matching. Also, there always exists a unique matching $\mathcal{E}_m$ corresponding to any $\mathbf{Z}^1 \subseteq \mathcal{Z}^1$ such that $\mathbf{Z}(\mathcal{E}_m) = \mathbf{Z}^1$. Therefore, $\mathbf{Z}(\mathcal{E}_m)$ is a one-to-one mapping between the set $\mathcal{Z}^1$ and the set of all matchings in the above bipartite graph. Further, from~\eqref{EQ:objective1} and~\eqref{wt}, it follows that the weight of a matching $\mathcal{E}_m$ is equal to the utility of the corresponding allocation $\mathbf{Z}(\mathcal{E}_m)$, \emph{i.e.,} $W(\mathcal{E}_m)= U_{\alpha,\tau}(\mathbf{Z}(\mathcal{E}_m))$. Therefore, $\mathbf{Z}(\mathcal{E}_m^*) \in \mathcal{Z}^1,$ is the allocation in $\mathcal{Z}^1$ with the highest utility if $\mathcal{E}_m^*$ is the matching with maximum weight.

Now, we want to show that under Condition~\ref{cond_inf}, $\mathbf{Z}(\mathcal{E}_m^*)$ is the allocation with the highest utility in $\mathcal{Z}$. Note that if a subchannel $n$ is allocated to more than one MS in an allocation $\mathbf{Z} \in \mathcal{Z},$ then the contribution to the network utility $U_{\alpha,\tau}(\mathbf{Z})$ by each of those MSs to which subchannel $n$ is allocated, will be $\log(\tau)$ if $\alpha=1$ and $\frac{\tau^{1-\alpha}}{1-\alpha}$ if $\alpha \neq 1$ by~\eqref{EQ:objective1},~\eqref{EQ:UnZ} and Condition~\ref{cond_inf}. This contribution is equal to the contribution of an MS to which no subchannel is assigned. Therefore, by~\eqref{EQ:objective1}, an allocation $\mathbf{Z}^1 \in \mathcal{Z}^1$, such that $U_{\alpha,\tau}(\mathbf{Z}^1)= U_{\alpha,\tau}(\mathbf{Z})$, can be derived from $\mathbf{Z}$ by deallocating each subchannel $n$ which is allocated to two or more MSs in $\mathbf{Z}$ from all the
MSs to which it was allocated. Since $\mathbf{Z}(\mathcal{E}_m^*)$ is the allocation in $\mathcal{Z}^1$ with the highest utility, it follows that under Condition~\ref{cond_inf},  $\mathbf{Z}(\mathcal{E}_m^*)$ is the allocation in $\mathcal{Z}$ with the highest utility.

Thus, if Condition~\ref{cond_inf} is satisfied, then the optimal solution of Problem~\ref{PB:ICIC:fixed:powers} is the allocation $\mathbf{Z}(\mathcal{E}_m^*)$ corresponding
to the maximum weight matching $\mathcal{E}_m^*$ in the above
bipartite graph. The Hungarian algorithm can be used to solve the bipartite matching problem in $O(d^3)$ time for a bipartite graph with $d$ nodes~\cite{hung}. Hence, Problem~\ref{PB:ICIC:fixed:powers} can be solved in $O((M+N)^3)$ time. Finally, an algorithm to optimally solve Problem~\ref{PB:ICIC:fixed:powers} when Condition~\ref{cond_inf} holds is provided in Fig.~\ref{algo}.
\end{IEEEproof}
\begin{figure}[!hbt]
\mbox{}\hrulefill \\
\begin{scriptsize}
\begin{algorithmic}[1]
\STATE{Suppose $\mathcal{M}$ and $\mathcal{N}$ are the two partitions of a given bipartite graph, and for $j \in \mathcal{M}_a \subseteq \mathcal{M}$, $n \in \mathcal{N}$, calculate the weight of edge $(j,n)$ using \eqref{wt}.}
\STATE{Using the Hungarian algorithm, solve the bipartite matching problem for this graph and obtain a maximum weight matching $\mathcal{E}_m^*$.}
\STATE{Return the allocation $z_{a,j}^n = \left\{ \begin{array}{ll}
1, & \mbox{if } (j,n) \in \mathcal{E}_m^*, \\
0, & \mbox{else}. 
\end{array} \right.$} 
\end{algorithmic}
\end{scriptsize}
\mbox{}\hrulefill
\caption{\label{bipartite_algo} The algorithm for optimally solving  Problem~\ref{PB:ICIC:fixed:powers} when, depending on the value of $\alpha,$ one of the Conditions~\ref{cond1},~\ref{cond2} and~\ref{cond3} is satisfied .}
\label{algo}
\end{figure}

\subsubsection{Finite Crosstalk Coefficients}
\label{finite}
We now consider the realistic case where the channel gains $H_{i,j}^n$ are finite for all $i \in \mathcal{B}, j \in \mathcal{M}, n \in \mathcal{N}$, as in practice.
%For BS $a \in \mathcal{B}$, MS $j \in \mathcal{M}_a$ and subchannel $n \in \mathcal{N}$, let:
%\begin{equation}
%\label{eta}
%\eta(a,j,n) = \frac{P H_{a,j}^n}{N_0}, 
%\end{equation}
%and
%\begin{equation}
%\label{EQ:beta}
%\beta(a,j,n) = \frac{\min_{b \in \mathcal{B} \backslash \{ a \}} H_{b,j}^n}{H_{a,j}^n}. 
%\end{equation}
%Consider the following conditions:
%\begin{condition}
%\label{cond1}
%\begin{equation}
%\beta(a,j,n) \geq \frac{\alpha}{\tau}+\frac{1}{\eta(a,j,n)},\;\mbox{and}\;\; \tau < \alpha.
%\end{equation}
%%for all $a \in \mathcal{B}$, $j \in \mathcal{M}_a, n \in \mathcal{N}$ and $\alpha > 0$. 
%\end{condition}
%\begin{condition}
%\label{cond2}
%\begin{equation}
%\frac{\alpha(2^{(\alpha-1)}-1)}{\tau(\alpha-1)}>1,\; \log(1+\eta(a,j,n)) \leq \tau \; \mbox{and} \;\; \beta(a,j,n) \geq \nonumber
%\end{equation}
%\begin{equation}
% \max\left(\frac{\alpha-1}{\tau-\frac{\left(\tau+\log \left(1 + \eta(a,j,n) \right)\right)^{1-\alpha}}{\tau^{-\alpha}}},\frac{\alpha(2^{(\alpha-1)}-1)}{\tau(\alpha-1)}+\frac{1}{\eta(a,j,n)} \right) 
%\end{equation}
%\end{condition}
%\begin{condition}
%\label{cond3}
%\begin{equation}
% \tau < 1 \; \; \mbox{and},\nonumber
%\end{equation}
%\begin{equation}
%\beta(a,j,n) \geq \max\left(\frac{1}{\eta(a,j,n)}+ \frac{1}{\tau}, \frac{1}{\tau \log(1+ \frac{\log(1+\eta(a,j,n))}{\tau})}\right).
%\end{equation}
%for all $a \in \mathcal{B}$, $j \in \mathcal{M}_a, n \in \mathcal{N}$ and $\alpha,\tau > 0$. 
%\end{condition}

%\begin{IEEEproof}
%The proof is relegated to the Appendix.
%\end{IEEEproof} 
To prove Theorem~\ref{thm_cond}, we will  start with an optimal allocation $\mathbf{\hat{Z}} \in \mathcal{Z}$, and  deallocate some MSs from one or
more subchannels, if necessary,  to obtain an allocation $\mathbf{Z}^* \in \mathcal{Z}^1$, such that $U_{\alpha,\tau}(\mathbf{Z}^*) \geq U_{\alpha,\tau}(\mathbf{\hat{Z}})$. From this and the fact that $\mathbf{\hat{Z}}$ is an optimal allocation, it will follow that there exists an allocation $\mathbf{Z}^* \in \mathcal{Z}^1$ 
that maximizes the utility in~\eqref{EQ:objective1}. The allocation $\mathbf{Z}^*$ can be found using the algorithm provided in  Fig.~\ref{algo}, which will prove Theorem~\ref{thm_cond}. 

Now, consider two cases based on the value of $\alpha$.

\textbf{Case 1: $\alpha \neq 1$}\\
%$U_{\alpha,\tau}(\mathbf{Z})=
%\sum_{j \in \mathcal{M}}  \frac{(\tau+U_{j}(\mathbf{Z}))^{1-\alpha}}{1-\alpha},$\\
Let $\mathbf{\hat{Z}} = \{z_{a,j}^n: a \in \mathcal{B}, j \in \mathcal{M}_a, n \in \mathcal{N}\} \in \mathcal{Z}$ be an optimal allocation. For a given subchannel $n \in \mathcal{N}$,  let $(a^*(n), j^*(n))  =$
\small
\begin{equation}
  \argmax_{a \in \mathcal{B}, j \in \mathcal{M}_a: z_{a,j}^n = 1} 
\frac{\left(\tau+\log \left( 1 + \frac{P H_{a,j}^n}{P \sum_{i \in \mathcal{B} \backslash \{a\}: y_i^n = 1} H_{i,j}^n + N_0} \right)\right)^{1-\alpha}}{1-\alpha}.  
\label{EQ:aj:argmax1}
\end{equation}
\normalsize
It can be observed that the (BS, MS) pair $(a^*(n), j^*(n))$ contributes the highest to $U_{\alpha,\tau}(\mathbf{\hat{Z}})$ in the RHS of \eqref{EQ:objective1}. Suppose subchannel $n$ is allocated to $k_n$ MSs in the allocation $\mathbf{\hat{Z}}$, \emph{i.e.}:
\begin{equation}
\label{EQ:no:MSs:on:subchannel}
k_n = \left| \left\{ i \in \mathcal{B}: y_{i}^n =1 \right\} \right|.  
\end{equation}
If $k_n \geq 2$, then the allocation $\mathbf{Z}^*$ is obtained from $\mathbf{\hat{Z}}$ by deallocating all MSs other than $j^*(n)$  from subchannel $n$~\textsuperscript{4}~\footnote{$^4$Similar deallocations of MSs from subchannels other than $n$ are performed.}. We can write $U_{\alpha,\tau}(\mathbf{\hat{Z}})$
\small
\begin{eqnarray}
  & \hspace{-.6cm}\leq & \hspace{-.4cm}\sum_{n \in \mathcal{N}} k_n \frac{\left(\tau+\log \left( 1 + \frac{P H_{a^*(n),j^*(n)}^n}{P \sum_{i \in \mathcal{B} \backslash \{a^*(n)\}: y_i^n = 1} H_{i,j^*(n)}^n + N_0} \right)\right)^{1-\alpha}}{1-\alpha} \nonumber \\
& \hspace{-.6cm}\leq &\hspace{-.4cm} \sum_{n \in \mathcal{N}} k_n \frac{\left(\tau+\log \left( 1 + \frac{P H_{a^*(n),j^*(n)}^n}{(k_n-1)P H_{b,j^*(n)}^n + N_0} \right)\right)^{1-\alpha}}{1-\alpha}\nonumber \\
& \hspace{-.6cm}\leq & \hspace{-.4cm}\sum_{n \in \mathcal{N}} k_n \frac{\left(\tau+\log \left( 1 + \frac{\eta(a^*,j^*,n)}{(k_n-1)\eta(a^*,j^*,n)\beta(a^*,j^*,n) + 1} \right)\right)^{1-\alpha}}{1-\alpha}\nonumber\\
 \label{upper:bound:1} 
\end{eqnarray}
\normalsize
\begin{equation}
\label{EQ:Hbjn}
\mbox{where, } \ H_{b,j^*(n)}^n = \min_{i \in \mathcal{B} \backslash \{ a^*(n) \}} H_{i,j^*(n)}^n \nonumber. 
\end{equation}
In~\eqref{upper:bound:1}, the first inequality follows from~\eqref{EQ:aj:argmax1} and the last inequality follows from~\eqref{eta} and~\eqref{EQ:beta}. Hereafter, for simplicity, we replace $\eta(a^*,j^*,n)$ and $\beta(a^*,j^*,n)$ by $\eta$ and $\beta$ respectively.
We want to show that:
\small
\begin{eqnarray}
&& \hspace{-1cm}\sum_{n \in \mathcal{N}} \frac{k_n\left(\tau+\log \left( 1 + \frac{\eta}{(k_n-1)\eta\beta + 1} \right)\right)^{1-\alpha}}{1-\alpha} \nonumber\\
& \leq & \hspace{-0.3cm}\sum_{n \in \mathcal{N}}\left[ \frac{\left(\tau+\log \left( 1 + \eta \right)\right)^{1-\alpha}}{1-\alpha}+ (k_n-1)\frac{\tau^{1-\alpha}}{1-\alpha}\right] = U_{\alpha,\tau}(\mathbf{Z^*}),\nonumber\\
 \label{alpgen} 
\end{eqnarray}
%\begin{eqnarray}
%\sum_{n \in \mathcal{N}} \frac{k_n\left(\tau+\log \left( 1 + \frac{\eta}{(k_n-1)\eta\beta + 1} \right)\right)^{1-\alpha}}{1-\alpha} 
%& \leq & \sum_{n \in \mathcal{N}}\left[ \frac{\left(\tau+\log \left( 1 + \eta \right)\right)^{1-\alpha}}{1-\alpha}+ (k_n-1)\frac{\tau^{1-\alpha}}{1-\alpha}\right]
% \label{alpgen} 
%\end{eqnarray}
\normalsize
%\begin{eqnarray}
%U_{\alpha,\tau}(\mathbf{Z})  & \leq & \sum_{n \in \mathcal{N}} k_n \frac{\left(\tau+\log \left( 1 + \frac{\eta}{(k_n-1)\eta\beta + 1} \right)\right)^{1-\alpha}}{1-\alpha} \nonumber \\
%& \leq & \sum_{n \in \mathcal{N}}\left[ \frac{\left(\tau+\log \left( 1 + \eta \right)\right)^{1-\alpha}}{1-\alpha}+ (k_n-1)\frac{\tau^{1-\alpha}}{1-\alpha}\right]
% \label{alpgen} 
%\end{eqnarray}
where the equality follows from the definition of $\mathbf{Z^*}$. Let
\begin{equation}
\label{EQ:fx}
f(x)= x \frac{\left(\tau+\log \left( 1 + \frac{\eta}{(x-1)\eta\beta + 1} \right)\right)^{1-\alpha}}{1-\alpha}-(x-1)\frac{\tau^{1-\alpha}}{1-\alpha}.
\end{equation}
To prove~\eqref{alpgen}, it suffices to show that $f(x) \leq f(1)\; \forall \; x \geq 1.$\\
\begin{lemma}
Let $\eta, \; \beta$ and $\tau$ be positive numbers such that for $\alpha \in (0,2)\backslash \{1\}$ (respectively, $\alpha \geq 2$), Condition~\ref{cond1} (respectively, Condition~\ref{cond2}) is satisfied and $f(x)$ be as in \eqref{EQ:fx}.
Then $f(x) \leq f(1)\; \forall \; x \geq 1$.
\label{l1}
\end{lemma}
\begin{IEEEproof}
The proof is relegated to Appendix A. 
\end{IEEEproof}

The inequality in \eqref{alpgen} follows from  Lemma~\ref{l1}. By \eqref{upper:bound:1} and \eqref{alpgen}, we can write:
\small
\begin{eqnarray}
U_{\alpha,\tau}(\mathbf{\hat{Z}})&\leq &\sum_{n \in \mathcal{N}}\left[ \frac{\left(\tau+\log \left( 1 + \eta \right)\right)^{1-\alpha}}{1-\alpha}+ (k_n-1)\frac{\tau^{1-\alpha}}{1-\alpha}\right] \nonumber\\
&= &U_{\alpha,\tau}(\mathbf{Z^*}).
 \label{alpp} 
\end{eqnarray}
\normalsize
%\begin{equation}
%\hspace{-5.4cm} \Leftrightarrow U_{\alpha,\tau}(\mathbf{\hat{Z}})\leq U_{\alpha,\tau}(\mathbf{Z^*})
% \label{alpp} 
%\end{equation}
Hence, from the fact that $\mathbf{\hat{Z}}$ is an optimal allocation and \eqref{alpp}, it follows that $\mathbf{Z^*}$ is also an optimal solution of Problem~\ref{PB:ICIC:fixed:powers}. Note that $\mathbf{Z^*} \in \mathcal{Z}^1$. Also, recall from the proof of Theorem~\ref{TH:infinite:crosstalk:coefficients} that $\mathbf{Z}(\mathcal{E}_m^*)$ is the allocation in $\mathcal{Z}^1$ with the highest utility. Thus, $\mathbf{Z}(\mathcal{E}_m^*)$ is also an optimal solution of Problem~\ref{PB:ICIC:fixed:powers}. Finally, recall that $\mathbf{Z}(\mathcal{E}_m^*)$  can be found by finding the maximum weight matching $\mathcal{E}_m^*$ in the
bipartite graph defined in the proof of Theorem~\ref{TH:infinite:crosstalk:coefficients} and finding the corresponding allocation. Hence, the algorithm in Fig.~\ref{algo} can be used to optimally solve Problem~\ref{PB:ICIC:fixed:powers} in  $\mathcal{O}((M+N)^3)$
time when Condition~\ref{cond1} holds for $\alpha \in [0,2)\backslash \{1\}$ and Condition~\ref{cond2} holds for $\alpha \geq 2$.

\textbf{Case 2: $\alpha =1$}\\
Let $\mathbf{\hat{Z}} = \{z_{a,j}^n: a \in \mathcal{B}, j \in \mathcal{M}_a, n \in \mathcal{N}\} \in \mathcal{Z}$ be an optimal allocation. For a given subchannel $n \in \mathcal{N}$, let $(a^*(n), j^*(n))=$
\small
\begin{equation}
 \argmax_{a \in \mathcal{B}, j \in \mathcal{M}_a: z_{a,j}^n = 1} 
\log\left(\tau+\log \left( 1 + \frac{P H_{a,j}^n}{P \sum_{i \in \mathcal{B} \backslash \{a\}: y_i^n = 1} H_{i,j}^n + N_0} \right)\right).  
\label{EQ:aj:argmax2}
\end{equation}
\normalsize
It can be observed that the (BS, MS) pair $(a^*(n), j^*(n))$ contributes the highest to $U_{\alpha,\tau}(\mathbf{\hat{Z}})$ in the RHS of \eqref{EQ:objective1}. Suppose subchannel $n$ is allocated to $k_n$ MSs in the allocation $\mathbf{\hat{Z}}$, \emph{i.e.}:
\begin{equation}
\label{EQ:no:MSs:on:subchannel:n}
k_n = \left| \left\{ i \in \mathcal{B}: y_{i}^n =1 \right\} \right|.  
\end{equation}
If $k_n \geq 2$, then the allocation $\mathbf{Z}^*$ is obtained from $\mathbf{\hat{Z}}$ by deallocating all MSs other than $j^*(n)$  from subchannel $n$~\textsuperscript{4}. We can write, $U_{\alpha,\tau}(\mathbf{\hat{Z}})$
\small
\begin{eqnarray}
  &\hspace{-0.5cm} \leq &\hspace{-0.5cm} \sum_{n \in \mathcal{N}} \hspace{-0.1cm}k_n \log \left(\tau+\log \left( 1 + \frac{P H_{a^*(n),j^*(n)}^n}{P \sum_{i \in \mathcal{B} \backslash \{a^*(n)\}: y_i^n = 1} H_{i,j^*(n)}^n + N_0} \right)\right) \nonumber \\
&\hspace{-0.5cm} \leq &\hspace{-0.4cm} \sum_{n \in \mathcal{N}} \hspace{-0.1cm}k_n \log \left(\tau+\log \left( 1 + \frac{P H_{a^*(n),j^*(n)}^n}{(k_n-1)P H_{b,j^*(n)}^n + N_0} \right)\right)\nonumber \\
&\hspace{-0.5cm} \leq &\hspace{-0.5cm} \sum_{n \in \mathcal{N}} \hspace{-0.1cm}k_n \log \left(\tau+\log \left( 1 + \frac{\eta(a^*,j^*,n)}{(k_n-1)\eta(a^*,j^*,n)\beta(a^*,j^*,n) + 1} \right)\right)\nonumber\\
 \label{EQ:UnZ:upper:bound:1} 
\end{eqnarray}
\normalsize
%\begin{eqnarray}
%U_{\alpha,\tau}(\mathbf{Z})  & \leq & \sum_{n \in \mathcal{N}} k_n \log \left(\tau+\log \left( 1 + \frac{P H_{a^*(n),j^*(n)}^n}{P \sum_{i \in \mathcal{B} \backslash \{a^*(n)\}: y_i^n = 1} H_{i,j^*(n)}^n + N_0} \right)\right) \nonumber \\
%& \leq & \sum_{n \in \mathcal{N}} k_n \log \left(\tau+\log \left( 1 + \frac{P H_{a^*(n),j^*(n)}^n}{(k_n-1)P H_{b,j^*(n)}^n + N_0} \right)\right)\nonumber \\
%& \leq & \sum_{n \in \mathcal{N}} k_n \log \left(\tau+\log \left( 1 + \frac{\eta(a^*,j^*,n)}{(k_n-1)\eta(a^*,j^*,n)\beta(a^*,j^*,n) + 1} \right)\right)
% \label{EQ:UnZ:upper:bound:1} 
%\end{eqnarray}
\normalsize
\begin{equation}
\mbox{where, } \ H_{b,j^*(n)}^n = \min_{i \in \mathcal{B} \backslash \{ a^*(n) \}} H_{i,j^*(n)}^n \nonumber. 
\end{equation}
In~\eqref{EQ:UnZ:upper:bound:1}, the first inequality follows from~\eqref{EQ:aj:argmax2} and the last inequality follows from~\eqref{eta} and~\eqref{EQ:beta}. Hereafter, for simplicity, we replace $\eta(a^*,j^*,n)$ and $\beta(a^*,j^*,n)$ by $\eta$ and $\beta$ respectively.
 We want to show that
\begin{eqnarray}
&&\hspace{-1cm} \sum_{n \in \mathcal{N}} k_n \log \left(\tau+\log \left( 1 + \frac{\eta}{(k_n-1)\eta\beta + 1} \right)\right)\nonumber\\
 &\leq & \hspace{-0.3cm}\sum_{n \in \mathcal{N}} \left[\log \left(\tau+\log \left( 1 + \eta \right)\right)+(k_n-1)\log \tau\right]= U_{\alpha,\tau}(\mathbf{Z^*}),\nonumber\\
 \label{alp1} 
\end{eqnarray}
%\begin{eqnarray}
%U_{\alpha,\tau}(\mathbf{Z})& \leq & \sum_{n \in \mathcal{N}} k_n \log \left(\tau+\log \left( 1 + \frac{\eta}{(k_n-1)\eta\beta + 1} \right)\right)\nonumber \\
%& \leq & \sum_{n \in \mathcal{N}} \log \left(\tau+\log \left( 1 + \eta \right)\right)+(k_n-1)\log \tau\nonumber \\
% \label{alp1} 
%\end{eqnarray}
where the equality follows from the definition of $\mathbf{Z^*}$. Let
\begin{equation}
 f_1(x)= x \log \left(\tau+\log \left( 1 + \frac{\eta}{(x-1)\eta\beta + 1}\right)\right)-(x-1)\log\tau.
 \label{EQ:f1x}
\end{equation}
To prove~\eqref{alp1}, it suffices to show that $f_1(x) \leq f_1(1)\; \forall \; x \geq 1.$\\
\begin{lemma}
Let $\eta, \; \beta$ and $\tau$ be positive numbers such that for $\alpha=1$, Condition~\ref{cond3} is satisfied and $f_1(x)$ be as in~\eqref{EQ:f1x}. 
Then $f_1(x) \leq f_1(1)\; \forall \; x \geq 1$.
\label{l2}
\end{lemma}
\begin{IEEEproof}
The proof is relegated to Appendix B. 
\end{IEEEproof}

The inequality in~\eqref{alp1} follows from Lemma~\ref{l2}. By~\eqref{EQ:UnZ:upper:bound:1} and~\eqref{alp1}, we can write:
\begin{eqnarray}
U_{\alpha,\tau}(\mathbf{\hat{Z}})&\leq &\sum_{n \in \mathcal{N}} \log \left(\tau+\log \left( 1 + \eta \right)\right)+(k_n-1)\log \tau \nonumber \\
&= &U_{\alpha,\tau}(\mathbf{Z^*}).
\label{alpp1}
\end{eqnarray}
%\begin{equation}
%\hspace{-5.1cm} \Leftrightarrow U_{\alpha,\tau}(\mathbf{\hat{Z}})\leq U_{\alpha,\tau}(\mathbf{Z^*})
% \label{alpp1} 
%\end{equation}
Hence, from the fact that $\mathbf{\hat{Z}}$ is an optimal allocation and \eqref{alpp1}, it follows that $\mathbf{Z^*}$ is also an optimal solution of Problem~\ref{PB:ICIC:fixed:powers}. Note that $\mathbf{Z^*} \in \mathcal{Z}^1$. Also, recall from the proof of Theorem~\ref{TH:infinite:crosstalk:coefficients} that $\mathbf{Z}(\mathcal{E}_m^*)$ is the allocation in $\mathcal{Z}^1$ with the highest utility. Thus, $\mathbf{Z}(\mathcal{E}_m^*)$ is also an optimal solution of Problem~\ref{PB:ICIC:fixed:powers}. Finally, recall that $\mathbf{Z}(\mathcal{E}_m^*)$  can be found by finding the maximum weight matching $\mathcal{E}_m^*$ in the
bipartite graph defined in the proof of Theorem~\ref{TH:infinite:crosstalk:coefficients} and finding the corresponding allocation. Hence, the algorithm in Fig.~\ref{algo} can be used to optimally solve Problem~\ref{PB:ICIC:fixed:powers} in  $\mathcal{O}((M+N)^3)$
time when Condition~\ref{cond3} holds for $\alpha = 1$.
This completes the proof of Theorem~\ref{thm_cond}.

%\vspace{-1em}
\section{ $\tau-\alpha-$Fair Distributed Subchannel Allocation Algorithm}
\label{SC:algorithms}
%\vspace{-.5em}
To approximately solve the NP-Complete Problem~\ref{PB:ICIC:fixed:powers} defined in Section~\ref{SC:model:objective}, we propose a simple, distributed subchannel allocation algorithm in this section. This algorithm is a generalization of an algorithm proposed in our prior work~\cite{Report2017} to solve the ICIC with fixed transmit power problem with the objective of maximizing the sum of throughputs of all the MSs in the network. 

%\subsection{Distributed $\alpha-$Fair Algorithm}
%\label{algorithm}
Let $\mathcal{B}_a \subseteq \mathcal{B}$ be the set of neighboring BSs of BS $a$. Every BS $a$ is directly connected to each of its neighboring BSs via high-speed links; these links are used to exchange information during the algorithm execution. For example in LTE systems, X2 interfaces~\cite{RF:ghosh:fundamentals:of:lte} are used to connect neighboring BSs.

The proposed algorithm proceeds as explained below:

During the initialization phase, the channel gain values are estimated as discussed in Section~\ref{SC:model:objective}. Each BS $a \in \mathcal{B}$ obtains channel gain information $\{H_{b,j}^n: j \in \mathcal{M}_a, b \in \mathcal{B}_a, n \in \mathcal{N}\}$ from its neighboring BSs in $\mathcal{B}_a$. In practice, each BS has a small number of neighboring BSs; therefore, the amount of information exchanged would be small.

After the initialization phase, the algorithm executes in iterations and each BS $a \in \mathcal{B}$ updates the variables $\{\hat{z}_{a,j}^n: j \in \mathcal{M}_a, n \in \mathcal{N}\}$, $\hat{y}_a^n$ and $\hat{y}_b^n$, $b \in \mathcal{B}_a$ after each iteration. Note that the temporary values of $z_{a,j}^n$ and $y_a^n$, specified in Section~\ref{SC:model:objective}, are contained in the variables $\hat{z}_{a,j}^n$ and $\hat{y}_a^n$ respectively after each iteration. Each BS $a$ initializes $\hat{z}_{a,j}^n = 0$, $\hat{y}_a^n = 0$ and $\hat{y}_b^n = 0$ for all $j \in \mathcal{M}_a, n \in \mathcal{N}$, $b \in \mathcal{B}_a$ at the beginning of the first iteration, and in subsequent iterations, if MS $j \in \mathcal{M}_a$ is allocated subchannel $n$, then BS $a$ assigns $\hat{z}_{a,j}^n = 1$ and correspondingly calculates the variable $\hat{y}_a^n = \sum_{j \in \mathcal{M}_a} \hat{z}_{a,j}^n$. The following operations are executed during each iteration $r = 1, 2, 3, \ldots$:
\begin{enumerate}[(1)]
\item 
\label{EN:algo:step1}
At the beginning of an iteration $r$, each BS $a \in \mathcal{B}$ computes and conveys $p_a$ to all the BSs in $\mathcal{B}_a$. For a BS $a$, $p_a$ is defined as:
\vspace{-1em}
\small
\begin{equation} 
 \hspace{-.5cm} \max_{j \in \mathcal{M}_a: \hat{z}_{a,j}^m = 0 \ \forall m \in \mathcal{N}} \max_{n \in \mathcal{N}: \hat{y}_a^n=0} \left\{ \log \left( 1 + \frac{P H_{a,j}^n}{P \displaystyle \sum_{b \in \mathcal{B}_a} H_{b,j}^n \hat{y}_b^n + N_0} \right)  \right\} 
\label{EQ:pa}
\end{equation}
%Similarly, compute $p_b \; \; \forall b \in \mathcal{B}_a$.
\normalsize
\item
\label{EN:algo:step2}
Let $j$ and $n$ be the maximizers in~\eqref{EQ:pa}. If $p_a \geq p_b  \ \forall b \in \mathcal{B}_a$, then the MS $j \in \mathcal{M}_a$ is assigned the subchannel $n$, and BS $a$ updates both the variables $\hat{z}_{a,j}^n$ and $\hat{y}_a^n$ to 1. Note that it is possible that multiple BSs allocate subchannels to their associated MS simultaneously in an iteration. 

\item
Each BS $a \in \mathcal{B}$ conveys the information of the subchannel, if any, allocated to one of its associated MSs in Step~\ref{EN:algo:step2}, say $n$, to all the BSs in $\mathcal{B}_a$ and updates the values of $\hat{y}_b^n \; \forall b \in \mathcal{B}_a, n \in \mathcal{N}$.  
\end{enumerate}

Each BS $a$ executes the above steps until at least one of the following conditions is fulfilled:
\begin{enumerate}[(i)]
\item
All the MSs in $\mathcal{M}_a$ are allocated subchannels.
\item
All the subchannels in $\mathcal{N}$ have been allocated to the MSs in $\mathcal{M}_a$.
\item
$p_a < p_{0}$, where $p_{0}$ is given by~\eqref{p} and~\eqref{alp} in Section~\ref{SC:simulations}.
\end{enumerate}
As soon as the algorithm terminates at BS $a$, its allocation is obtained using ${z}_{a,j}^n = \hat{z}_{a,j}^n, \forall j \in \mathcal{M}_a$ and $n \in \mathcal{N}$.

During each iteration, the distributed algorithm adopts a greedy approach in~\eqref{EQ:pa} and step~\ref{EN:algo:step2} to choose \emph{(MS, subchannel) pairs with high throughputs}. From~\eqref{EQ:pa} and the rule to update the variables $(\hat{y}_{i}^n : i \in \mathcal{B}, n \in \mathcal{N})$, $p_a$ either decreases or remains unchanged for each BS $a \in \mathcal{B}$ during each iteration. Also from~\eqref{p} and~\eqref{alp}, the higher the value of $\alpha$, the lower the value of $p_0$. By condition (iii) above for termination, when $\alpha$ is high, the distributed algorithm operates for a longer duration and hence subchannels are allocated to more MSs, which leads to high interference. Due to the increased interference, the total throughput is lower (see Sections~\ref{opt thr} and~\ref{tradeoff}), but the allocation is fairer since resources (subchannels) are allocated to more MSs. In summary, condition (iii) above for termination ensures that \emph{higher the value of $\alpha$, greater the degree of fairness and lower the total throughput of the allocation found by the above algorithm}. This is confirmed by the simulation results in Section~\ref{SC:simulations}. 
%\vspace{-1em}
\section{Simulations}
\label{SC:simulations}
%\vspace{-.5em}
In this section, we provide simulation results to investigate the trade-off between the total throughput and fairness achieved using the exhaustive search algorithm and the proposed $\tau-\alpha-$fair distributed subchannel allocation algorithm in Section~\ref{SC:algorithms}.

We consider the following scenario throughout our simulations. Suppose that $K$ BSs and $M$ MSs are placed uniformly at random in a square area of dimension 1$\times$1 $\mbox{unit}^2$. However, any two BSs must be at least $d_{min}$ distance apart from each other, where $d_{min}$ is a parameter. Let $d_{min} = 0.1$ units and suppose all the BSs which are within a radius of 0.4 units from BS $a$ are considered as the neighboring BSs of $a$ (\emph{i.e.,} in the set $\mathcal{B}_a$). Further, suppose the MS-BS association is distance dependent, \emph{i.e.,} each MS associates with the BS that is nearest to it.

To account for the effects of fast fading, shadow fading and the path loss phenomenon, we consider that the channel gains are given by $H_{i,j}^{n}=\frac{k S_{ij}X_{ij}^{n}}{d_{ij}^{\gamma}}$, where $d_{ij}$ denotes the distance between BS $i$ and MS $j$, $\gamma$ denotes the path loss exponent which can take values in the range $(2,4)$ and $k$ is a constant~\cite{Rapp}. To model the effect of shadow fading, a log-normal random variable $S_{ij}$ is considered. For distinct pairs $(i,j)$, $S_{ij}$ are independent and identically distributed (\emph{iid}) random variables. Similarly, Rayleigh distributed \emph{iid} random variables $X_{ij}^{n}$ are considered to model the effect of fast fading.

Next, we consider Jain's fairness index as a fairness metric which is defined as follows~\cite{Jain}:
\vspace{-.5em}
\begin{equation}
\label{fi}
\mbox{FI}= \frac{(\sum_{j=1}^{M} U_{j}(\mathbf{Z}))^2}{M (\sum_{j=1}^{M} U_{j}^{2}(\mathbf{Z}))},
\end{equation} 
where $U_{j}(\mathbf{Z})$ is given by~\eqref{EQ:UnZ}. The value of FI lies between 0 and 1. Also, it increases with the degree of fairness of the distribution of throughput; if all MSs get exactly equal throughput, it takes value 1 and it equals $\frac{n}{M}$ when exactly $n$ out of $M$ MSs have equal throughput and the remaining $(M-n)$ MSs have 0 throughput~\cite{Jain}. See~\cite{Jain} for further properties of the fairness index. 
%\vspace{-1em}
\subsection{Trade-off Between the Total Throughput and Fairness Index Under the Exhaustive Search Algorithm and Selection of $\tau$}
\label{exhaust}
%\vspace{-1em}
%\vspace*{-1em}
First, for different values of $\alpha$ we found the allocation that maximizes the system utility function in~\eqref{EQ:objective1} by exhaustive search over all possible combinations of subchannel allocation to all the MSs of the system. Then, the total throughput and fairness index FI were calculated for the obtained allocation using~\eqref{EQ:objective} and~\eqref{fi} respectively. Figs.~\ref{extv} and~\ref{extv1}\subref{3a} plot~\footnote{For all the plots in Figs.~\ref{extv} to~\ref{trd}, each data point was obtained by averaging
across 50 runs with different random seeds.} the variation of the total throughput and fairness index FI with $\alpha$, each for different values~\footnote{Note that for all the plots in Figs.~\ref{extv},~\ref{extv1} and~\ref{tau}, only small values of the parameters $K, N$ and $M$ were used since it is computationally prohibitive to execute the exhaustive search algorithm with large values of $K, N$ and $M$.} of parameters $K, N$ and $M$ and for three different values of $\tau$. In Figs.~\ref{extv} and~\ref{extv1}\subref{3a}, the total throughput decreases and fairness index FI increases with $\alpha$.
%\begin{figure}[!hbt]
%\begin{minipage}[b]{0.484\columnwidth}
%%\resizebox{1.14\columnwidth}{!}{\includegraphics{optconv_N11.eps}}
%\begin{subfigure}[t]{1in}
%\centering
%\includegraphics[width=4.6cm, height=4cm]{exhst3_3_11nww.eps}
%\subcaption{\tiny $K=3,N=3,M=11$}
%\label{1a}
%\end{subfigure}
%\end{minipage}
%\hspace{0.2mm}
%\begin{minipage}[b]{0.484\columnwidth}
%\begin{subfigure}[t]{1in}
%\centering
%%\resizebox{1.15\columnwidth}{!}{\includegraphics{optconv_N11.eps}}
%\includegraphics[width=4.6cm, height=4cm]{exst2_4_7nww.eps}
%\subcaption{\tiny $K=2, N=4, M=7$}\label{1b}
%\end{subfigure}
%\end{minipage}
%\centering
%%\vspace{-1em}
%\caption{The figure plots the total throughput and fairness index (FI) values obtained by exhaustive search over all possible subchannel allocations with $\alpha$ for different $K, N$ and $M$.}
%%\vspace{-1.5em}
%\label{extv}
%\end{figure}
%\vspace{-2em}
\begin{figure}%
    \centering
    \subfloat[For $K=3, N=3, M=11$]{\label{1a}{\includegraphics[width=4.2cm, height=4cm]{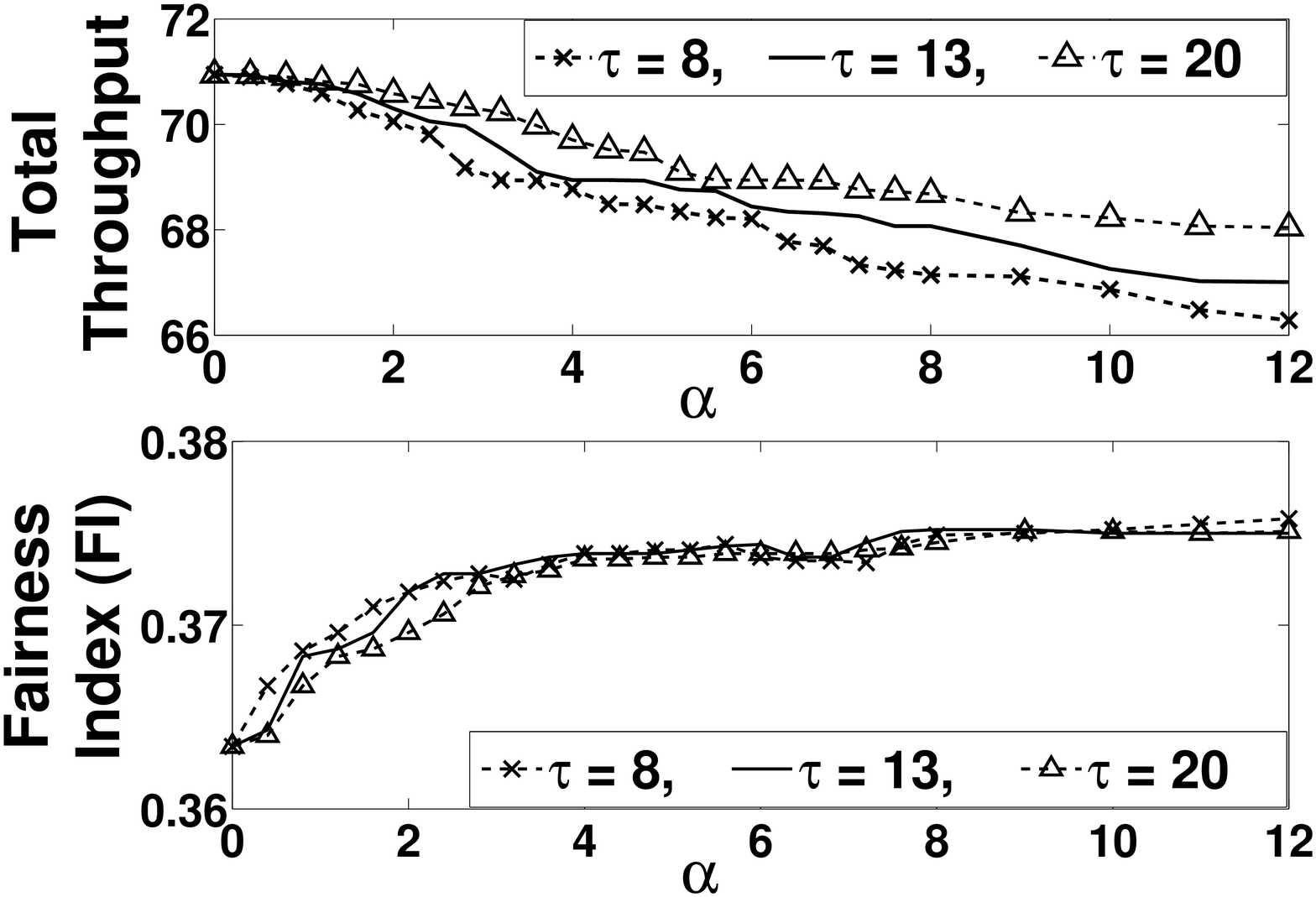} }}%
    \hspace{.1cm}
    \subfloat[For $K=2, N=4, M=7$]{\label{1b}{\includegraphics[width=4.2cm, height=4cm]{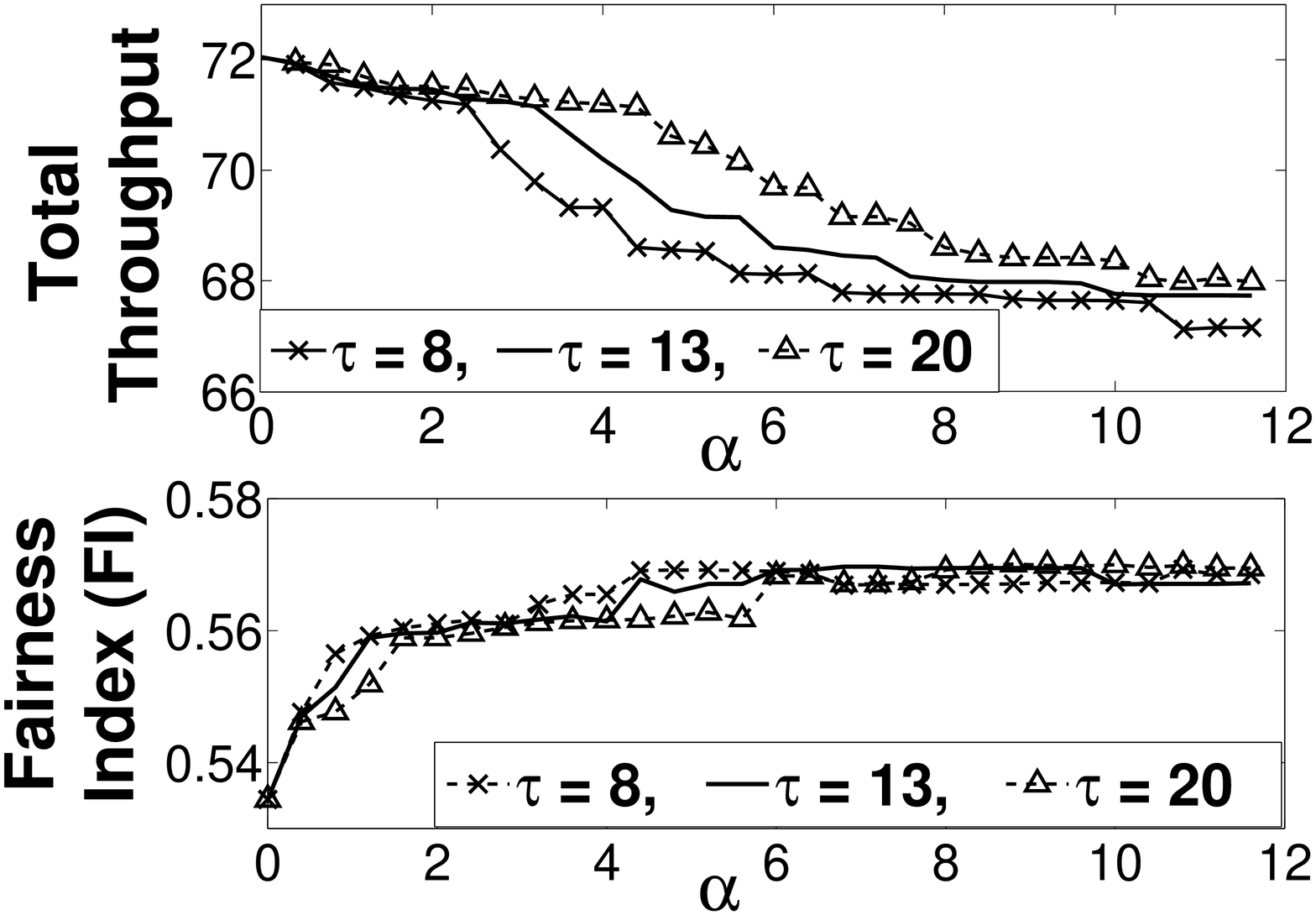} }}%
   % \hspace{.05cm}
    %\subfloat[For $K=3, N=3, M=11$]{\label{1c}{\includegraphics[width=4.5cm, height=4.5cm]{exhst3_3_11nww.eps} }}%
    \caption{The figure plots the total throughput and fairness index (FI) values obtained by exhaustive search over all possible subchannel allocations with $\alpha$ for different $K, N$ and $M$.}%
     \vspace{-1em}
    \label{extv}%
\end{figure} 
%\vspace{-1.5em}
%\begin{figure}[!hbt]
%\begin{minipage}[b]{0.484\columnwidth}
%%\resizebox{1.14\columnwidth}{!}{\includegraphics{optconv_N11.eps}}
%\begin{subfigure}[t]{1in}
%\centering
%\includegraphics[width=4.6cm, height=4cm]{exstv_nw.eps}
%\caption{\hspace*{-2.5cm}}\label{3a}
%\end{subfigure}
%\end{minipage}
%\hspace{0.2mm}
%\begin{minipage}[b]{0.484\columnwidth}
%\begin{subfigure}[t]{1in}
%\centering
%%\resizebox{1.15\columnwidth}{!}{\includegraphics{optconv_N11.eps}}
%\includegraphics[width=4.6cm, height=4cm]{var_tau4310nw.eps}
%\caption{\hspace*{-2.5cm}}\label{3b}
%\end{subfigure}
%\end{minipage}
%\centering
%%\vspace{-1em}
%\caption{The figure (a) (respectively, (b)) plots the total throughput and fairness index (FI) values obtained by exhaustive search over all possible subchannel allocations with $\alpha$ (respectively, $\tau$) for different $K, N$ and $M$.}
%%\vspace{-1.5em}
%\label{extv1}
%\end{figure}
%\vspace{-2em}
\begin{figure}%
    \centering
    \subfloat[For $K=3, N=4, M=10$]{\label{3a}{\includegraphics[width=4cm, height=4cm]{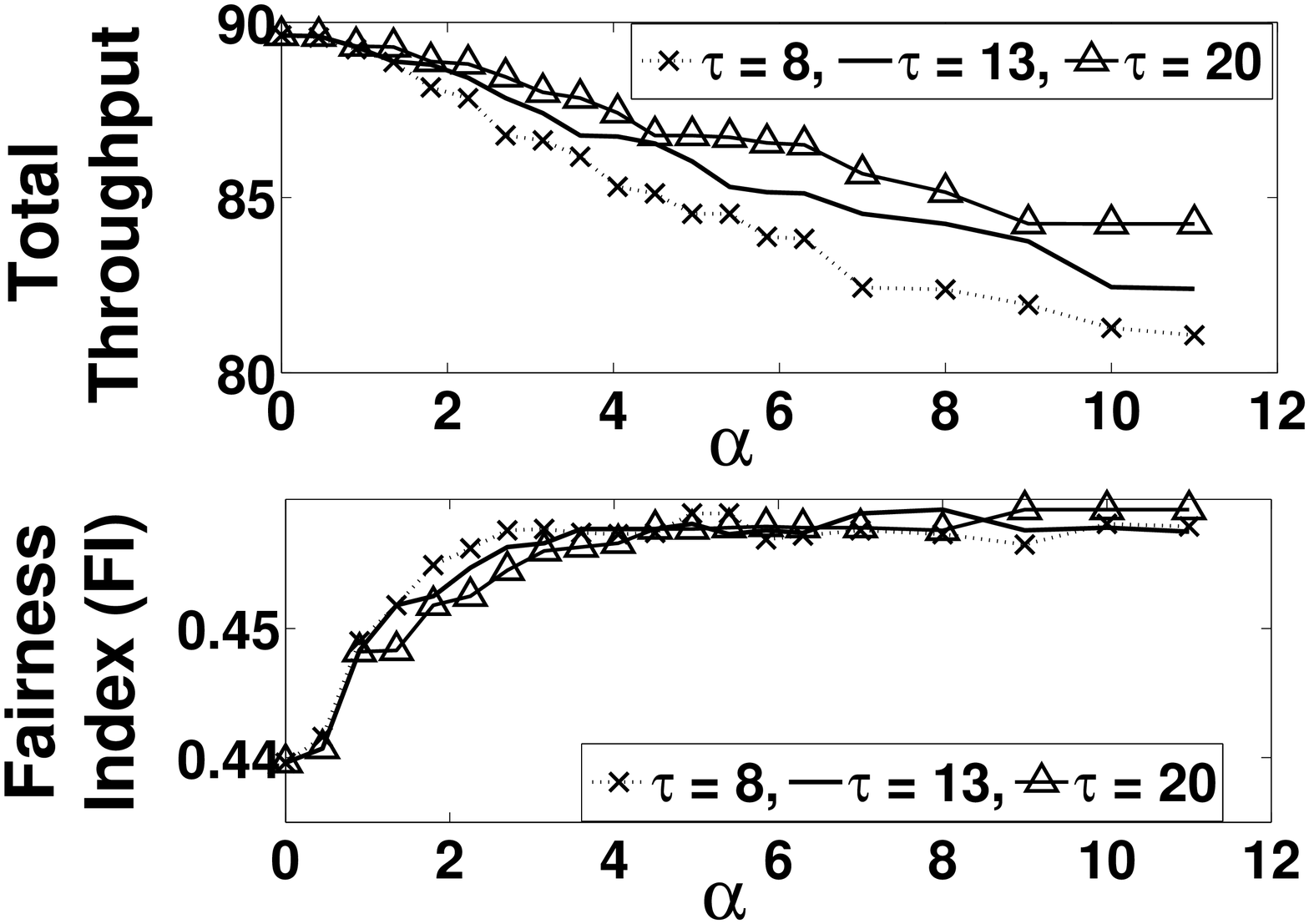} }} %
    \hspace{.1cm}
    \subfloat[For $K=3, N=4, M=10$]{\label{3b}{\includegraphics[width=4cm, height=4cm]{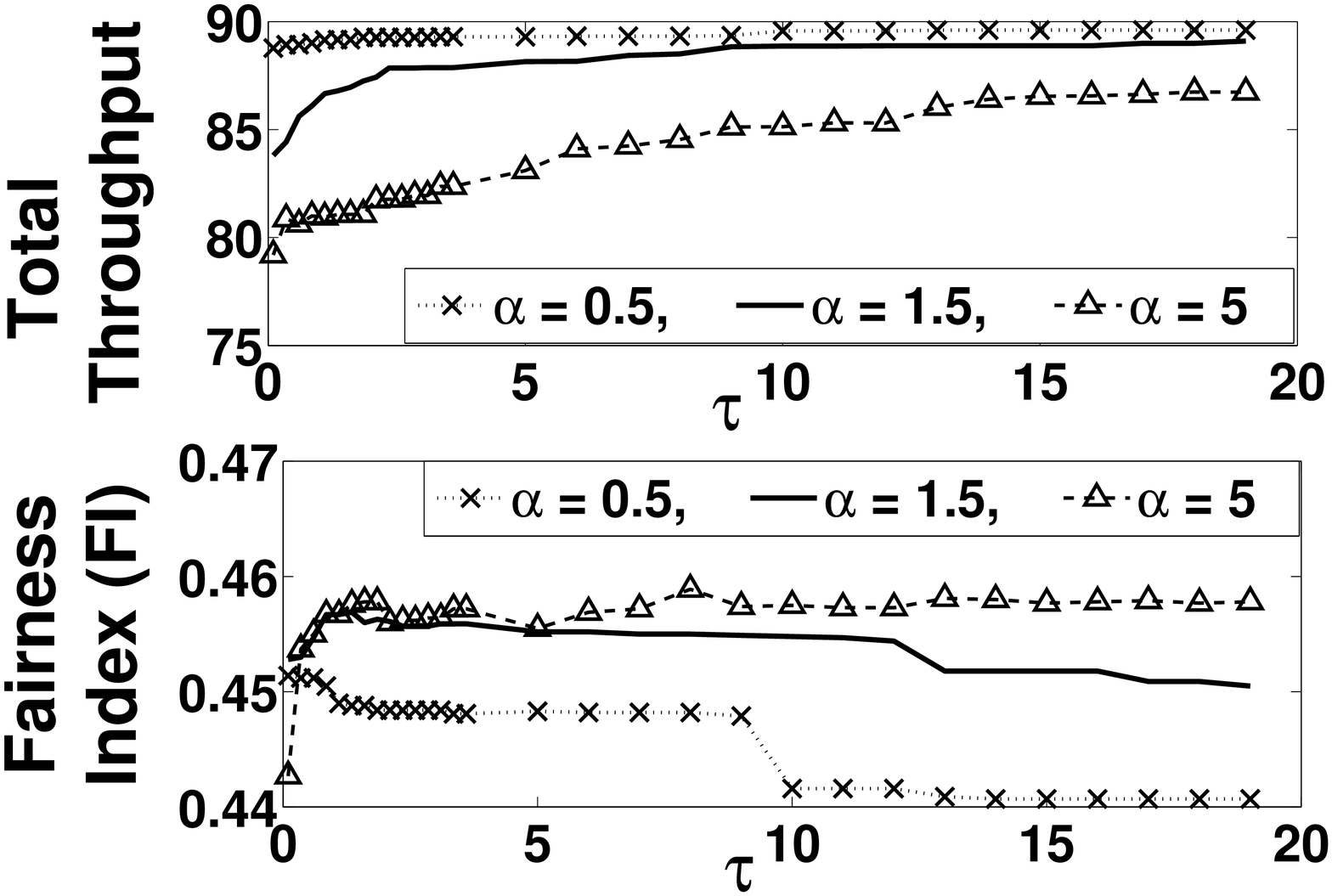} }} %
    \caption{The figure (a) (respectively, (b)) plots the total throughput and fairness index (FI) values obtained by exhaustive search over all possible subchannel allocations with $\alpha$ (respectively, $\tau$) for different $K, N$ and $M$.}%
     
    \label{extv1}%
\end{figure}
%\vspace{-.5cm}
%\label{tau3}
%\end{figure}

Next, we address the question of how the value of $\tau$ should be selected. Figs.~\ref{extv1}\subref{3b} and~\ref{tau} show the variation of the total throughput and fairness index FI with $\tau$, for different values of the parameters $K, N$ and $M$ and for three different values of $\alpha$. %In Figs.~\ref{extv},~\ref{extv1} and~\ref{extv2}, we have considered the values of $(K,N,M)$ as $(3,4,10), (2,4,7)$ and $(3,3,11)$ respectively. 
In Figs.~\ref{extv1}\subref{3b} and~\ref{tau}, the total throughput first increases and then approximately saturates as $\tau$ increases. Also, in most cases in Figs.~\ref{extv1}\subref{3b} and~\ref{tau}, the fairness index FI slightly decreases as $\tau$ increases. Similar to the trends in Figs.~\ref{extv} and~\ref{extv1}\subref{3a}, for fixed $\tau$, the total throughput decreases and fairness index FI increases with $\alpha$. Figs.~\ref{extv1}\subref{3b} and~\ref{tau} show that the values of total throughput and fairness index FI are not very sensitive to the value of $\tau$. Nevertheless, the figure shows that the choice $\tau \in [8,9]$ results in large total throughput and large FI. From Figs.~\ref{extv},~\ref{extv1} and~\ref{tau}, it can be concluded that by solving Problem~\ref{PB:ICIC:fixed:powers} with a fixed value $\tau > 0$ and different values of $\alpha \in [0,\infty)$, allocations that achieve various trade-offs between the total throughput and degree of fairness can be obtained.
%\vspace{-1em}
  %$\tau \in (0,1)$ if $\alpha \in (0,1)$
%\vspace{-1em}
  \begin{figure}%
    \centering   
    \subfloat[For $K=2, N=4, M=7$]{\label{1btau}{\includegraphics[width=4.2cm, height=4cm]{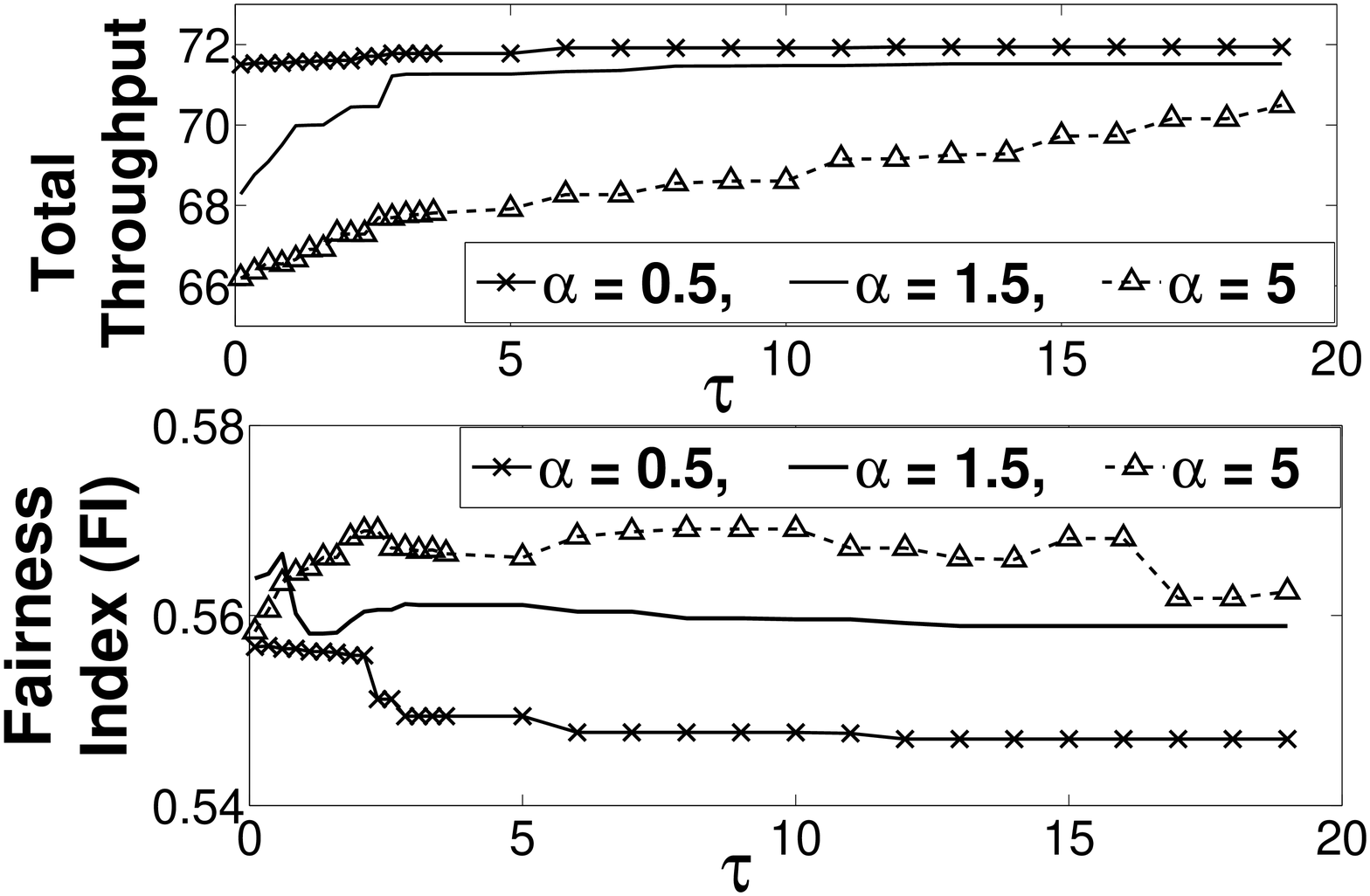} }}%
    \hspace{.1cm}
    \subfloat[For $K=3, N=3, M=11$]{\label{1c}{\includegraphics[width=4.2cm, height=4cm]{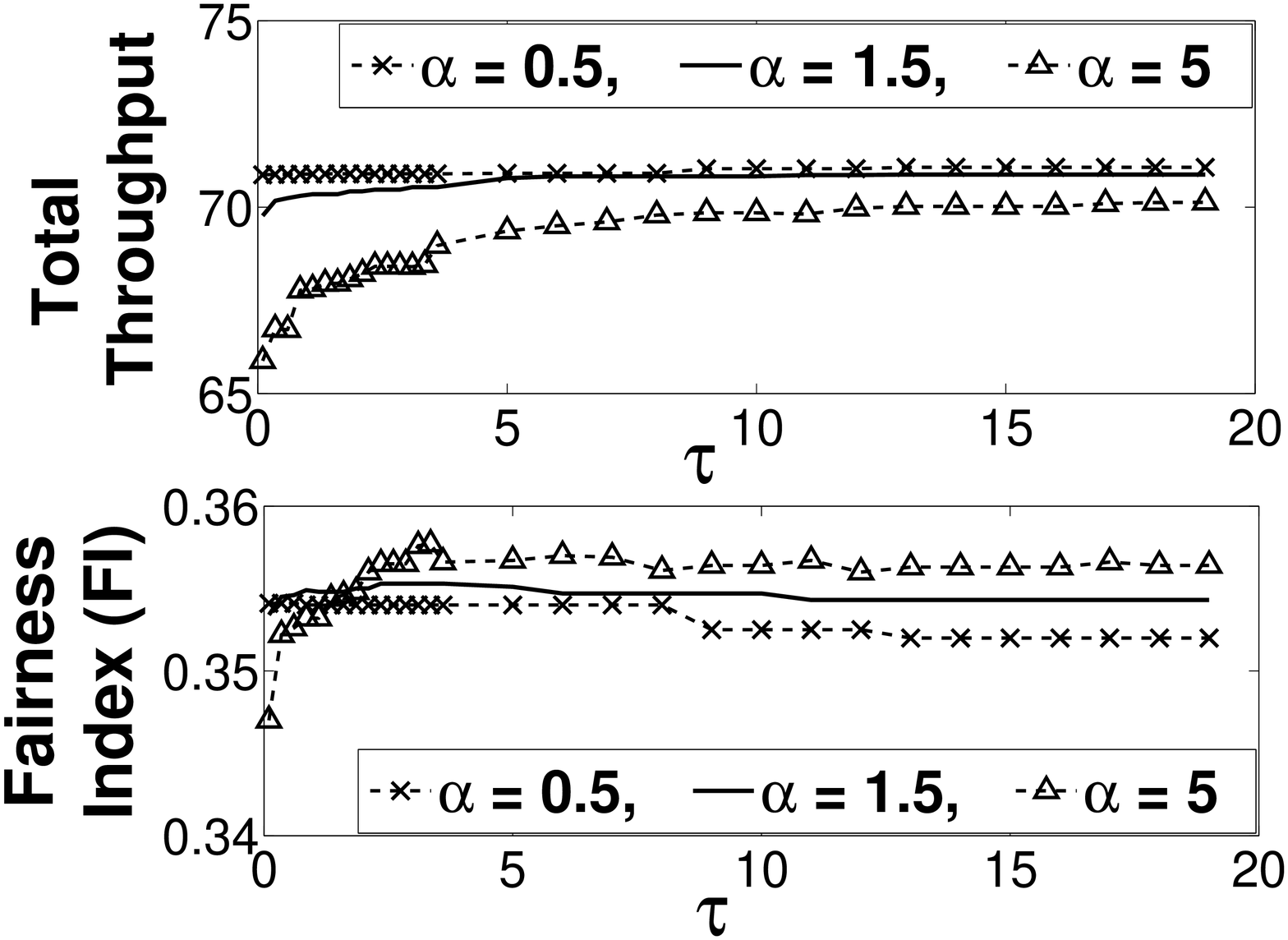} }}%
    \caption{The figure plots the total throughput and fairness index (FI) values obtained by exhaustive search over all possible subchannel allocations versus $\tau$ for different $K, N$ and $M$.}%  
    \vspace{-1.5em}
    \label{tau}%
\end{figure}
\subsection{To Obtain the Value of $p_{0}$ that Maximizes the Total Throughput}
\label{opt thr}

For the distributed $\tau-\alpha-$fair subchannel allocation algorithm, we want to first find the value of the parameter $p_0$ (see the condition (iii) for termination of the algorithm in Section~\ref{SC:algorithms}), say $p_{0}^{*}$, that results in the maximum total throughput under the allocation found by the algorithm. The value $p_{0}^{*}$ will later be used in Section~\ref{tradeoff} to investigate as to how $p_{0}$ should be selected as a function of $\alpha$ such that the higher the value of $\alpha$, the lower the total throughput and higher the degree of fairness under the allocation found by the algorithm.
%\vspace{-2em}
\begin{figure}%
    \centering
    \subfloat[For $N=20, M=300$]{\label{1aa}{\includegraphics[width=4.2cm, height=3cm]{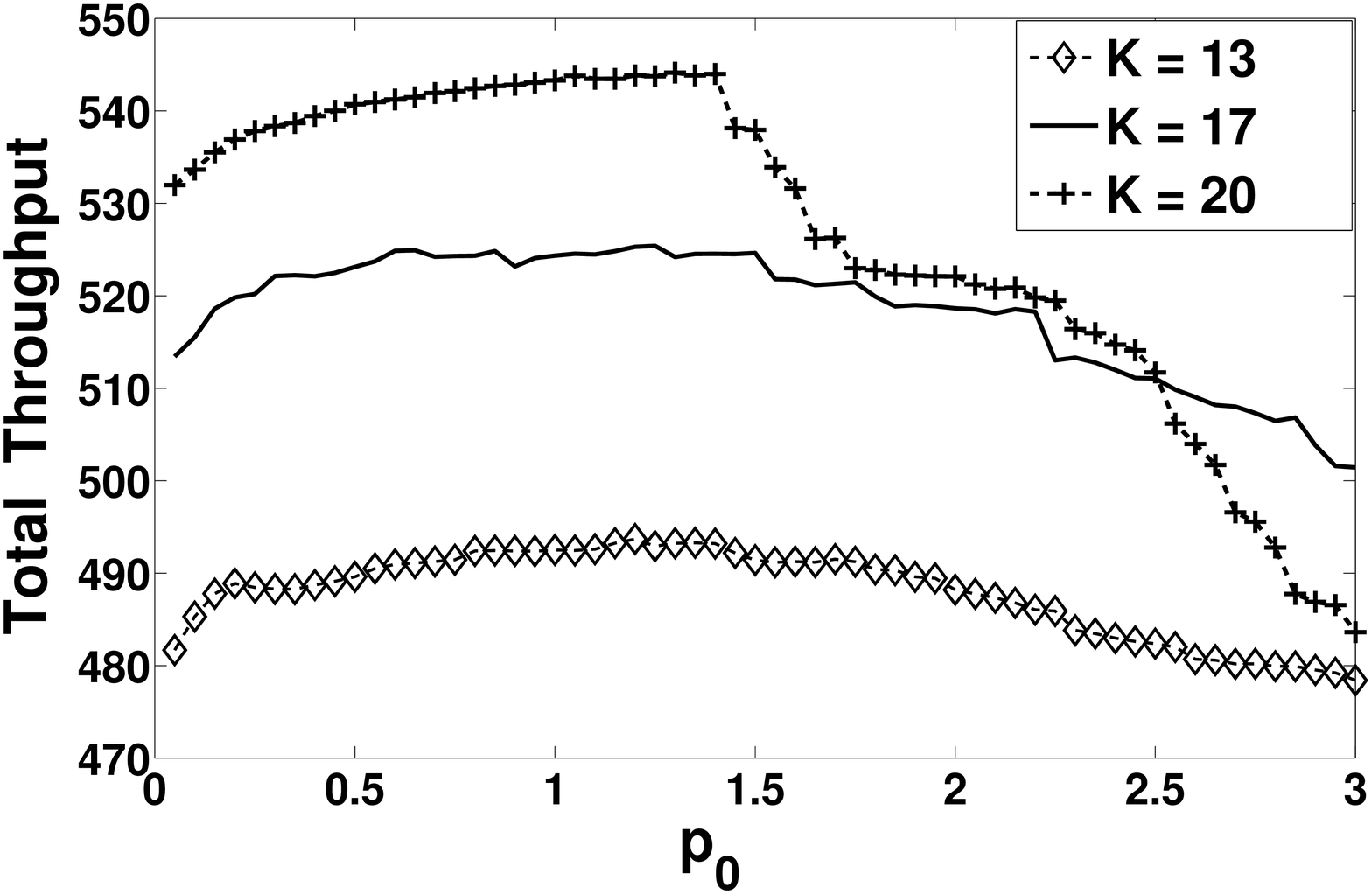}}} %
    \hspace{.2cm}
    \subfloat[For $K=15, M=240$]{\label{1bb}{\includegraphics[width=4.2cm, height=3cm]{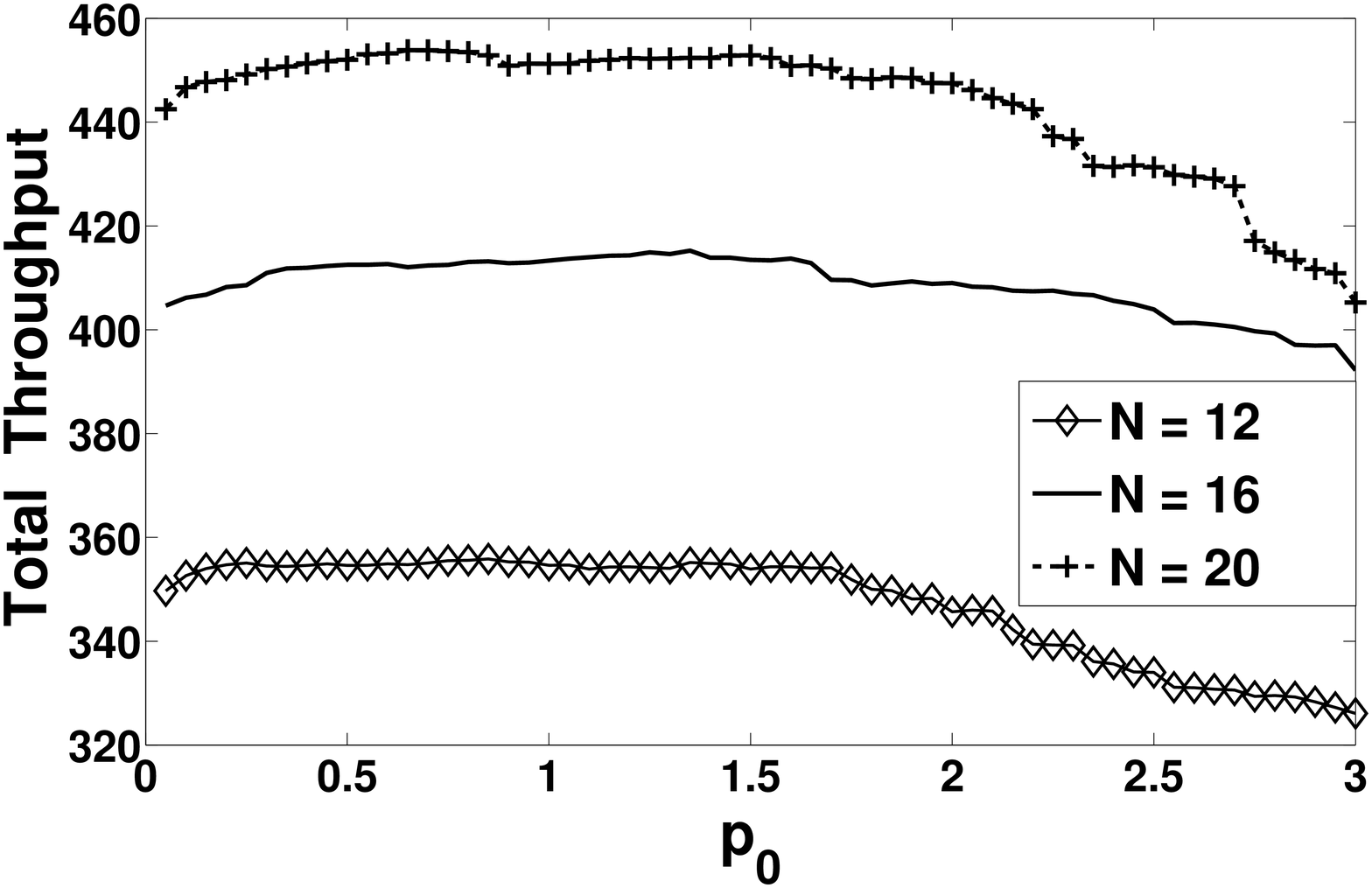} }}%
%    \hspace{.05cm}
%    \subfloat[For $N=16, K=12$]{\label{1c}{\includegraphics[width=4.5cm, height=4cm]{thr_vs_M_1.eps} }}%
    \caption{The figure (a) (respectively, (b)) plots the total throughput under the distributed $\tau-\alpha-$fair algorithm versus $p_0$ for different $K$ (respectively, $N$).}%
    \vspace{-1.5em} 
    \label{p0}%
\end{figure} 
%\vspace{-1em}
The variation of the total throughput with the parameter $p_0$ is depicted in Figs.~\ref{p0}\subref{1aa}, \ref{p0}\subref{1bb} and~\ref{p00}\subref{1a0} for different values of $K, N$ and $M$  respectively. %We have considered $N=20, M=300$ in Fig.~\ref{p0}(\subref{1a}), $N=16, K=12$ in Fig.~\ref{p0}(\subref{1b}) and $K=15, M=240$ in Fig.~\ref{p0_1}(\subref{2a}).
In Figs.~\ref{p0} and~\ref{p00}\subref{1a0}, the total throughput is maximized for medium values of $p_0$. Intuitively, this is because for too low values of $p_0$, the proposed algorithm allocates subchannels to a large number of MSs (see condition (iii) for termination of the algorithm in Section~\ref{SC:algorithms}), which results in high interference and low total throughput. Similarly, for too high values of $p_0$ the algorithm does not allocate subchannels to enough of MSs, which results in low total throughput. Therefore, the total throughput first increases then decreases as $p_0$ increases.
%\vspace{-1em}
% from Fig.~\ref{p0}(\subref{1a}) that system throughput is high for large $k$ only for low values of $p_0$ because algorithm allocates a large number of MSs for high $p_0$ value, which increases interference and reduces the throughput. 
Figs.~\ref{p00}\subref{1b0},~\ref{p0_1}\subref{1a_1} and~\ref{p0_1}\subref{1b_1} present the variation of FI with  $p_0$ for different values of $M, K$ and $N$ respectively.
 %We have considered $N=20, M=300$ in Fig.~\ref{p0_1}(\subref{2b}), $N=16, K=12$ in Fig.~\ref{p0_2}(\subref{3a}) and $K=15, M=240$ in Fig.~\ref{p0_2}(\subref{3b}). 
Figs.~\ref{p00}\subref{1b0} and~\ref{p0_1} show that the FI decreases as $p_0$ increases. Intuitively, this is because as $p_0$ decreases, the algorithm runs for a longer duration and allocates subchannels to more number of MSs which increases fairness. After extensive simulations, we empirically found that the value of the parameter $p_0$ (say $p_{0}^{*}$) which gives close to maximum total throughput in terms of the parameters $K, M$ and $N$ is given by the following expression: 
\begin{equation}
\label{p}
p_{0}^{*}=\left\{ 
\begin{array}{ll}
1+\frac{M}{2(N K)}, & \mbox{if} \; \; M \leq K \times N, \\
1+\frac{log (NK)}{2\;log M}, & \mbox{otherwise}. \\
\end{array}
\right.
\end{equation}
%\vspace{-1cm}
\begin{figure}%
    \centering
%    \subfloat[For $N=20, M=300$]{\label{1a}{\includegraphics[width=4.4cm, height=4cm]{thr_vs_K_1.eps}}} %
%    \hspace{.05cm}
%    \subfloat[For $K=15, M=240$]{\label{1b}{\includegraphics[width=4.5cm, height=4cm]{thr_vs_N_1.eps} }}%
%    \hspace{.05cm}
    \subfloat[For $K=12, N=16$]{\label{1a0}{\includegraphics[width=4.2cm, height=3cm]{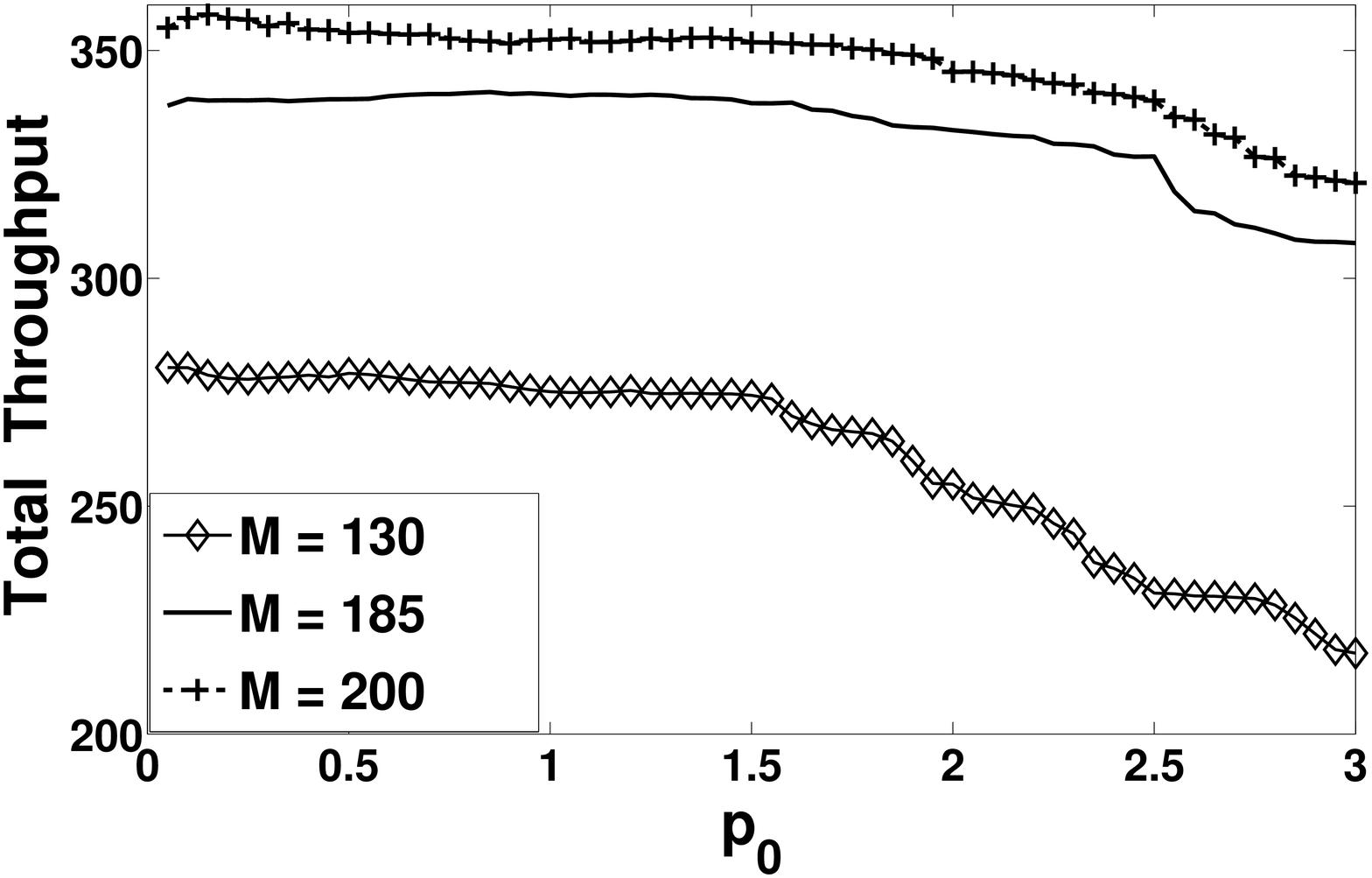} }}%
    \hspace{.1cm}
    \subfloat[For $K=12, N=16$]{\label{1b0}{\includegraphics[width=4.2cm, height=3cm]{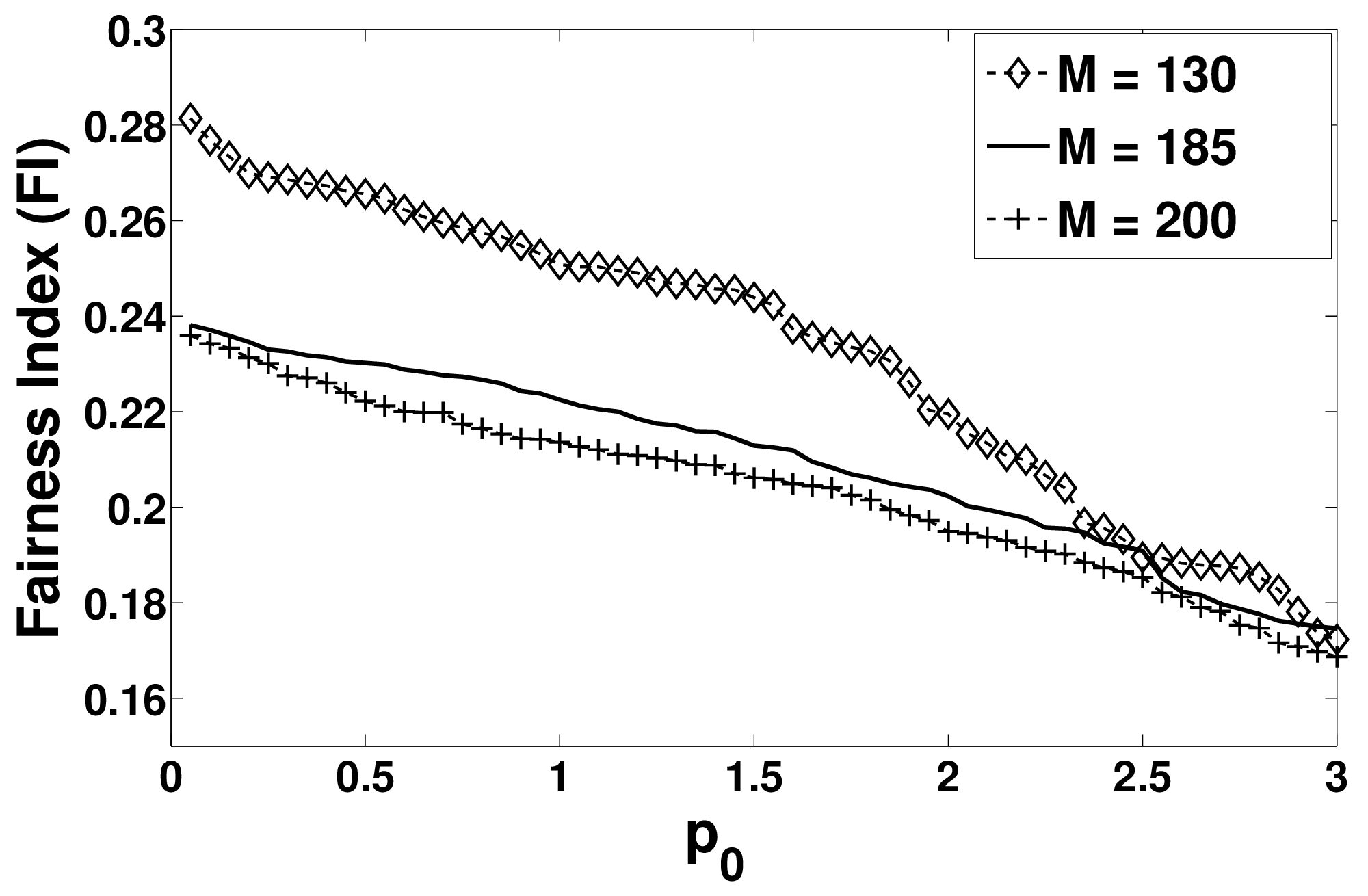} }}%
    \caption{The figure (a) (respectively, (b)) plots the total throughput  (respectively, fairness index (FI)) under the distributed $\tau-\alpha-$fair algorithm versus $p_0$ for different $M$.}%
     \vspace{-1.5em}
    \label{p00}%
\end{figure}
%\vspace{-1cm}
 \begin{figure}%
    \centering
    \subfloat[For $N=20, M=300$]{\label{1a_1}{\includegraphics[width=4.2cm, height=3cm]{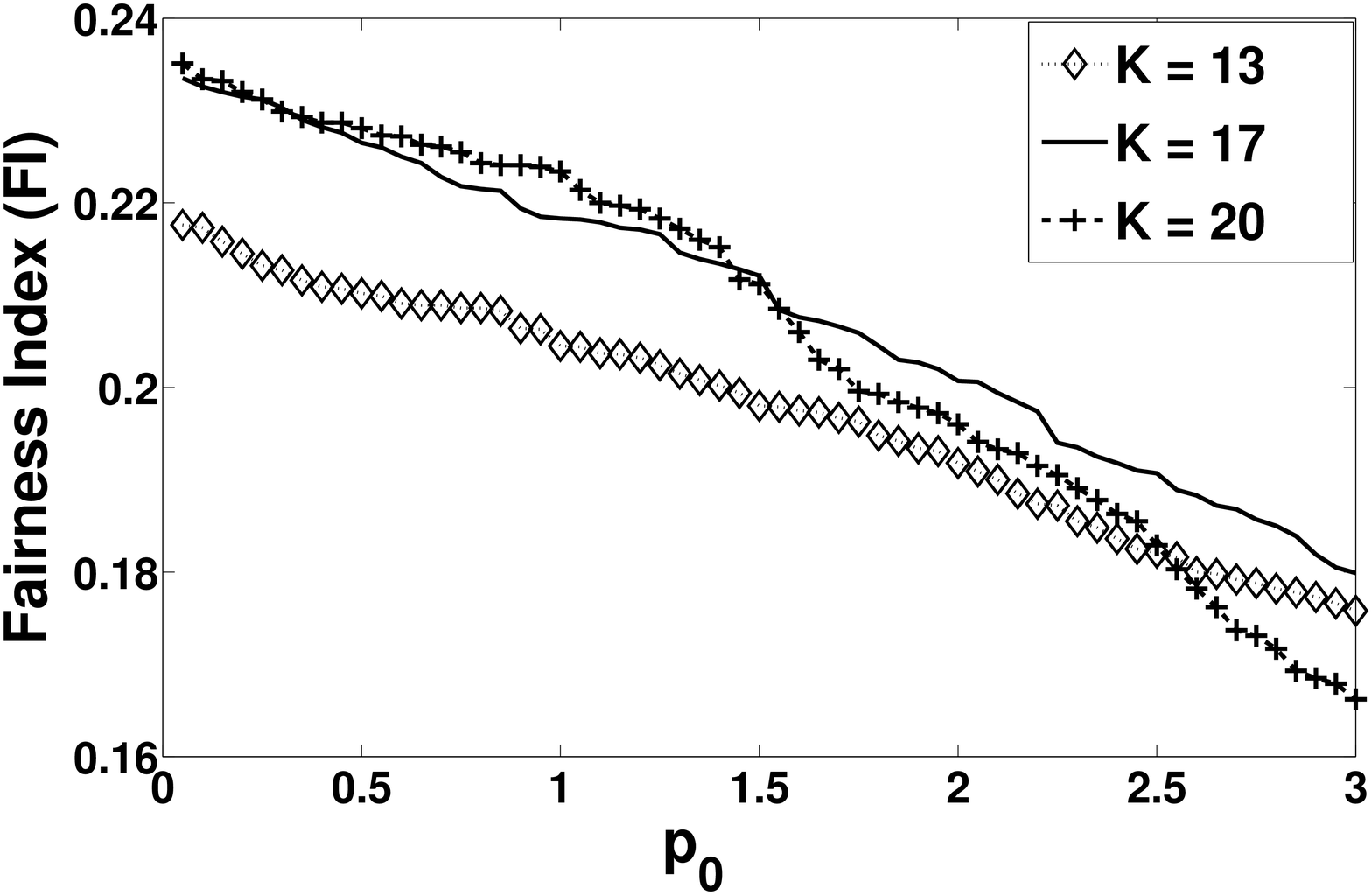} }} %
    \hspace{.1cm}
    \subfloat[For $K=15, M=240$]{\label{1b_1}{\includegraphics[width=4.2cm, height=3cm]{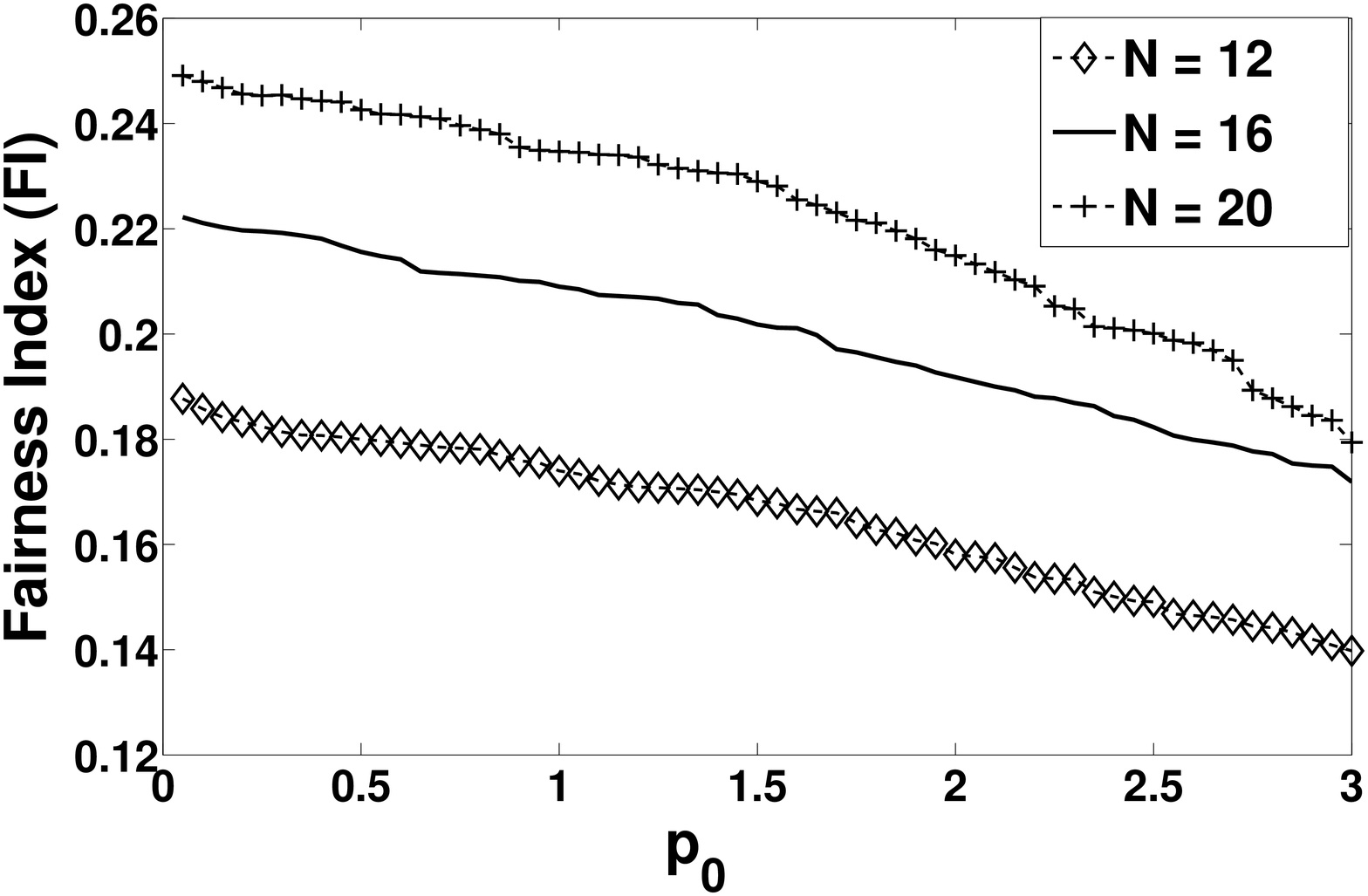} }}%
%    \hspace{.05cm}
%    \subfloat[For $K=12, N=16$]{\label{1c}{\includegraphics[width=4.4cm, height=4cm]{FI_po_M_1.eps} }}%
    \caption{The figure (a) (respectively, (b)) plots fairness index (FI) values obtained by exhaustive search over all possible subchannel allocations versus $p_0$ for different $K$ (respectively, $N$).}%
     \vspace{-1em}
    \label{p0_1}%
\end{figure}
%\vspace{-1cm}
%\vspace{1em}

\subsection{Selection of the Value of $p_0$ as a Function of $\alpha$}
\label{tradeoff}
From Figs.~\ref{p0},~\ref{p00} and~\ref{p0_1}, it can be concluded that there is a tradeoff between the total throughput and degree of fairness when the parameter $p_0$ is in the range $[0, p_{0}^{*}]$. In particular, within the range $p_0 \in [0, p_{0}^{*}]$, the total throughput (respectively, fairness) is maximized at $p_0=p_{0}^{*}$ (respectively, $p_0=0$). However, recall that $\alpha=0$ (respectively, $\alpha=\infty$) corresponds to maximum total throughput (respectively, fairness) and minimum fairness (respectively, total throughput). This motivates us to set $p_0$, in terms of $\alpha$, as:
\begin{equation}
\label{alp}
p_0= \frac{1}{\frac{1}{p_{0}^{*}}+\alpha}.
\end{equation}
In summary, the choice of $p_0$ in~\eqref{alp} ensures that as $\alpha$ increases from 0 to $\infty$, the total throughput (respectively, degree of fairness) of the allocation found using the algorithm described in Section~\ref{SC:algorithms} decreases (respectively, increases).
%%%%\vspace{-2em}

%%%%\vspace{-3em}
\subsection{Performance Evaluation of the Proposed Distributed Algorithm}
For different sets of values of $K, M$ and $N$ and for different values of $\alpha$, $p_0$ was computed using~\eqref{p} and~\eqref{alp}. Using the calculated value of $p_0$, the proposed distributed $\tau-\alpha-$fair subchannel allocation algorithm was run and a subchannel allocation was obtained. The total throughput and FI under the obtained allocation
were calculated using~\eqref{EQ:objective} and~\eqref{fi} respectively. Figs.~\ref{trd}\subref{1at},~\ref{trd}\subref{1bt} and~\ref{trd}\subref{1ct} depict the variation of the total throughput and FI with $\alpha$ for different values of $K, N$ and $M$ respectively. In Fig.~\ref{trd}, the total throughput decreases and fairness index FI increases as $\alpha$ increases.
Therefore, it can be verified from Fig.~\ref{trd} that the distributed  $\tau-\alpha-$fair subchannel allocation algorithm proposed in Section~\ref{SC:algorithms} and the expressions for $p_0^*$ and $p_0$ in~\eqref{p} and~\eqref{alp} provide the required trade-off between the total throughput and degree of fairness.
\begin{figure*}%
    \centering
    %\hspace{-2em}
    \subfloat[For $N=20$ and $M=300$]{\label{1at}{\includegraphics[width=5.5cm, height=4.5cm]{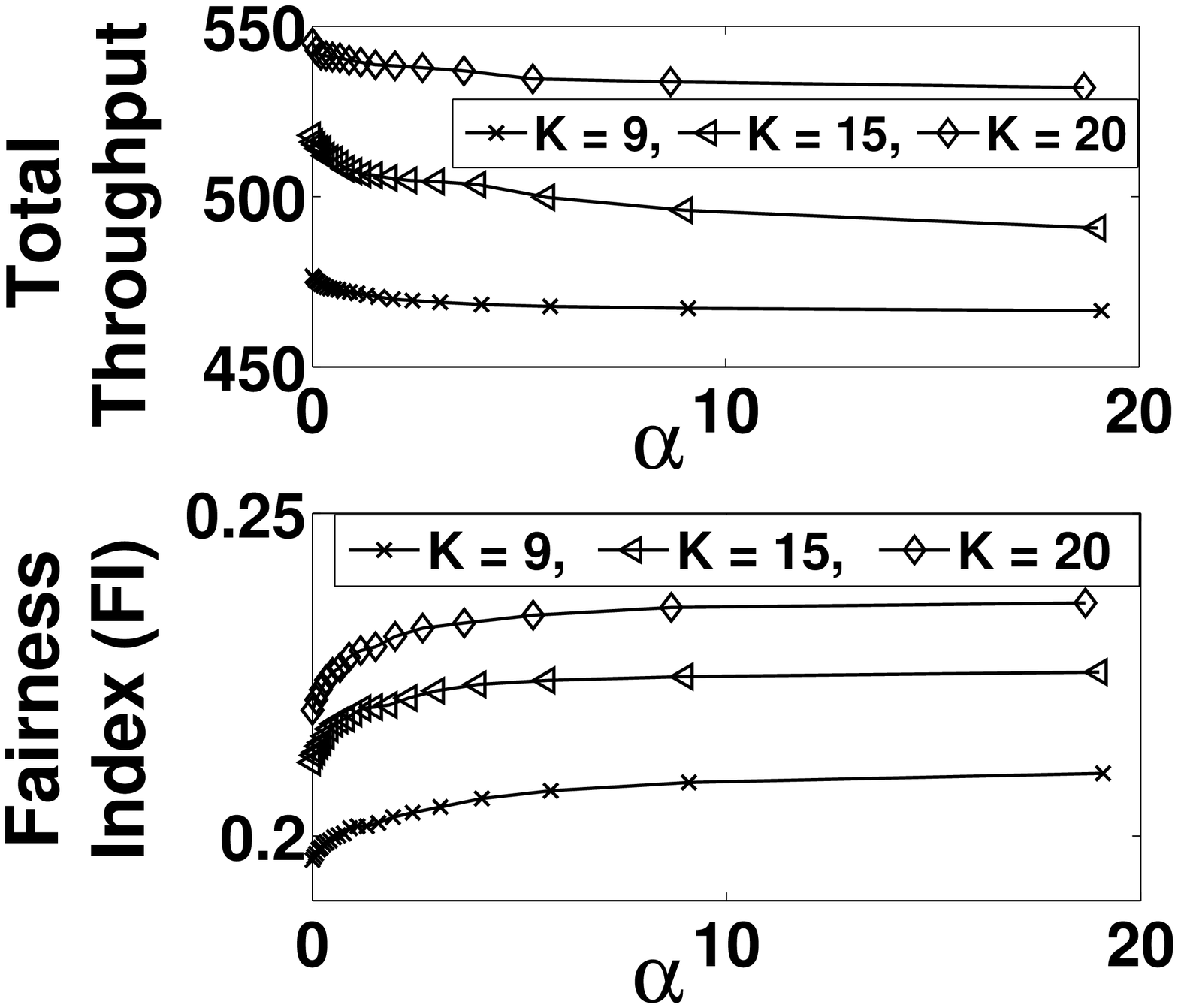} }} %
    %\hspace{-1em}
    \subfloat[For $K=12$ and $M=300$]{\label{1bt}{\includegraphics[width=5.5cm, height=4.5cm]{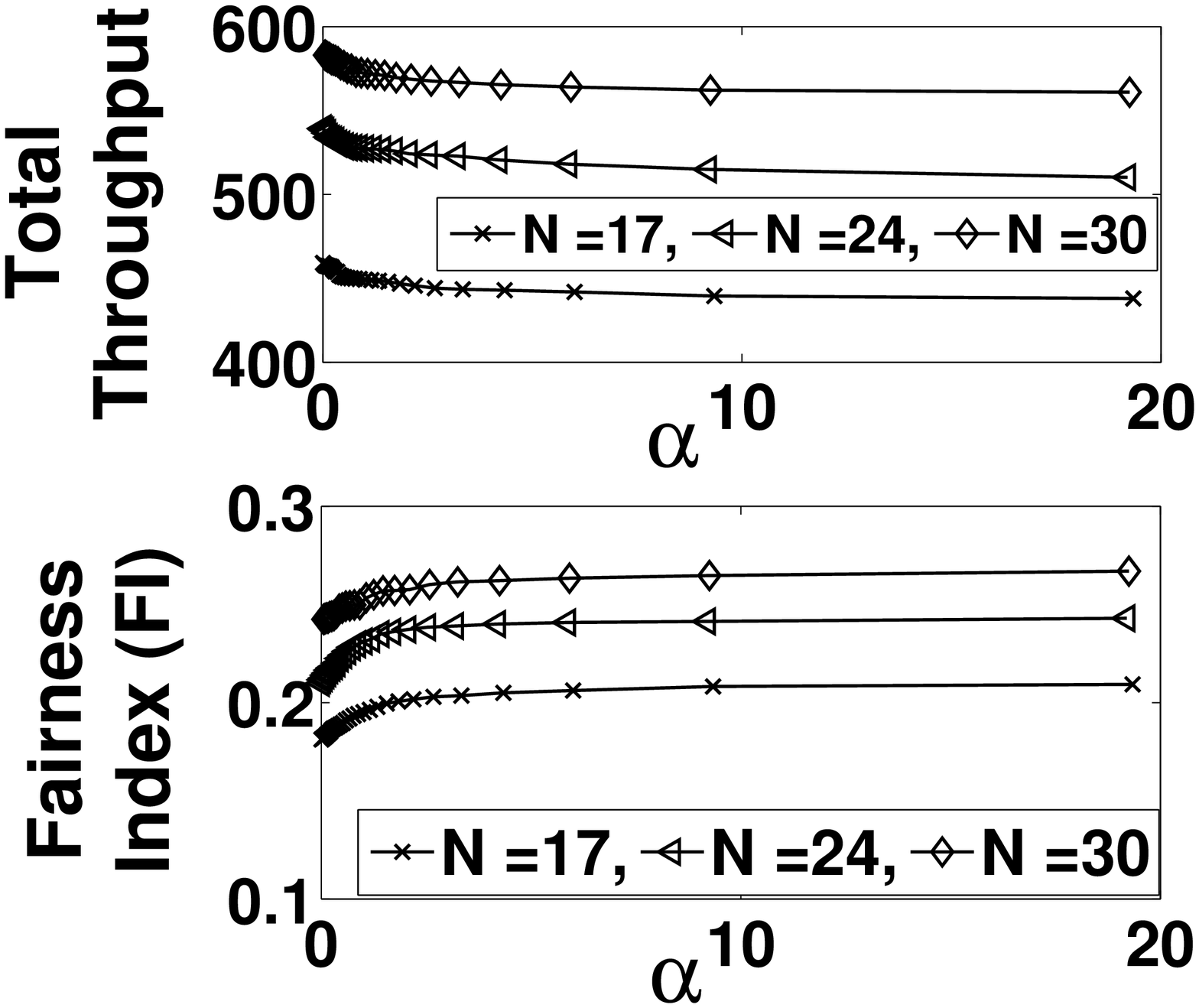} }}%
   %\hspace{.05cm}
    \subfloat[For $K=15$ and $N=20$]{\label{1ct}{\includegraphics[width=5.5cm, height=4.5cm]{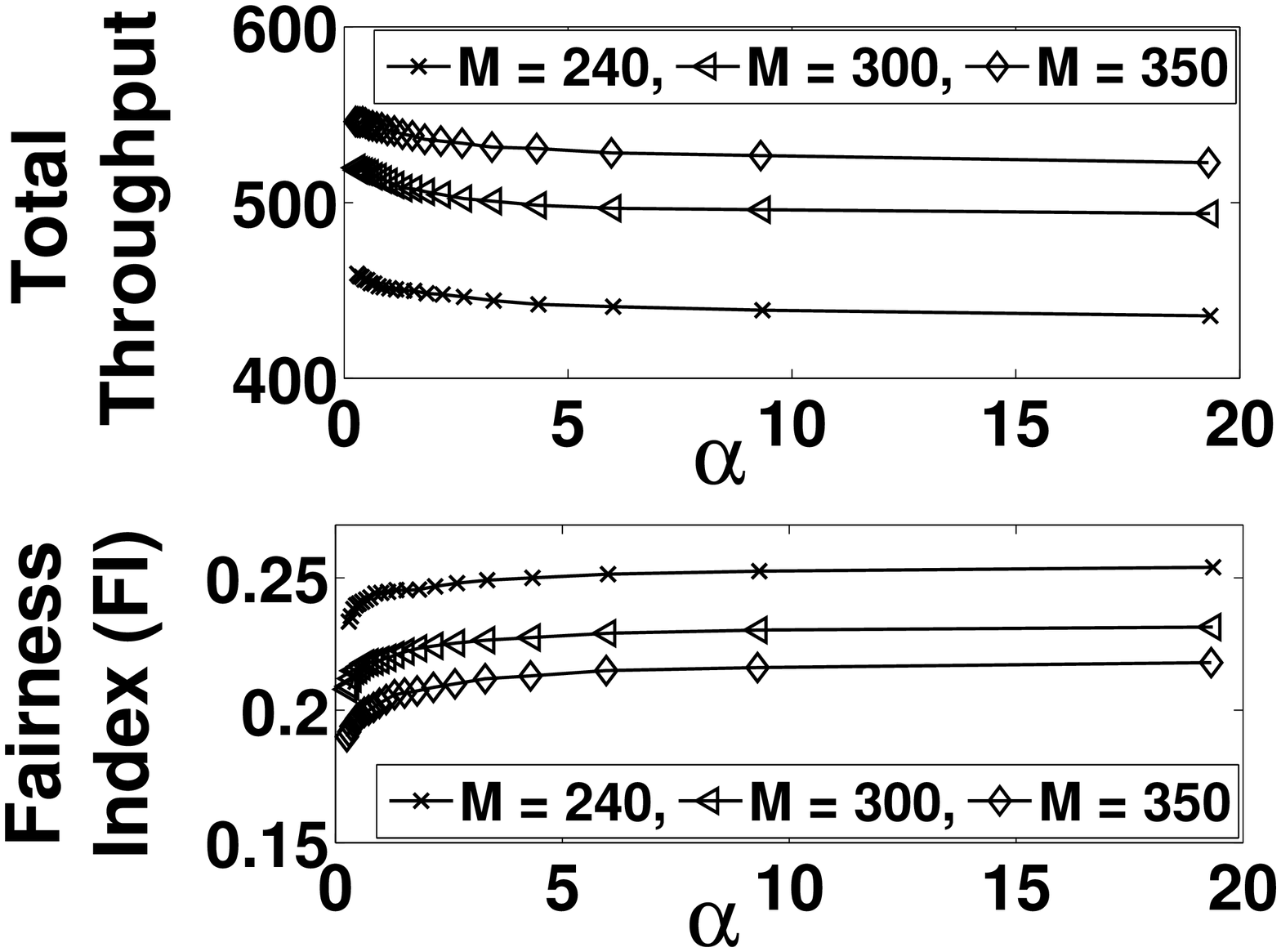} }}%
    \caption{The figure (a) (respectively, (b) and (c)) plots the total throughput and fairness index under the distributed $\tau-\alpha-$fair algorithm versus $\alpha$ for different values of $K$ (respectively, $N$ and $M$).}%
     \vspace{-1em}
    \label{trd}%
\end{figure*}
%\begin{figure*}[ht]%
%    \centering
%    %\hspace{-2em}
%    \subfloat[For $N=20$ and $M=300$]{\label{1a}{\includegraphics[width=4.15cm, height=4.2cm]{trd_dis_k223.eps} }} %
%    %\hspace{-1em}
%    \subfloat[For $K=12$ and $M=300$]{\label{1b}{\includegraphics[width=4.15cm, height=4.2cm]{trd_dis_N223.eps} }}%
%   %\hspace{.05cm}
%    \subfloat[For $K=15$ and $N=20$]{\label{1c}{\includegraphics[width=4.15cm, height=4.2cm]{trd_dis_M223.eps} }}%
%    \caption{The figure (a) (respectively, (b) and (c)) plots the total throughput and fairness index under the distributed $\tau-\alpha-$fair algorithm versus $\alpha$ for different values of $K$ (respectively, $N$ and $M$).}%
%     
%    \label{trd}%
%\end{figure*}
\vspace{-1em}
\section{Conclusions and Future Work}
\label{conc}
%%%%\vspace{-1em}
In this paper, we introduced the concept of $\tau-\alpha-$fairness in the context of the ICIC with fixed transmit power problem by modifying the  concept of $\alpha-$fairness. The concept of $\tau-\alpha-$fairness allows us to achieve arbitrary trade-offs between the total throughput and degree of fairness by selecting an appropriate value of $\alpha$ in $[0,\infty)$. We showed that for every $\alpha \in [0,\infty)$ and every $\tau > 0$, the problem of finding a $\tau-\alpha-$fair allocation  is NP-Complete. Next, we showed that for every $\alpha \in [0, \infty)$, there exist thresholds such that if the potential interference levels experienced by each MS on every subchannel are above the threshold values, then the problem can be optimally solved in polynomial time by reducing it to the bipartite graph matching problem. Also, we proposed a simple, distributed subchannel allocation algorithm for the ICIC problem, which is flexible, requires a small amount of time to operate, and requires information exchange among only neighboring BSs. We investigated via simulations as to how the algorithm parameters should be selected so as to achieve any desired trade-off between the total throughput and fairness. Our analytical results provide insight into the structure of the ICIC with fixed transmit power problem, with the objective of achieving arbitrary throughput-fairness trade-offs, which would be useful to future work on the design of approximation algorithms with a provable approximation ratio for the problem.
\vspace{-1em}

\appendix
\begin{IEEEproof}[\textbf{A. Proof of Lemma~\ref{l1}}] 
First, we will show that the function $f(x)$ is quasi-convex on the domain $x \geq 1$.
\begin{property}
\label{p1}
A function $Q(.)$ is quasi-convex if $Q''(z)>0$ whenever $Q'(z)=0$~\cite{boyd}. 
\end{property}
Let 
\begin{equation}
\label{parachange}
y=(x-1)\beta+ \frac{1}{\eta}.
\end{equation}
 Then, 
 \begin{equation}
\label{fg}
f(x)=g(y),
\end{equation}
 where $g(y)= $
 \begin{equation}
 \left(\frac{y}{\beta}+1-\frac{1}{\eta\beta}\right)\frac{\left(\tau+\log \left( 1 + \frac{1}{y} \right)\right)^{1-\alpha}}{1-\alpha}-(\frac{y}{\beta}-\frac{1}{\eta\beta})\frac{\tau^{1-\alpha}}{1-\alpha}.\nonumber
 \end{equation} 

Let $p= \left(\tau+\log \left( 1 + \frac{1}{y} \right)\right)$. Then,
\begin{equation}
g'(y)= \frac{1}{\beta}\left(\frac{p^{1-\alpha}}{1-\alpha}-\frac{(y+\beta-\frac{1}{\eta})\left(p \right)^{-\alpha}}{y(y+1)}-\frac{\tau^{1-\alpha}}{1-\alpha} \right).
\label{g_desh}
\end{equation}
\begin{equation}
g'(y)=0 \Leftrightarrow y+\beta-\frac{1}{\eta}= \frac{y(y+1)(p-p^{\alpha}\tau^{1-\alpha})}{1-\alpha}.
\label{gdesh}
\end{equation}
Further, $g''(y)$
%\small
%\begin{equation}
%\frac{1}{\beta}\left[-\frac{1}{y(y+1)p^{\alpha}}-\left( \frac{y(y+1)p^{\alpha}-(y+\beta-\frac{1}{\eta})((2y+1)p^{\alpha}-\alpha p^{\alpha-1})}{(y(y+1)p^{\alpha})^{2}}\right) \right]\nonumber
%\end{equation}
%\normalsize
\begin{equation}
\hspace{-0.1cm}=\frac{1}{\beta y(y+1)p^{\alpha}}\left( -2+ \frac{y+\beta-\frac{1}{\eta}}{y(y+1)p^{\alpha}}\left((2y+1)p^{\alpha}-\alpha p^{\alpha-1} \right)\right)\nonumber
\end{equation}
%\begin{eqnarray}
%&\hspace{-2.9cm}=&\hspace{-1.3cm}\frac{1}{\beta (y(y+1)p^{\alpha})^{2}}\left(-yp^{\alpha}-\alpha yp^{\alpha-1}+\left(\beta-\frac{1}{\eta}\right)\nonumber
%& &\hspace{-.3cm}\left((2y+1)p^{\alpha}-\alpha p^{\alpha-1}\right)\right)
%\end{eqnarray}
So $g''(y) > 0 $\\
$\Leftrightarrow py(2(\beta-\frac{1}{\eta})-1)-\alpha(y+\beta-\frac{1}{\eta})+p(\beta-\frac{1}{\eta})) > 0$\\
Substituting from~\eqref{gdesh} in the above inequality, we get:\\
$py(2(\beta-\frac{1}{\eta})-1)-\alpha y(y+1)\frac{p-p^{\alpha}\tau^{1-\alpha}}{1-\alpha}+p(\beta-\frac{1}{\eta})) > 0$ \\
\small
\begin{equation}
\Leftrightarrow y(2(\beta-\frac{1}{\eta})-1)-\alpha y(y+1)\frac{1-\left(1+\frac{\log(1+\frac{1}{y})}{\tau}\right)^{\alpha-1}}{1-\alpha}+(\beta-\frac{1}{\eta}) > 0 \\
\label{fnleq}
\end{equation}
\normalsize
Now, we find sufficient conditions for~\eqref{fnleq} to hold for three different values of $\alpha$:\\ 
\textbf{(a) $\alpha <1:$}\\
Because $(1+x)^r \geq 1+ rx \;\; \forall x \geq -1, \;r \in \; \mathcal{R} \backslash (0,1)$~\cite{Mitri}, a sufficient condition for~\eqref{fnleq} to hold is\\
$y(2(\beta-\frac{1}{\eta})-1)-\alpha \frac{y(y+1)}{1-\alpha}[1-\{1+(\alpha-1)\frac{\log(1+\frac{1}{y})}{\tau}\}]+(\beta-\frac{1}{\eta})) > 0$.\\
Because $\log(1+x) \leq x \; \forall x \geq -1$~\cite{Logineq}, a sufficient condition for the above inequality to hold is\\
$y\tau(2(\beta-\frac{1}{\eta})-1)-\alpha (y+1)+\tau(\beta-\frac{1}{\eta})) > 0.$\\
$ \Leftrightarrow y(2\tau(\beta-\frac{1}{\eta})-\tau-\alpha)-\alpha +\tau(\beta-\frac{1}{\eta})) > 0$.\\
A sufficient condition for the above inequality to hold is\\
\begin{equation}
\label{condnn1}
\beta \geq \frac{\alpha}{\tau}+\frac{1}{\eta}\;\; \mbox{and} \; \;\tau < \alpha.
\end{equation}
Hence, when~\eqref{condnn1} holds and $\alpha < 1$, then $g''(y)> 0$ when $g'(y) = 0$.

 \textbf{(b) $1< \alpha < 2:$}\\
From~\eqref{fnleq},\\
$y(2(\beta-\frac{1}{\eta})-1)+\alpha y(y+1)\frac{1-\left(1+\frac{\log(1+\frac{1}{y})}{\tau}\right)^{\alpha-1}}{\alpha-1}+(\beta-\frac{1}{\eta}) > 0. $\\
Because $(1+x)^r \leq 1+ rx \;\; \forall x \geq -1, \;r \in \; (0,1)$~\cite{Mitri}, a sufficient condition for the above inequality to hold is\\
$y(2(\beta-\frac{1}{\eta})-1)+\alpha \frac{y(y+1)}{\alpha-1}[1-\{1+(\alpha-1)\frac{\log(1+\frac{1}{y})}{\tau}\}]+(\beta-\frac{1}{\eta})) > 0.$\\ 
Because $\log(1+x) \leq x \; \forall x \geq -1$~\cite{Logineq}, a sufficient condition for the above inequality to hold is\\
$y\tau(2(\beta-\frac{1}{\eta})-1)-\alpha (y+1)+\tau(\beta-\frac{1}{\eta})) > 0$\\
$ \Leftrightarrow y(2\tau(\beta-\frac{1}{\eta})-\tau-\alpha)-\alpha +\tau(\beta-\frac{1}{\eta})) > 0.$\\
A sufficient condition for the above inequality to hold is
\begin{equation}
\label{condnn2}
\beta \geq \frac{\alpha}{\tau}+\frac{1}{\eta}\;\; \mbox{and} \; \; \tau < \alpha, \nonumber
\end{equation}
which is the same as \eqref{condnn1}.
Hence, when~\eqref{condnn1} holds and $1< \alpha < 2$, then $g''(y)> 0$ when $g'(y) = 0$.

\textbf{(c) $\alpha \geq 2:$}\\
From~\eqref{fnleq},
\begin{eqnarray}
\label{al2}
& y(2(\beta-\frac{1}{\eta})-1)+\alpha y(y+1)\frac{1-\left(1+\frac{\log(1+\frac{1}{y})}{\tau}\right)^{\alpha-1}}{\alpha-1}\nonumber \\
& \hspace{-5cm}+(\beta-\frac{1}{\eta}) > 0, 
\end{eqnarray}
because $(1+x)^r \leq 1+ (2^r-1)x \;\; \forall x \in [0,1], \;r \in \; \mathcal{R} \backslash (0,1)$~\cite{Mitri}. \\
Now,
\begin{equation}
\label{pro}
0 \leq \log(1+\eta) \leq \tau \Rightarrow \log(1+\frac{1}{y}) \leq \tau \;(\mbox{since}\; y \geq \frac{1}{\eta}).
\end{equation}
Next, if~\eqref{pro} holds, a sufficient condition for~\eqref{al2} to hold is\\
$y(2(\beta-\frac{1}{\eta})-1)+\alpha y(y+1)\frac{1-\left(1+(2^{(\alpha-1)}-1)\frac{\log(1+\frac{1}{y})}{\tau}\right)}{\alpha-1}+(\beta-\frac{1}{\eta}) > 0 $\\
Using the fact that $\log(1+x)\leq x \;\;\forall x \geq -1$~\cite{Logineq}, a sufficient condition for the above inequality to hold is\\
$y\tau\left(2(\beta-\frac{1}{\eta})-1\right)- \frac{\alpha(y+1)(2^{(\alpha-1)}-1)}{\alpha-1}+\tau(\beta-\frac{1}{\eta})> 0$\\
$\Leftrightarrow y\left(2\tau(\beta-\frac{1}{\eta})-\tau-\frac{\alpha(2^{(\alpha-1)}-1)}{\alpha-1}\right)- \frac{\alpha(2^{(\alpha-1)}-1)}{\alpha-1}+\tau(\beta-\frac{1}{\eta})> 0$.\\
The above inequality holds if the following two inequalities hold:\\
\begin{equation}
\beta-\frac{1}{\eta}>\frac{\alpha(2^{(\alpha-1)}-1)}{\tau(\alpha-1)}
\label{condn44}
\end{equation}
 and $2\tau(\beta-\frac{1}{\eta})-\tau-\frac{\alpha(2^{(\alpha-1)}-1)}{\alpha-1}>0$. \\
 Using~\eqref{condn44}, a sufficient condition for the above inequality to hold is
 \begin{eqnarray}
 & \hspace{2.5cm} 2\tau\frac{\alpha(2^{(\alpha-1)}-1)}{\tau(\alpha-1)}-\tau-\frac{\alpha(2^{(\alpha-1)}-1)}{\alpha-1}>0.\nonumber\\ 
% \begin{equation}
% \label{condnn3}
 & \Leftrightarrow \frac{\alpha(2^{(\alpha-1)}-1)}{\tau(\alpha-1)}>1.
 \label{condnn3}
 \end{eqnarray}
 Hence, when~\eqref{pro},~\eqref{condn44} and~\eqref{condnn3} hold and $\alpha \geq 2$, then $g''(y)> 0$ when $g'(y) = 0$. Therefore, it follows from Property~\ref{p1} that $g(.)$ is quasi-convex when~\eqref{condnn1} holds (respectively,~\eqref{pro},~\eqref{condn44} and~\eqref{condnn3} hold) and $\alpha \in [0,2)\backslash \{1\}$ (respectively, $\alpha \geq 2$).

Now, it follows from~\eqref{parachange} and~\eqref{fg} that $f'(x)=\beta g'(y)$ and $f''(x)=\beta^2 g''(y).$ Therefore, $f(.)$ also satisfies the condition in Property~\ref{p1} whenever $g(.)$ satisfies it. Hence, $f(.)$ is quasi-convex when~\eqref{condnn1} holds (respectively,~\eqref{pro},~\eqref{condn44} and~\eqref{condnn3} hold) and $\alpha \in [0,2)\backslash \{1\}$ (respectively, $\alpha \geq 2$).

 Also, $\lim_{x \rightarrow \infty} f(x) $
 \small
 \begin{equation}
 = \lim_{x \rightarrow \infty}  \frac{\tau^{1-\alpha}}{1-\alpha}\left[x\left( 1+\frac{\log \left(1 + \frac{\eta}{(x-1)\eta\beta + 1} \right)}{\tau}\right)^{1-\alpha}-(x-1) \right] \nonumber
\end{equation}
\normalsize
 \begin{equation}
 = \lim_{x \rightarrow \infty}  \frac{\tau^{1-\alpha}}{1-\alpha}\left[\frac{\left(\left( 1+\frac{\log \left(1 + \frac{\eta}{(x-1)\eta\beta + 1} \right)}{\tau}\right)^{1-\alpha}-1\right)}{\frac{1}{x}}+1 \right] 
\end{equation}
Using L'Hopital's rule,
\begin{equation}
\label{lh1}
\lim_{x \rightarrow \infty} f(x)  =  \frac{\tau^{-\alpha}(1+\beta\tau-\alpha)}{\beta(1-\alpha)}.
\end{equation}
Now, let
\begin{equation}
\label{bt1}
\beta > \frac{1-\alpha}{\frac{\left(\tau+\log \left(1 + \eta \right)\right)^{1-\alpha}}{\tau^{-\alpha}}-\tau}.
\end{equation}
From~\eqref{lh1} and~\eqref{bt1}, it is easy to show that for $\alpha \in [0, \infty) \backslash \{1\}$:
\begin{equation}
\lim_{x \rightarrow \infty} f(x) < f(1) .
\label{fx2}
\end{equation}
Now, consider the sublevel set:
\[
S = \{x > 1 | f(x) < f(1) \}. 
\]
By \eqref{fx2}, there exists $x_0 > 1$ such that $x \in S$ for all $x > x_0$. Also, clearly $1 \in S$. Since $f(\cdot)$ is quasi-convex, the set $S$ is convex~\cite{boyd}; so $x \in S$ for all $x > 1$. That is, when~\eqref{pro},~\eqref{condn44},~\eqref{condnn3} and~\eqref{bt1} hold (respectively, \eqref{condnn1} and~\eqref{bt1} hold) for $\alpha \geq 2$ (respectively, $\alpha \in [0,2) \backslash \{1\}$), then $f(x) < f(1)$ for all $x > 1$ and the result follows.

\end{IEEEproof}
%\textcolor{blue}{

\begin{IEEEproof}[\textbf{B. Proof of Lemma~\ref{l2}}] 

First, we will show that the function $f_1(x)$ is quasi-convex on the domain $x \geq 1$.
Let 
\begin{equation}
\label{parachange1}
y=(x-1)\beta+ \frac{1}{\eta}.
\end{equation}
 Then, 
 \begin{equation}
\label{fg1}
f_1(x)=g_1(y).
\end{equation}
 where $g_1(y)=$
 \begin{equation}
  \left(\frac{y}{\beta}+1-\frac{1}{\eta\beta}\right)\log\left(\tau+\log \left( 1 + \frac{1}{y} \right)\right)-(\frac{y}{\beta}-\frac{1}{\eta\beta})\log\tau. \nonumber
  \end{equation} 
Now, $g_1'(y)= \frac{1}{\beta}\left[ \log p-\frac{y+\beta-\frac{1}{\eta}}{y(y+1)p}-\log \tau \right],$ \\
where, $p= \left(\tau+\log \left( 1 + \frac{1}{y} \right)\right).$\\
\begin{equation}
\hspace{-2.5cm} g_1'(y)=0 \Leftrightarrow y+\beta-\frac{1}{\eta}= y(y+1)p \log \frac{p}{\tau}.
\label{gdesh2}
\end{equation}
Now, $g_1''(y)$\\
$=\frac{1}{\beta}\left[-\frac{1}{y(y+1)p}-\left( \frac{y(y+1)p-(y+\beta-\frac{1}{\eta})((2y+1)p-1)}{(y(y+1)p)^{2}}\right) \right]$\\
 \begin{equation}
=\frac{1}{\beta (y(y+1)p)^{2}}(-yp-y+2yp(\beta-\frac{1}{\eta})+(p-1)(\beta-\frac{1}{\eta})).\nonumber
\end{equation}
So $g_1''(y) > 0$\\
$ \Leftrightarrow py(2(\beta-\frac{1}{\eta})-1)-(y+\beta-\frac{1}{\eta})+p(\beta-\frac{1}{\eta})) > 0.$\\
Substituting from~\eqref{gdesh2} in the above inequality, we get:\\
$py(2(\beta-\frac{1}{\eta})-1)-y(y+1)p \log (1+\frac{\log(1+\frac{1}{y})}{\tau})+p(\beta-\frac{1}{\eta})) > 0$ \\
As $\log(1+x) \leq x \;\;\forall x \geq -1$~\cite{Logineq}, a sufficient condition for the above inequality to hold is\\
$y(2(\beta-\frac{1}{\eta})-1)-y(y+1)\frac{1}{y\tau}+(\beta-\frac{1}{\eta})) > 0.$ \\
$\Leftrightarrow y(2\tau(\beta-\frac{1}{\eta})-1-\tau)-1+\tau(\beta-\frac{1}{\eta})) > 0.$ \\
A sufficient condition for the above inequality to hold is\\
\begin{equation}
\label{condnn2_1}
\beta \geq \frac{1}{\tau}+\frac{1}{\eta}\;\; \mbox{and} \; \; \tau < 1.
\end{equation}
 Hence, under the condition in~\eqref{condnn2_1}, $g_1''(y)> 0$ when $g_1'(y) = 0$ for $\alpha =1.$ Therefore, it follows from Property~\ref{p1} that $g_1(.)$ is quasi-convex.

Now, it follows from~\eqref{parachange1} and~\eqref{fg1} that $f_1'(x)=\beta g_1'(y)$ and $f_1''(x)=\beta^2 g_1''(y).$ Therefore, $f_1(.)$ also satisfies the condition in Property~\ref{p1} whenever $g_1(.)$ satisfies it. Hence, $f_1(.)$ is quasi-convex when \eqref{condnn2_1} holds.

 Also, $\lim_{x \rightarrow \infty} f_1(x)  $
\begin{equation}
= \lim_{x \rightarrow \infty} x\log \left[\tau \left(1+ \frac{\log(1+\frac{\eta}{(x-1)\eta\beta+1})}{\tau} \right) \right]-x\log \tau+\log \tau \nonumber 
\end{equation}
\begin{equation}
 \hspace{-1.7cm} = \lim_{x \rightarrow \infty} x\log \left(1+ \frac{\log(1+\frac{\eta}{(x-1)\eta\beta+1})}{\tau} \right) +\log \tau \nonumber 
\end{equation}
\begin{equation}
 \hspace{-2.5cm} = \lim_{x \rightarrow \infty} \frac{\log \left(1+ \frac{\log(1+\frac{\eta}{(x-1)\eta\beta+1})}{\tau} \right)}{\frac{1}{x}} +\log \tau .
\end{equation}
Using L'Hopital's rule,
\begin{equation}
\lim_{x \rightarrow \infty} f_1(x)  =  \frac{1}{\beta\tau}+\log \tau .
\label{lh} 
\end{equation}
%We want to show that $f(x) \leq f(1) \; \forall x >1$ therefore,\\
%$\lim_{x \rightarrow \infty} f(x)  =  \frac{1}{\beta\tau}+\log \tau \leq f(1)$\\ $\Leftrightarrow \frac{1}{\beta\tau}+\log \tau \leq \log(\tau + \log(1+\eta))$
Now, let
\begin{equation}
\label{fcond_1}
 \beta > \frac{1}{\tau \log(1+ \frac{\log(1+\eta)}{\tau})}.
\end{equation}
From~\eqref{lh} and~\eqref{fcond_1}, it is easy to show that 
\begin{equation}
\label{fx1}
\lim_{x \rightarrow \infty} f_1(x)  < f_1(1).
\end{equation}
Now, consider the sublevel set:
\[
S = \{x > 1 | f_1(x) < f_1(1) \}. 
\]
By \eqref{fx1}, there exists $x_0 > 1$ such that $x \in S$ for all $x > x_0$. Also, clearly $1 \in S$. Since $f_1(\cdot)$ is quasi-convex, the set $S$ is convex~\cite{boyd}; so $x \in S$ for all $x > 1$. That is, when~\eqref{condnn2_1} and~\eqref{fcond_1} hold, then $f_1(x) < f_1(1)$ for all $x > 1$ and the result follows.\\
% From~\eqref{condnn2_1} and~\eqref{fcond_1}, the final condition for $f(x) \leq f(1) \; \forall x \geq 1$ for $\alpha=1$ is:
% \begin{equation}
% \label{ffcond_1}
%\beta \geq \max\left(\frac{1}{\eta}+ \frac{1}{\tau}, \frac{1}{\tau \log(1+ \frac{\log(1+\eta)}{\tau})}\right)\;\; \mbox{and}\;\; \tau<1.
% \end{equation} 
\end{IEEEproof}

\bibliographystyle{IEEEtran}
\bibliography{ref}

\begin{thebibliography}{10}
\providecommand{\url}[1]{#1}
\csname url@samestyle\endcsname
\providecommand{\newblock}{\relax}
\providecommand{\bibinfo}[2]{#2}
\providecommand{\BIBentrySTDinterwordspacing}{\spaceskip=0pt\relax}
\providecommand{\BIBentryALTinterwordstretchfactor}{4}
\providecommand{\BIBentryALTinterwordspacing}{\spaceskip=\fontdimen2\font plus
\BIBentryALTinterwordstretchfactor\fontdimen3\font minus
  \fontdimen4\font\relax}
\providecommand{\BIBforeignlanguage}[2]{{%
\expandafter\ifx\csname l@#1\endcsname\relax
\typeout{** WARNING: IEEEtran.bst: No hyphenation pattern has been}%
\typeout{** loaded for the language `#1'. Using the pattern for}%
\typeout{** the default language instead.}%
\else
\language=\csname l@#1\endcsname
\fi
#2}}
\providecommand{\BIBdecl}{\relax}
\BIBdecl

\bibitem{netgcoop}
V.~K. Gupta and G.~S. Kasbekar, ``Achieving arbitrary throughput--fairness
  trade-offs in the inter-cell interference coordination with fixed transmit
  power problem,'' in \emph{Network Games, Control, and Optimization}.\hskip
  1em plus 0.5em minus 0.4em\relax Cham: Springer International Publishing,
  2019, pp. 17--35.

\bibitem{RF:ghosh:fundamentals:of:lte}
A.~Ghosh, J.~Zhang, J.~G. Andrews, and R.~Muhamed, \emph{Fundamentals of LTE},
  1st~ed.\hskip 1em plus 0.5em minus 0.4em\relax Upper Saddle River, NJ, USA:
  Prentice Hall Press, 2010.

\bibitem{RF:Kosta}
C.~{Kosta}, B.~{Hunt}, A.~U. {Quddus}, and R.~{Tafazolli}, ``On interference
  avoidance through inter-cell interference coordination (icic) based on ofdma
  mobile systems,'' \emph{IEEE Communications Surveys Tutorials}, vol.~15,
  no.~3, pp. 973--995, Third 2013.

\bibitem{RF:eICIC:lopezperez}
D.~{Lopez-Perez}, I.~{Guvenc}, G.~{de la Roche}, M.~{Kountouris}, T.~Q.~S.
  {Quek}, and J.~{Zhang}, ``Enhanced intercell interference coordination
  challenges in heterogeneous networks,'' \emph{IEEE Wireless Communications},
  vol.~18, no.~3, pp. 22--30, June 2011.

\bibitem{Andrews20141065}
J.~G. {Andrews}, S.~{Buzzi}, W.~{Choi}, S.~V. {Hanly}, A.~{Lozano}, A.~C.~K.
  {Soong}, and J.~C. {Zhang}, ``What will 5g be?'' \emph{IEEE Journal on
  Selected Areas in Communications}, vol.~32, no.~6, pp. 1065--1082, June 2014.

\bibitem{BinSediq2}
A.~{Bin Sediq}, R.~{Schoenen}, H.~{Yanikomeroglu}, and G.~{Senarath},
  ``Optimized distributed inter-cell interference coordination (icic) scheme
  using projected subgradient and network flow optimization,'' \emph{IEEE
  Transactions on Communications}, vol.~63, no.~1, pp. 107--124, Jan 2015.

\bibitem{Report2017}
V.~K. {Gupta}, A.~{Nambiar}, and G.~S. {Kasbekar}, ``Complexity analysis,
  potential game characterization and algorithms for the inter-cell
  interference coordination with fixed transmit power problem,'' \emph{IEEE
  Transactions on Vehicular Technology}, vol.~67, no.~4, pp. 3054--3068, April
  2018.

\bibitem{Kosta2012}
C.~{Kosta}, B.~{Hunt}, A.~U. {Quddus}, and R.~{Tafazolli}, ``A low-complexity
  distributed inter-cell interference coordination (icic) scheme for emerging
  multi-cell hetnets,'' in \emph{2012 IEEE Vehicular Technology Conference (VTC
  Fall)}, Sep. 2012, pp. 1--5.

\bibitem{RF:Rahman}
M.~{Rahman} and H.~{Yanikomeroglu}, ``Enhancing cell-edge performance: a
  downlink dynamic interference avoidance scheme with inter-cell
  coordination,'' \emph{IEEE Transactions on Wireless Communications}, vol.~9,
  no.~4, pp. 1414--1425, April 2010.

\bibitem{RF:wei:Yassin}
M.~YASSIN, ``Inter-cell interference coordination in wireless networks,'' Ph.D.
  dissertation, 11 2015.

\bibitem{mert}
M.~Yagcioglu and O.~Bayat, ``Next generation dynamic inter-cellular
  scheduler,'' \emph{International Journal of Electronics and
  Telecommunications}, vol. vol. 65, no. No 3, pp. 441--448, 2019.

\bibitem{HTch}
H.~T. {Cheng} and W.~{Zhuang}, ``An optimization framework for balancing
  throughput and fairness in wireless networks with qos support,'' \emph{IEEE
  Transactions on Wireless Communications}, vol.~7, no.~2, pp. 584--593,
  February 2008.

\bibitem{BinSediq}
A.~B. {Sediq}, R.~H. {Gohary}, and H.~{Yanikomeroglu}, ``Optimal tradeoff
  between efficiency and jain's fairness index in resource allocation,'' in
  \emph{2012 IEEE 23rd International Symposium on Personal, Indoor and Mobile
  Radio Communications - (PIMRC)}, Sep. 2012, pp. 577--583.

\bibitem{Jain}
\BIBentryALTinterwordspacing
R.~Jain, D.-M. Chiu, and W.~Hawe, ``A quantitative measure of fairness and
  discrimination for resource allocation in shared computer systems,''
  \emph{CoRR}, vol. cs.NI/9809099, 1998. [Online]. Available:
  \url{http://dblp.uni-trier.de/db/journals/corr/corr9809.html\#cs-NI-9809099}
\BIBentrySTDinterwordspacing

\bibitem{Sheikh}
S.~{Sheikh}, R.~{Wolhuter}, and H.~A. {Engelbrecht}, ``An adaptive congestion
  control and fairness scheduling strategy for wireless mesh networks,'' in
  \emph{2015 IEEE Symposium Series on Computational Intelligence}, Dec 2015,
  pp. 1174--1181.

\bibitem{Jmo}
J.~{Mo} and J.~{Walrand}, ``Fair end-to-end window-based congestion control,''
  \emph{IEEE/ACM Transactions on Networking}, vol.~8, no.~5, pp. 556--567, Oct
  2000.

\bibitem{Kelly}
F.~Kelly, A.~Maulloo, and D.~Tan, ``Rate control for communication
  networks:shadow prices, proportional fairness and stability,'' \emph{Journal
  of the Operational Research Society}, vol.~49, 02 1998.

\bibitem{Gall}
D.~Bertsekas and R.~Gallager, \emph{Data Networks (2Nd Ed.)}.\hskip 1em plus
  0.5em minus 0.4em\relax Upper Saddle River, NJ, USA: Prentice-Hall, Inc.,
  1992.

\bibitem{Tlan}
T.~{Lan}, D.~{Kao}, M.~{Chiang}, and A.~{Sabharwal}, ``An axiomatic theory of
  fairness in network resource allocation,'' in \emph{2010 Proceedings IEEE
  INFOCOM}, March 2010, pp. 1--9.

\bibitem{RF:kleinberg:algorithm}
J.~Kleinberg and E.~Tardos, \emph{Algorithm Design}.\hskip 1em plus 0.5em minus
  0.4em\relax Boston, MA, USA: Addison-Wesley Longman Publishing Co., Inc.,
  2005.

\bibitem{yoon}
J.~{Yoon} and G.~{Hwang}, ``Distance-based inter-cell interference coordination
  in small cell networks: Stochastic geometry modeling and analysis,''
  \emph{IEEE Transactions on Wireless Communications}, vol.~17, no.~6, pp.
  4089--4103, June 2018.

\bibitem{Kim}
S.~{Kim}, H.~{Jwa}, J.~{Moon}, and J.~{Na}, ``Achieving fair cell-edge
  performance: Low-complexity interference coordination in ofdma networks,'' in
  \emph{2018 20th International Conference on Advanced Communication Technology
  (ICACT)}, Feb 2018, pp. 6--11.

\bibitem{jiang}
L.~{Jiang} and R.~{Song}, ``A low-complexity resource allocation scheme for
  ofdma multicast systems with proportional fairness,'' \emph{China
  Communications}, vol.~15, no.~1, pp. 1--11, Jan 2018.

\bibitem{Xin}
X.~{Ge}, H.~{Jin}, and V.~{C. M. Leung}, ``Joint opportunistic user scheduling
  and power allocation: throughput optimisation and fair resource sharing,''
  \emph{IET Communications}, vol.~12, no.~5, pp. 634--640, 2018.

\bibitem{Shen}
Y.~{Shen}, X.~{Huang}, B.~{Yang}, S.~{Gong}, and S.~{Wang}, ``Fair resource
  allocation algorithm for chunk based ofdma multi-user networks,'' in
  \emph{2017 IEEE 86th Vehicular Technology Conference (VTC-Fall)}, Sep. 2017,
  pp. 1--5.

\bibitem{Miki}
N.~{Miki}, Y.~{Kanehira}, and H.~{Tokoshima}, ``Investigation on joint
  optimization for user association and inter-cell interference coordination
  based on proportional fair criteria,'' in \emph{2017 11th International
  Conference on Signal Processing and Communication Systems (ICSPCS)}, Dec
  2017, pp. 1--6.

\bibitem{pastore}
A.~{Pastore} and M.~{Navarro}, ``A fairness–throughput tradeoff perspective
  on noma multiresolution broadcasting,'' \emph{IEEE Transactions on
  Broadcasting}, vol.~65, no.~1, pp. 179--186, March 2019.

\bibitem{mik}
N.~Miki and Y.~Kanehira, ``Investigation on distributed optimization of user
  association and inter-cell interference coordination based on proportional
  fair criteria,'' 01 2019, pp. 1--5.

\bibitem{xhuang}
X.~Huang, D.~Zhang, S.~Tang, Q.~Chen, and J.~Zhang, ``Fairness-based
  distributed resource allocation in two-tier heterogeneous networks,''
  \emph{IEEE Access}, vol.~PP, pp. 1--1, 03 2019.

\bibitem{shahsavari}
\BIBentryALTinterwordspacing
S.~Shahsavari, N.~Akar, and B.~H. Khalaj, ``Joint cell muting and user
  scheduling in multi-cell networks with temporal fairness,'' \emph{Wireless
  Communications and Mobile Computing}, vol. 2018, 2018. [Online]. Available:
  \url{https://doi.org/10.1155/2018/4846291}
\BIBentrySTDinterwordspacing

\bibitem{jin}
Y.~{Jin} and M.~{Hayashi}, ``Trade-off between fairness and efficiency in
  dominant alpha-fairness family,'' in \emph{IEEE INFOCOM 2018 - IEEE
  Conference on Computer Communications Workshops (INFOCOM WKSHPS)}, April
  2018, pp. 391--396.

\bibitem{Rapp}
T.~Rappaport, \emph{Wireless Communications: Principles and Practice},
  2nd~ed.\hskip 1em plus 0.5em minus 0.4em\relax Upper Saddle River, NJ, USA:
  Prentice Hall PTR, 2001.

\bibitem{Tse}
D.~Tse and P.~Viswanath, \emph{Fundamentals of Wireless Communication}.\hskip
  1em plus 0.5em minus 0.4em\relax New York, NY, USA: Cambridge University
  Press, 2005.

\bibitem{hung}
C.~H. Papadimitriou and K.~Steiglitz, \emph{Combinatorial Optimization:
  Algorithms and Complexity}.\hskip 1em plus 0.5em minus 0.4em\relax Upper
  Saddle River, NJ, USA: Prentice-Hall, Inc., 1982.

\bibitem{boyd}
S.~Boyd and L.~Vandenberghe, \emph{Convex Optimization}.\hskip 1em plus 0.5em
  minus 0.4em\relax New York, NY, USA: Cambridge University Press, 2004.

\bibitem{Mitri}
\BIBentryALTinterwordspacing
D.~S. Mitrinovi{\'{c}} and J.~E. Pe{\v{c}}ari{\'{c}}, ``Bernoulli's
  inequality,'' \emph{Rendiconti del Circolo Matematico di Palermo}, vol.~42,
  no.~3, pp. 317--337, Oct 1993. [Online]. Available:
  \url{https://doi.org/10.1007/BF02844624}
\BIBentrySTDinterwordspacing

\bibitem{Logineq}
E.~Love, ``64.4 some logarithm inequalities,'' \emph{The Mathematical Gazette},
  vol.~64, no. 427, p. 55–57, 1980.

\end{thebibliography}

\end{document}